\documentclass[prb,twocolumn,groupeaddress]{revtex4-1}

\usepackage[latin1]{inputenc}

\usepackage{amsmath,amssymb,amsthm}
\usepackage{xspace}
\usepackage{graphicx}
\usepackage{grffile}
\usepackage{nicefrac}
\usepackage{color}

\usepackage{braket}

\usepackage{pgfplots}

\usepackage{tikz}
\usetikzlibrary{arrows, chains, matrix,
	        scopes, fit, calc, decorations,
	        shapes, decorations.pathreplacing, shadows, external,
                trees, mindmap, backgrounds, tikzmark }

\usepgfplotslibrary{groupplots}

\pgfplotsset{compat=newest}

\usepackage[colorinlistoftodos,prependcaption,textsize=tiny]{todonotes}

\newcommand{\myvec}[1]{{\textbf{#1}}}

\newcommand{\sumBeta}{\sum_\beta}

\newcommand{\tg}{$t_{2g}$}
\newcommand{\eg}{$e_{g}$}

\tikzset{%
  highlight/.style={rectangle,rounded corners,fill=red!15,draw,fill opacity=0.5,thick,inner sep=0pt}
}

\newcommand{\varphimat}{\phi}

\begin{document}

\title{Rigorous symmetry adaptation of multiorbital rotationally invariant 
slave-boson theory with application to Hund's rules physics}

\author{Christoph Piefke}
\affiliation{I. Institut f{\"u}r Theoretische Physik, Universit{\"a}t Hamburg, 
D-20355 Hamburg, Germany}
\author{Frank Lechermann}
\affiliation{I. Institut f{\"u}r Theoretische Physik, Universit{\"a}t Hamburg, 
D-20355 Hamburg, Germany}


\begin{abstract}
The theory of correlated electron systems on a lattice proves 
notoriously complicated because of the exponential growth of Hilbert space. Mean-field approaches 
provide valuable insight when the self-energy has a dominant local structure. Additionally,
the extraction of effective low-energy theories from the generalized many-body representation
is highly desirable. In this respect, the rotational-invariant slave boson (RISB) approach in 
its mean-field formulation enables versatile access to correlated lattice problems. 
However in its original form, due to numerical complexity, the RISB approach is limited to 
about three correlated orbitals per lattice site.
We thus present a thorough symmetry-adapted advancement of RISB theory, suited to 
efficiently deal with multi-orbital Hubbard Hamiltonians for complete atomic-shell manifolds. 
It is utilized to study the intriguing problem of Hund's physics for three- and especially 
five-orbital manifolds on the correlated lattice, including crystal-field terms as well as 
spin-orbit interaction. The well-known Janus-face phenomenology, i.e. strenghtening of 
correlations at smaller-to-intermediate Hubbard $U$ accompanied by a shift of the Mott 
transition to a larger $U$ value, has a stronger signature and more involved multiplet 
resolution for five-orbital problems. Spin-orbit interaction effectively reduces the critical 
local interaction strength and weakens the Janus-face behavior. Application to the realistic
challenge of Fe chalcogenides underlines the subtle interplay of the orbital degrees of freedom
in these materials.
\end{abstract}

\maketitle

\section{Introduction}

The dichotomy of higher-energy localization and lower-energy itinerancy poses a key 
challenge of correlated electron systems on lattices with spatial dimension $dim>1$. To cope with 
this problem on general grounds, many-body theory has to cover a large energy window, rendering 
standard perturbation theory or renormalization-group approaches difficult. Integrating out
degrees of freedom is notoriously complex. Only if the physics singles out a certain energy 
scale, e.g. low-energy in the Kondo problem, (numerical) exact theoretical methods become 
available. 

Auxiliary-particle or Gutzwiller-based~\cite{gut63} schemes approach the given problem in a 
simplified, but often qualitatively adequate way, that is especially useful to study 
low-temperature properties of correlated electron systems in the thermodynamic limit. 
Instead of aiming for a complete treatment of the lattice electrons' dichotomy with further 
necessary approximations in real- or reciprocal-space, temperature- or frequency range, 
key focus is on an approximate handling of the two-faced character of the electrons.
A certain protocol for liberating the itinerant from the localized degrees of freedom is common 
to all the various flavors of these schemes. In this work, we concentrate on the auxiliary- or
'slave'-particle methods, but in practise the Gutzwiller frameworks carry in principle the same 
physics~\cite{bue07}, using a different representation/language~\cite{bue98,hua12,lanata12}.

Originally introduced~\cite{bar76} to handle the Anderson model, the slave-boson concept was 
further developed in the context of mixed-valent and Kondo-lattice systems~\cite{col83,rea83}, 
and afterwards has been modified and extended in various directions over the 
years~\cite{kot86,aro88,kot88,li89,has97,fre97,flo04,med05,lec07,isi09,bue11,geo15}. The main 
idea is to distinguish between the localized and the delocalized character of an electron on 
the operator level. In its simplest one-orbital form at infinite local interaction strength 
$U$, one introduces a quasiparticle (QP) fermionic operator $f$ for the itinerant behavior, 
while a bosonic operator $\phi$ takes care of the stricly-local empty state on lattice site $i$. 
Therewith, the physical electron creation operator $c^\dagger$ may be reexpressed, and a 
straightforward  constraint to abandon doubly occupied lattice sites established, i.e.
\begin{equation}
  c^\dagger_{i\sigma}=f_{i\sigma}^\dagger \phi_i^{\hfill} \quad\wedge\quad
  \sum_{i\sigma}f_{i\sigma}^\dagger f_{i\sigma}^{\hfill}
  +\phi_i^\dagger \phi_i^{\hfill}=1\quad,
\label{eq:sbdef}
\end{equation}
whereby $\sigma=\uparrow,\downarrow$ marks the spin projection. This efficient route to select 
the physical states on an interacting lattice can be generalized by various means. Kotliar 
and Ruckenstein~\cite{kot86} increased the number of local bosons to describe finite-$U$ cases. 
Multi-orbital extensions thereof ask for a further increase of the bosonic 
variables~\cite{fre97}. In parallel, there are options to replace/modify the character of the 
auxiliary particle in order to strengthen or focus certain aspects of the correlated electron 
problem. For instance, the slave-rotor method~\cite{flo04} and the slave-spin 
framework~\cite{med05} are two such alternative theories. 

The present work deals with a new efficient realization of the rotational-invariant 
slave-boson (RISB) theory~\cite{li89,lec07,isi09}, an elaborate generalization of the original 
ideas for manifest multi-orbital problems. Rotational invariance in the 
theoretical description is essential to promote the simple one-orbital empty-state bosonic 
degree of freedom to an object that can address the intricate multiplet structure of a local 
quantum-chemical entity in full generality. {\sl As well as} its general coupling to the 
$k$-dependent quasiparticle degrees of freedom.
The RISB framework handles these issues properly and it has been successfully utilized to 
study various correlated condensed matter problems, both on the model level and using 
realistic dispersions in the context of concrete materials. Namely, applications to 
multi-orbital Mott transitions~\cite{lec07,fac17}, quasi-twodimensional 
lattices~\cite{rac06,lec09,fer092,maz14}, spin-orbit and related anisotropic 
interactions~\cite{beh12,schu12,ham15}, 3-orbital Hund's physics~\cite{beh15,fac17},  
multi-orbital superconductivity~\cite{isi09}, and real-space defect problems~\cite{beh15_2} 
were performed. In addition, inspired by a Gutzwiller-based scheme~\cite{schi10}, 
a time-dependent extension to address non-equilibrium electronic correlations 
has also been put forward~\cite{beh13,beh15,beh16}.

However the number of auxiliary bosons grows exponentially with the number of orbitals, as 
expected for a method coping with a faithful coverage of the generic quantum Hilbert space. 
Therefore, we here report an advancement of the RISB framework that makes rigorous
use of the various detailed symmetries of the lattice problem at hand. This allows us to
perform full transition-metal $d$-shell investigations with general Coulomb interactions
and including crystal fields and spin-orbit coupling. Moreover, the methodology may be 
combined with density functional theory (DFT) in a charge self-consistent manner. We apply 
the generalized scheme to study the prominent Hund's physics~\cite{wer08,hau09,med11,geo13} 
in 3- and 5-orbital model Hamiltonians and provide results within the materials context 
of FeSe and FeTe.

The paper is organized as follows. In the next section~\ref{sec:basics} the basic principles of
RISB are introduced for the canonical single-band case, in line with a brief classification of
the formalism in view of other many-body techniques. Section~\ref{sec:modham} discusses the
characteristics of the available multi-orbital model Hamiltonians. The technical 
section~\ref{sec:multsb} presents the symmetry-adapted RISB extension to many orbitals. A
compendious account of combining RISB with DFT to approach realistic
systems is given in section~\ref{sec:dftrisb}. Finally, section~\ref{sec:results} deals with 
the selected applications to model- and materials problems in the broader context of Hund's 
physics. 

\section{Brief Survey of\\
 Rotational-invariant Slave-Boson Theory\label{sec:basics}}
To set the stage, we first provide a short overview about the key methodological steps of the
RISB approach on the basis of the canonical single-band Hubbard model. This serves the goal
to introduce the principles of the theory, which is generalized to the symmetry-adapted 
multi-orbital case in section~\ref{sec:multsb}. Additionally, this then allows us to discuss 
the important mean-field (or saddle-point) approximation as well as comparisons to other
many-body schemes. For a general, more
detailed introduction to RISB see Ref.~\onlinecite{lec07}.

\subsection{Single-Orbital Formalism}
The Hamiltonian for the single-band Hubbard model with
nearest-neighbor hopping $t$ and local Coulomb repulsion $U$ reads~\footnote{In order not
to overburden the notation, we do not mark quantum operators explicitly. The difference to 
c-numbers should be obvious from the context, or is mentioned explicitly when needed.} 
\begin{equation}
  \mathcal{H}= -t\sum_{\ij\sigma}\, c^\dagger_{i \sigma}c^{\hfill}_{j
    \sigma}\, + U\sum_{i} n_{i \uparrow} n_{i \downarrow}\equiv
  \mathcal{H}^{\rm (kin)}+\sum_i\mathcal{H}^{\rm
    (loc)}_i\;, \label{eq:singham}
\end{equation}
with $i,j$ labeling lattice sites. For the following, the details of the
crystal lattice are irrelevant, and we assume a Bravais lattice in
spatial dimension $dim>1$.~\footnote{Onedimensional physics is
  in principle also accessible via RISB by invoking a cluster
  description (see section~\ref{sec:mf}). Yet within the scope of the present work, we are not 
  discussing such variations of the method.}

On a lattice site $i$, the four possible electron states are given by
\begin{equation}
\mathcal{A}=\{|E\rangle, |S_\downarrow\rangle,
  |S_\uparrow\rangle,|D\rangle\}\quad, \label{eqn:thefourstates}
\end{equation}
i.e., an empty site $|E\rangle$, a site $\sigma$ $|S_\sigma \rangle$
occupied by a single electron with spin projection and a site $|D\rangle$
occupied by two electrons of opposite spin, are represented in RISB in full 
generality through acting on the vacuum state $|{\rm vac}\rangle$ as follows
\begin{align}
  |E\rangle = | 0 \rangle = \phi_E^{\dagger}\,&|{\rm vac}\rangle\;,\\
  |S_\downarrow\rangle = |\downarrow\rangle =
  \frac{1}{2}\left\{\phi_{\downarrow\uparrow}^{\dagger}\,f_{\uparrow}^{\dagger}+
  \phi_{\downarrow\downarrow}^{\hfill}\,f_{\downarrow}^{\dagger}\right\}\,&|{\rm vac}\rangle
  \label{eq:spdn}\;,\\
  |S_\uparrow\rangle = |\uparrow\rangle = 
  \frac{1}{2}\left\{\phi_{\uparrow\uparrow}^{\dagger}\,f_{\uparrow}^{\dagger}+
  \phi_{\uparrow\downarrow}^{\hfill}\,f_{\downarrow}^{\dagger}\right\}\,&|{\rm vac}\rangle 
  \label{eq:spup}\;,\\
  |D\rangle = |\uparrow\downarrow\rangle = \phi_D^{\dagger}\,&|{\rm vac}\rangle\;.
\end{align}

Thus, the method introduces two fermionic QP operators and six bosonic
operators on every site, i.e.
\begin{equation}
  \mbox{site}\;i:\qquad
  f_\downarrow\;,\;f_\uparrow\;\;;\;\;\phi_E^{\hfill}\;,\;
  \left(\begin{array}{cc} \phi_{\downarrow\downarrow} &
    \phi_{\uparrow\downarrow} \\
    \phi_{\downarrow\uparrow} &
    \phi_{\uparrow\uparrow} \end{array}\right)\;,\;
  \phi_D^{\hfill}\;.
\end{equation}
The second index on the single-particle bosons refers to a QP degree
of freedom, whereas the first index is generally associated with the
local state. The physics of this higher-dimensional bosonic-operator
structure may be read off from eqns.~(\ref{eq:spdn},\ref{eq:spup}). A
low-energy QP excitation is not necessarily only connected to its
spin-{\sl identical} high-energy local counterpart, but may also
connect to other local configurations. In this simple case a state
with opposite spin configuration. This general structure renders it
possible to account for full rotational invariance in the
description.

Choosing these four slave-boson operators in the one-particle sector, 
derived by physical intuition, already accounts for a given symmetry of 
the system: particle-number symmetry is included, there exists no
slave-boson operator that mixes two states with different
number of particles. In fact, since we aim for a matrix-based formulation,
these four objects can be organized in a
larger slave-boson operator matrix on the set $\mathcal{A}$, i.e.,

\tikzset{external/export next=false}

\pgfdeclarelayer{background}
\pgfsetlayers{background,main}

\begin{equation}
  \Phi =
  \begin{tikzpicture}[baseline=(m.center)]
    \matrix[matrix of math nodes, left delimiter = (, right delimiter = ), row sep = 2pt, column sep = 2pt] (m)
           {
             \phi_E & 0 & 0 & 0 \\
             0 & \phi_{\downarrow \downarrow} & \phi_{\uparrow \downarrow} & 0 \\
             0 & \phi_{\downarrow \uparrow} & \phi_{\uparrow \uparrow} & 0 \\
             0 & 0 & 0 & \phi_D \\
           };
           \node[] at ($(m-1-1.north)+(0pt,16pt)$) {$E$};
           \node[] at ($(m-1-2.north)+(0pt,16pt)$) {$S_\downarrow$};
           \node[] at ($(m-1-3.north)+(0pt,16pt)$) {$S_\uparrow$};
           \node[] at ($(m-1-4.north)+(0pt,16pt)$) {$D$};
 
           \node[anchor=center, text width=1cm, align=center] at ($(m-1-4.east)+(30pt,0pt)$) {$E_{\phantom{\downarrow}}$};
           \node[anchor=center, text width=1cm, align=center] at ($(m-2-4.east)+(30pt,0pt)$) {$S_\downarrow$};
           \node[anchor=center, text width=1cm, align=center] at ($(m-3-4.east)+(30pt,0pt)$) {$S_\uparrow$};
           \node[anchor=center, text width=1cm, align=center] at ($(m-4-4.east)+(27pt,0pt)$) {$D_{\phantom{\downarrow}}$};

           \begin{pgfonlayer}{background}
             \draw[rounded corners,dotted,fill=red!50!white,fill opacity=0.1] (m-1-1.north west) rectangle (m-1-1.south east) {};
           \end{pgfonlayer}
           \begin{pgfonlayer}{background}
             \draw[rounded corners,dotted,fill=blue!50!white,fill opacity=0.1] (m-2-2.north west) rectangle (m-3-3.south east) {};
           \end{pgfonlayer}
           \begin{pgfonlayer}{background}
             \draw[rounded corners,dotted,fill=green!50!white,fill opacity=0.1] (m-4-4.north west) rectangle (m-4-4.south east) {};
           \end{pgfonlayer}
  \end{tikzpicture}
\end{equation}

Note that the matrix $\Phi$ is block-diagonal in
particle numbers with (as color coded) zero-particle sector
\tikzset{external/export next=false}
\begin{tikzpicture}{baseline={(current bounding box.center)}}
  \draw[dotted,rounded corners,fill=red!50!white,fill opacity=0.1] (0, 0) rectangle (9pt,9pt);
\end{tikzpicture}
one-particle sector
\tikzset{external/export next=false}
\begin{tikzpicture}{baseline={(current bounding box.center)}}
  \draw[dotted,rounded corners,fill=blue!50!white,fill opacity=0.1] (0, 0) rectangle (9pt,9pt);
\end{tikzpicture}, and 
two-particle sector
\tikzset{external/export next=false}
\begin{tikzpicture}{baseline={(current bounding box.center)}}
  \draw[dotted,rounded corners,fill=green!50!white,fill opacity=0.1] (0, 0) rectangle (9pt,9pt);
\end{tikzpicture}. Its matrix elements $\phi_{AB}$ with $A,B \in \mathcal {A}$ are labeled by all 
available local states.
Already here, be aware that troughout this work, we exclude the possibility for pairing 
instabilites and therefore do not couple different particle sectors by slave bosons. For a RISB
representation describing superconductivity, we refer to Ref.~\onlinecite{isi09}.

In order to select the true complete physical states, composed of slave bosons and quasiparticles,
the constraints
\begin{eqnarray}
1&=&\phi_E^\dagger\phi_E^{\hfill}+\sum_{\sigma\sigma'}
\phi^\dagger_{\sigma\sigma'}\phi_{\sigma'\sigma}^{\hfill}+
\phi^\dagger_D\phi_D^{\hfill}\label{eq:con1}\\ f^\dagger_\sigma
f_\sigma^{\hfill}&=&\phi_D^\dagger\phi_D^{\hfill}
+\sum_{\sigma'}\phi^\dagger_{\sigma\sigma'}\phi_{\sigma'\sigma}^{\hfill}\label{eq:con2}\\ 
f^\dagger_{\sigma}f_{\bar{\sigma}}^{\hfill}&=&
\sum_{\sigma'}\phi^\dagger_{\bar{\sigma}\sigma'}\phi_{\sigma'\sigma}^{\hfill}\label{eq:con3}
\end{eqnarray}
have to be enforced on each site $i$, whereby $\bar{\sigma}$ denotes the opposite spin projection
to $\sigma$. The full electron operator is expressed through
\begin{eqnarray}
\underline{c}_{i\sigma}^\dagger&=&\frac{1}{\sqrt{2}}
\sum_{\sigma'}\left\{\phi_{i\sigma\sigma'}^\dagger\phi_{iE}^{\hfill}-
(-1)^{\delta_{\sigma\sigma'}}\phi^\dagger_{iD}\phi_{i\bar{\sigma}\bar{\sigma}'}^{\hfill}\right\}
\,f_{i\sigma'}^\dagger\nonumber \\ &\equiv&\sum_{\sigma'}
R_{i\sigma'\sigma}^{\dagger}\,f_{i\sigma'}^\dagger\quad.\label{eq:cop}
\end{eqnarray}
The kinetic Hamiltonian $\mathcal{H}^{\rm (kin)}$ is then readily
written in RISB as
\begin{equation}
  \underline{\mathcal{H}}^{\rm
    (kin)}=-t\sum_{ij}\sum_{\sigma\sigma'\sigma''}
  R_{i\sigma'\sigma}^{\dagger}\,R_{j\sigma\sigma''}^{\hfill}
  \,f_{i\sigma'}^\dagger f_{j\sigma''}^{\hfill}\quad.\label{eq:hkin}
\end{equation}
Depending on the bosons, the $R_i$ matrix relates the QP character to
the full electron excitation on site $i$.
To represent the local Hamiltonian, one uses the key fact that any local operator
$\mathcal{O}$ may be written in quadratic terms of the bosonic degrees
of freedom on the enlarged local Hilbert space, e.g. the four states in 
$\mathcal{A}$ as defined in~(\ref{eqn:thefourstates}). The general RISB form,
with $( A|\mathcal{O}|A' )$ represented as a matrix element in the basis set 
$\mathcal{A}$, is written as
\begin{equation}
\underline{\mathcal{O}}=\sum_{AA'} ( A|\mathcal{O}|A' )
\sum_{n}\phi^\dagger_{nA}\phi^{\hfill}_{A'n}\quad.
\label{eqn:gen-risb-form}
\end{equation}
For the local Hubbard interaction $\mathcal{H}^{\rm (loc)}_i$ on each
site, i.e.  $\underline{\mathcal{O}}=U\underline{n}_{i \uparrow}
\underline{n}_{i \downarrow}$, the slave-boson representation
$\mathcal{H}_{\rm U}=U\phi_D^\dagger\phi_D^{\hfill}$ is readily
obtained. Together with the kinetic part, this completes the RISB
single-band Hubbard Hamiltonian representation
\begin{eqnarray}
  \underline{\mathcal{H}}=-t\hspace*{-0.2cm}\sum_{ij,\sigma\sigma'\sigma''}\hspace*{-0.2cm}
  R_{i\sigma'\sigma}^{\dagger}\,R_{j\sigma\sigma''}^{\hfill}
  \,f_{i\sigma'}^\dagger f_{j\sigma''}^{\hfill}
  +U\sum_i\phi^\dagger_{iD}\phi_{iD}^{\hfill}\;.
\end{eqnarray}
It is noted that as common in usual auxiliary-particle theories, there are
inherent gauge symmetries. This can already be 
illustrated~\cite{wol95,nay00,leenag06} using the most-simplest slave-boson 
introduction from eq. (\ref{eq:sbdef}) by marking the U$(1)$ gauge symmetry
\begin{equation}
\phi_i\rightarrow{\rm e}^{i\theta_i}\phi_i\quad,\qquad
f_{i\sigma}\rightarrow{\rm e}^{i\theta_i}f_{i\sigma}\quad.
\end{equation}
In specific cases, this symmetry may be used to gauge away the phases of the 
bosonic fields~\cite{fre12}. Furthermore within single-orbital RISB on a given 
lattice site, an arbitrary SU$(2)$ rotation of the QP operators provides some 
freedom in the representation of the corresponding QP indices. This hold also 
in the multi-orbital case, but as shown in Ref.~\onlinecite{lec07}, physical 
observables remain of course generally gauge invariant. Let us mention that in 
this regard, Lanat{\`a} {\sl et al.}~\cite{lan17} recently proposed an alternative 
RISB representation.

\subsection{Saddle-Point Approximation and Comparison to other Many-Body Techniques\label{sec:mf}}
Enforcing the constraints eqns. (\ref{eq:con1}-\ref{eq:con3}) on each site in the 
thermodynamic limit, while keeping the operator character of the introduced electronic
degrees of freedom appears unfeasible. Therefore in most cases, slave-boson theory
is actually practised in its simplest non-trivial form, namely within the saddle-point
approximation.
Its realization amounts to three essential steps~\cite{lec07}. First, the bosons are condensed 
to $c$ numbers $\varphi_{An}^{\hfill}\equiv\langle\phi^{\hfill}_{An}\rangle$. Second, the 
constraints are treated on average by introducing Lagrange multipliers in a free-energy 
functional (see below). And third, the representation (\ref{eq:cop}) of the physical electron
operator has to be modified by a proper normalization in order to recover the correct 
non-interacting limit within the given mean-field picture.

With inverse temperature $\beta=1/T$ and transformation of the kinetic part to $k$-space
with dispersion $\varepsilon_{\bf k}$, the saddle-point is obtained from the free-energy 
functional~\cite{lec07}
\begin{eqnarray}
  &&\Omega[\{\varphi^{\hfill}_{An}\};\Lambda,\lambda_0]\,=\,\nonumber\\ 
  &&=-\frac{1}{\beta} \sum_{\bf k} \rm{tr}
  \ln\left[1+e^{-\beta\left(\mathbf{R}^\dagger(\varphi)
      \mathbf{\varepsilon}_{\bf k}\mathbf{R}(\varphi)
      +\Lambda\right)}
    \right]-\lambda_0+\nonumber \\
  &&\hspace*{0.4cm}+\sum_{AA'nn'}\varphi^*_{An'}\,
  \bigl\{\,\delta_{nn'}\delta_{AA'}\,\lambda_0+\delta_{nn'}
  ( A|\mathcal{H}^{\rm{loc}}|A')-\nonumber\\
  &&\hspace*{1.4cm}-\,\delta_{AA'}\sum_{\sigma\sigma'} \Lambda_{\sigma\sigma'}
  ( n|f^\dagger_\sigma f_{\sigma'}^{\hfill} |n')\,\bigr\}\,
  \varphi_{A'n}^{\hfill}\quad,
  \label{eq:gpot}
\end{eqnarray}
through extremalizing over the set $\{\varphi_{An}^{\hfill}\}$ and the Lagrange multipliers. 
Note that $\lambda_0$ is associated with the constraint (\ref{eq:con1}) and $\Lambda$ deals 
with the remaining constraints (\ref{eq:con2}-\ref{eq:con3}).

Importantly, the physical self-energy takes on the form
\begin{equation}
  \mathbf{\Sigma}(\omega)=\omega\left(1-[\mathbf{R R^\dagger}]^{-1}\right)
  +[\mathbf{R}^\dagger]^{-1}\mathbf{\Lambda}\mathbf{R}^{-1}-[\varepsilon^0]\;,
  \label{eq:Sigma_physical}
\end{equation}
with $\varepsilon^0$ as the onsite part of the dispersion. Hence the self-energy
consists of a term linear in frequency as well as a static part, and the local QP weight 
$Z\equiv\left[1-\frac{\partial}{\partial\omega}\Sigma\,\right]_{\omega=0}^{-1}$ 
is here generally given in matrix form via 
${\bf Z}^{\hfill}={\bf R}^{\hfill}{\bf R}^{\dagger}$.  
For the rest of the paper, we will discuss RISB by assuming that the mean-field limit is 
taken in the final equations. In this respect, to simplify notations, we thus also keep the 
$\phi$ notation for the slave bosons throughout the writing and will not furthermore 
highlight the difference to the condensed $\varphi$ quantity.

We now try to classify briefly the performance of RISB on a qualitative level in view of
some other available many-body techniques on the lattice. The self-energy is local, but carries
frequency dependence, contrary to simplest Hartree-Fock for the Hubbard model. The optimal
local many-body theory is given by dynamical mean-field theory (DMFT) (see 
e.g. Refs.~\onlinecite{geo96,kot04} for reviews). This theory describes a most general $\omega$ 
dependence within the context of a mapping of the correlated lattice problem onto the problem
an quantum-impurity residing within a self-consistent bath. The full-frequency information may 
be extracted via e.g. quantum-Monte-Carlo or exact-diagonalization impurity solvers. The 
linear-frequency restriction of RISB allows one to study only Fermi-liquid regimes and 
spectral-weight transfer to e.g. Hubbard bands is not accessible from the spectral function.
Yet importantly, via an inspection of the occupation of the local states, still relevant 
information on the high-energy physics may be obtained. In principle, the slave-boson
approach can also be interpreted as a simplified impurity solution within DMFT, since the 
RISB formalism may also be implemented within a quantum-impurity-in-bath scope~\cite{fer092}.
Let us note here, that there are also differences concerning the level of approximation
among the different available auxiliary-particle schemes, since this often 
raises some confusion. For instance, interacting 5-orbital Hamiltonians may be straightforwardly 
encountered by mean-field representations of Kotliar-Ruckenstein slave bosons~\cite{kot86} or  
slave spins~\cite{med05}, since those approaches enable a simplified treatment of the problem. In 
the more elaborate RISB method, such large orbital manifolds asks for special adavancements, 
as discussed in the present work.

Going beyond the local-self-energy concept is generally tough and we here do not want to enter
this branch of many-body theory by general means. Let us note that there are rare slave-boson 
formulations beyond the mean-field limit, usually by invoking Gaussian fluctuations around 
the saddle-point (see e.g. Refs.~\onlinecite{rai93,zim97}) in selected single-orbital problems. Such 
formulations are capable of describing $k$-dependent parts of the self-energy. But to our 
knowledge so far no such advancement has been undertaken for the general RISB framework. 
Deriving and putting into practise a generic multi-orbital slave-boson scheme beyond 
saddle-point proves technically very demanding. Cluster self-energies coping with short-range
non-local correlations have been introduced to extended DMFT (see e.g. Ref.~\onlinecite{mai05} for a 
review), and the RISB method can indeed also be formulated within a cluster 
scheme~\cite{lec07,fer092,maz14}. It yields very good results in comparison to more elaborate
cluster-DMFT.~\cite{fer092} Here however, we remain throughout the paper within the single-site 
framework.

\section{Multi-Orbital Hamiltonians\label{sec:modham}}
In this study, the multi-orbital Hamiltonians are composed of a kinetic part and a local
part that includes the electron-electron interaction, thus again of the general 
form (\ref{eq:singham}), $\mathcal{H}= 
\mathcal{H}^{\rm (kin)}+\sum_i\mathcal{H}^{\rm (loc)}_i \label{eq:multham}$.
We here will model orbital manifolds with angular momentum $l$=1,2, i.e. $p$- and $d$-shell
systems.
\begin{figure}[b]
\centering
\includegraphics*[width=4cm]{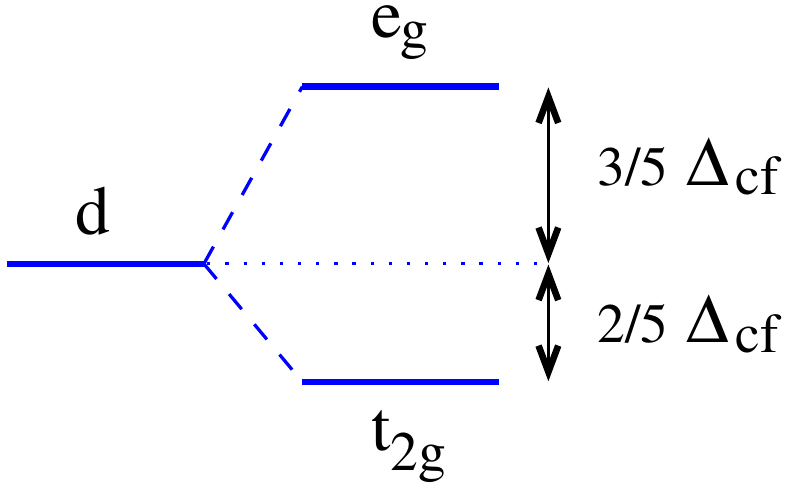}
\caption{\label{fig:comic-o-cfs} Example for crystal-field splitting in the case of a
$d$-level in cubic symmetry. It splits into twofold $e_g=\{z^2, x^2-y^2\}$ and threefold
$t_{2g}=\{xz, yz, xy\}$. The level center is not changed.}
\end{figure}

\subsection{Kinetic Hamiltonian\label{sec:kinham}}
Electrons in $M$ orbitals of Wannier type on the $dim=3$-dimensional 
simple-cubic lattice are considered. The kinetic Hamiltonian with only nearest-neighbor 
hopping $t$ reads
\begin{align}
 \mathcal{H}^{\rm (kin)}=-t\sum_{\langle ij\rangle m\sigma} 
c_{im\sigma}^\dagger c^{\hfill}_{jm\sigma}\quad,
\end{align}
using $m=1,\ldots M$. Note that we do not allow hopping between different orbital 
flavors. For the rest of the paper, the half bandwidth $W/2$ is the unit of 
energy. Since the present kinetic Hamiltonian is diagonal in orbital space with identical
eigenvalues for all orbitals, it remains invariant under orbital rotations.

\subsection{Local Hamiltonian}
On the local level of a single lattice site, the respective Hamiltonian part is given by
\begin{equation}\label{eq:locham}
\mathcal{H}^{\rm (loc)}=\mathcal{H}^{\rm (cf)}+\mathcal{H}^{\rm (int)}+
\mathcal{H}^{\rm (soc)}\quad,
\end{equation}
whereby the first term describes the crystal-field (cf) term, 
the second the Coulomb interaction (int) and the third the spin-orbit coupling (soc). 
The single-particle crystal-field Hamiltonian takes care of a possible onsite energy 
splitting $\Delta_m$ between the orbitals (see Fig.~\ref{fig:comic-o-cfs}), reading
\begin{equation}\label{eq:cfham}
\mathcal{H}^{\rm (cf)}=\sum_{m\sigma}\Delta_m\,c_{m\sigma}^\dagger c^{\hfill}_{m\sigma}\quad.
\end{equation}

\subsubsection{Slater-Condon form of the local interaction}
For the case of a complete rotational-invariant treatment on the local level, the corresponding
two-particle interaction is described by the Slater-Condon (SC) Hamiltonian
\begin{align}
\mathcal{H}^{\rm (int)}=\frac{1}{2}\sum_{\substack{m_1 m_2 \\m_3 m_4}}\sum_{\sigma\sigma'} 
U_{m_1 m_2 m_3 m_4} c^\dagger_{m_1\sigma} c^\dagger_{m_2\sigma{'}}
c^{\hfill}_{m_4\sigma{'}} c^{\hfill}_{m_3\sigma}\;.
\label{eqn:slatco}
\end{align}
Since we aim at a canonical modelling, we assume spherical symmetry of the electron-electron
interaction troughout this work, i.e. no orbital-dependent screening mechanism is allowed.
Then the Coulomb matrix element is expressed via standard Slater integrals $F^{k}$ through
\begin{align}
U_{m_1 m_2 m_3 m_4}   = \sum_{k=0}^{2l} a_k(m_1,m_2,m_3,m_4)\,F^k\quad,
  \label{eqn:SC-parametrisation}
\end{align}
with expansion coefficients $a_k$ given by
\begin{align}
  a_k(m_1,m_2,m_3,m_4) =& \sum_{q=-k}^k (2l+1)^2 (-1)^{m_1+q+m_2}\nonumber \\ 
  &\hspace*{-3.5cm}\times
  \begin{pmatrix}
    l & k & l \\
    0 & 0 & 0
  \end{pmatrix}^2
  \begin{pmatrix}
    l & k & l \\
    -m_1 & q & m_3
  \end{pmatrix}
  \begin{pmatrix}
    l & k & l \\
    -m_2 & -q & m_4
  \end{pmatrix}\;.
\end{align}
In the case of $p$- and $d$-electrons, the relevant Slater integrals may be parametrized by 
averaged Coulomb integrals, namely the Hubbard $U$ and the Hund's exchange $J_{\rm H}$ as 
\begin{eqnarray}
l=1&:&\quad F^{0}=U\;,\;\,F^{2}=5J_{\rm H}\\
l=2&:&\quad F^{0}=U\;,\;\,F^{2}=\frac{14}{1+r}J_{\rm H}\;,\;\,F^{4}=rF^{2}\quad.
\end{eqnarray}
The $F^{4}/F^{2}$ Slater-integral ratio is here chosen as $r=0.625$, which is adequate for 
transition-metal atoms.

\subsubsection{Slater-Kanamori form of the local interaction}
Treating full rotational invariance in the interactions was a longstanding problem in 
many-body techniques. Therefore, simpler versions of interacting Hamiltonians have been
introduced. In the context of local Coulomb interactions, the Slater-Kanamori (SK) 
Hamiltonian is the most prominent one. It reads 
\begin{eqnarray}
\mathcal{H}^{\rm (int)}&=& U\sum_{m} n_{m\uparrow}n_{m\downarrow}+\nonumber\\
&&\hspace*{-0.75cm}+\,\frac 12 \sum \limits _{m \ne m',\sigma}\hspace*{-0.2cm}
\Big\{\left(U\hspace*{-0.1cm}-\hspace*{-0.1cm}2J_{\rm H}\right) \, n_{m \sigma} n_{m' \bar \sigma}
+ \left(U\hspace*{-0.1cm}-\hspace*{-0.1cm}3J_{\rm H}\right) \,n_{m \sigma}n_{m' \sigma}\nonumber\\
&&\hspace*{-0.75cm}+\left.J_{\rm H}\left(c^\dagger_{m \sigma} c^\dagger_{m' \bar\sigma}
c^{\hfill}_{m \bar \sigma} c^{\hfill}_{m' \sigma}
+c^\dagger_{m \sigma} c^\dagger_{m \bar \sigma}
 c^{\hfill}_{m' \bar \sigma} c^{\hfill}_{m' \sigma}\right)\right\}\;,\\
 &&\hspace*{-0.75cm}= \left(U\hspace*{-0.1cm}-\hspace*{-0.1cm}3J_{\rm H}\right) 
\frac{N(N-1)}{2}+\frac{5}{2}J_{\rm H}N -2J_{\rm H} 
S^2-\frac{1}{2}J_{\rm H}L^2\;. \nonumber
\end{eqnarray}
Here $n=c^\dagger c$, and $N$ marks the total-particle operator, $S$ the 
spin operator and $L$ the orbital-momentum operator.
This form of the local interaction is obtained from the general Slater-Condon form 
(\ref{eqn:slatco}) via the restriction to only two-orbital interaction terms and the setting
\begin{eqnarray}
U_{m m m m} = U \quad&,&\quad U_{m m' m m'} = U-2J_{\rm H} \nonumber\\
U_{m m' m' m} = J_{\rm H} \quad&,&\quad U_{m m m' m'} = J_{\rm H}\quad.
  \label{eqn:kanamori-parametrisation}
\end{eqnarray}
Though reduced, the Slater-Kanamori Hamiltonian is rotational invariant in the case of 
the $p$-shell as well as the \eg- and \tg-subshell of $d$-electron manifolds. However it 
differs in the scaling of the local Coulomb interaction when compared to the SC form.
In fact, the SC Hamiltonian takes on the SK operator structure, if the 
$F^{4}/F^{2}$ Slater-integral ratio is set to a formal, rather unphysically large value 
$r=1.8$.

\subsection{Spin-Orbit Coupling}
Spin and orbital momentum of an electron are coupled due to relativistic effects. The 
Hamiltonian expressing this coupling reads
\begin{equation}
  \mathcal{H}^{\rm(soc)}=\frac{\lambda}{2} \sum_{k=x,y,z}
\sum_{\mu \mu'} \sum_{\sigma \sigma'} 
\tilde{L}_{\mu \mu'}^k\,{\cal S}_{\sigma \sigma'}^k\,c^\dagger_{\mu \sigma} 
c^{\hfill}_{\mu' \sigma'}\quad,
\end{equation}
whereby $\tilde{L}$ denotes the angular-momentum representation on the $l_z$ states and
${\cal S}$ carries the Pauli matrices as components and $\lambda$ is the 
spin-orbit interaction constant. Because of the mixing of 
spin- and orbital degrees of freedom, the spin-orbit Hamiltonian does generally not
commutate with $S^2$ and $L^2$ in many-electron systems. However it commutates 
with the total angular-momentum operator $J^2$ as well as its z-component 
$J_z$.

\section{Multi-Orbital Slave Bosons\label{sec:multsb}}

\subsection{States and Operators}
In the following, three building blocks of the local multi-orbital
rotational-invariant representation of the slave-boson formalism are
introduced. First, the local many-body Hilbert space in Fock-basis
representation. Second, a set of commutating quantum operators, which can be
represented on that space which fully and uniquely determine the different
states. And third a matrix containing all state-connecting variational
parameters. The latter will be determined such that it follows the
given symmetries of the system and may ultimately be used as the
slave-boson operator matrix $\Phi$.

\begin{figure}[b]
\centering
\includegraphics*[width=4cm]{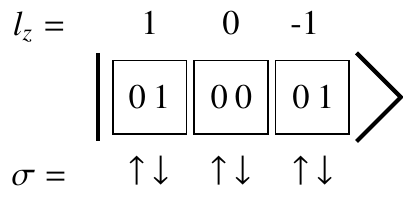}
\caption{\label{fig:example-states} Example for an initial state for
  a 3-orbital $p$-shell model. This state represents two
  electrons, one in the $l_z=1$-spin-down orbital, the second one in
  the $l_z=-1$-spin-down orbital. By construction, these states
  diagonalize the operators $L_z$ and $S_z$.}
\end{figure}

The first building block of the $M$-orbital rotationally invariant
representation is the local many-body Hilbert space at one space-point entity 
$i$ spanned by the set 
$\mathcal{Q}=\{\myvec{v}_1,\dots,\myvec{v}_a,\dots,\myvec{v}_{Q}\}$
of $Q=2^{2M}$ vectors,
such that renormalizations between adjacent space-points are excluded.
It contains for example all possible local many-body $d$-states or - in a 
cluster-scheme - all possible many-electron configurations of a four-site 
plaquette.
The vectors $\myvec{v}_a$ describe two spin-resolved orbitals for
every local-orbital degree of freedom. Each one may either be occupied 
(represented by 1) or empty, (represented by 0), which is nothing but a 
Fock-space occupation-number representation of all configurations over 
a body of binary numbers $\mathbb{B}$, i.e.,
$\myvec{v}_a \in \mathbb{B}^{2M}.$
Thus each vector has $2M$ entries, which e.g. in the case of an
atomic shell with angular momentum $l$ amounts to $2(2l+1)$. An example
for $M=3$, a $p$-shell state, is shown in Fig.~\ref{fig:example-states}. 
Since the size of $\mathcal{Q}$ scales like $Q=2^{2M}$, matrices represented 
in this space become large for $M>3$ and numerical computation becomes costly 
in time and memory. It is worthwile, to use symmetries to rule out states that do
not take part in local interaction processes and hence reduce the size
of the problem.

This leads to the second building block of the advanced RISB formalism, namely
the set $\mathfrak{S}$ of local-commutating operators. In the absence of
spin-orbit-coupling, it is chosen to consist of the
particle-number operator $N$, spin-square $S^2$, and
spin-z-component operator $S_z$, the seniority-operator
$\Xi$ which measures the number of unpaired spins in a given
state, the orbital-angular-momentum $L^2$, and
orbital-angular-momentum-z-component operator $L_z$, i.e.
$\mathfrak{S}:=\{N,S^2,S_z,\Xi,L^2,L_z\}$.
The set $\mathfrak{S}$ is easily represented on $\mathcal{Q}$ via
explicitly considering the operator action on the state vectors. In the
following, we always assume the initial Fock-space orbitals to be
labeled with magnetic quantum numbers $l_z$ and spin quantum number
$s_z$ from highest to lowest value (e.g. $l=1$ $\rightarrow$
$1\uparrow, 1\downarrow, 0\uparrow, 0\downarrow, -1\uparrow,
-1\downarrow$), thus diagonalizing the operators $L_z$ and
$S_z$ by construction (cf. Fig~\ref{fig:example-states}). The 
particle-number operator is most easily represented on these states. 
It is by construction also diagonal in the
occupation-number basis and counts the number of occupied orbitals, thus, electrons. 
For the matrix representation of the $S^2$-operator
between vectors $\myvec{v}_a,\myvec{v}_b$, the expression
\begin{eqnarray}
\left(S^2\right)_{ab}=&&\hspace*{-0.0cm}\langle\myvec{v}_a | \sum_{k=x,y,z}
\sum_{\sigma_1\sigma_2\sigma_3\sigma_4}\nonumber\\
&&\hspace*{-0.3cm} \sum_{\mu_1\mu_2}
{\cal S}_{\sigma_1\sigma_2}^k {\cal S}_{\sigma_3\sigma_4}^k
c^\dagger_{\mu_1\sigma_1} c_{\mu_1\sigma_2} c^\dagger_{\mu_2\sigma_3} 
c_{\mu_2\sigma_4} | \myvec{v}_b \rangle
\end{eqnarray}
holds, with local orbital indeces $\mu_1,\mu_2$ and Pauli matrices
${\cal S}$. For the $S_z$-operator it follows
\begin{align}
(S_z)_{ab}=\langle\myvec{v}_a | \sum_\mu \sum_{\sigma_1 \sigma_2} 
{\cal S}_{\sigma_1 \sigma_2}^z c^\dagger_{\mu\sigma_1} 
c_{\mu\sigma_2} | \myvec{v}_b \rangle\quad.
\end{align}
For the $L^2$ and $L_z$ operators hold similar relations
as for the spin operators, namely
\begin{eqnarray}
\left(L^2\right)_{ab}=&&\hspace*{-0.0cm}\langle\myvec{v}_a | \sum_{k=x,y,z}
\sum_{\mu_1\mu_2\mu_3\mu_4}\nonumber\\
&&\hspace*{-0.3cm} \sum_{\sigma} \tilde{L}_{\mu_1\mu_2}^k 
\tilde{L}_{\mu_3\mu_4}^k
c^\dagger_{\mu_1\sigma} c_{\mu_2\sigma} c^\dagger_{\mu\sigma} 
c_{\mu\sigma} | \myvec{v}_b \rangle
\end{eqnarray}
and
\begin{align}
(L_z)_{ab}=\langle\myvec{v}_a | \sum_\sigma \sum_{\mu_1 \mu_2} 
\tilde{L}_{\mu_1 \mu_2}^z c^\dagger_{\mu_1\sigma} 
c_{\mu_2\sigma} | \myvec{v}_b \rangle\quad.
\end{align}

Writing out the matrix elements for the seniority operator $\Xi$ in 
second quantization is a bit more involved and can be found in the
literature~\cite{drake06,rudz07} in terms of vector-coupled ladder
operators of two spins at zero net momentum.

A common eigenbasis $\mathcal{A}$, the so-called adapted basis, for all
operators in $\mathfrak{S}$ has to be generated. It is spanned by the 
vectors $\myvec{A}_A$, of which there are again $Q$ with index 
$A \in [1,\dots,Q]$. They are conveniently labeled as 
$|A\rangle = |\nu,s,\sigma,\xi,l,l_z\rangle \in \mathbb{C}^{2M}$. These
six quantum numbers are sufficient to label the states of the local
Hilbert space up to five orbitals ($M=5$) unambigously. After the
diagonalization process, the states are ordered with left
quantum numbers varying slowest, right fastest, starting from lowest
values to highest. The basis is stored as a unitary transformation
matrix ${\cal U}_\mathcal{A}$, which transforms all operators represented 
in Fock space $\mathcal{Q}$ to the adapted basis as a linear combination
of the previous occupation-number representation with coefficients of
complex-number kind.

The central building block, the slave-boson operator $\Phi$, may
now be be seen as a transition operator, that mediates between the
quasi-particle- and local-exitation degree of freedom. 
Conveniently, it can be expressed with the same set of possible local
states $|A\rangle = |\nu_A,s_A,\sigma_A,\xi_A,l_A,l_{zA}\rangle$ and
$|B\rangle = |\nu_B,s_B,\sigma_B,\xi_B,l_B,l_{zB}\rangle$. Hence, it 
it carries two indeces and is represented as a matrix on the local 
Hilbert space of the adapted-basis-set states, i.e.
$\Phi_{AB} \in \mathcal{C}^{Q \times Q}$.
The specific action of the operator $\Phi_{AB}$ is however unknown a
priori and has to be determined self-consistenly in the
saddle-point approximation, with its matrix elements as
parameters. Thus, the number of slave-boson parameters in the
RISB calculation scales like $n_\phi=Q^2$.

\subsection{Rotationally Invariant Representation}

Let us for the following use a common index $\alpha=\{m,\sigma\}$
for orbital- and spin-projection degree of freedom. By rearranging 
eqn.~(\ref{eqn:gen-risb-form}) with the help of the unitary transformation
matrix ${\cal U}_\mathcal{A}$, one obtains a representation 
which is by construction basis free, hence obviously invariant under 
unitary rotations~\footnote{The trace operation 'Tr' is throughout this 
work generally understood as the summation of the diagonal elements of 
the involved objects.}:

\begin{align}
  \underline{\mathcal{O}} & =\sum_{AA'} ( A|\mathcal{O}|A' )
  \sum_{n}\phi^\dagger_{nA}\phi^{\hfill}_{A'n} =\sum_{nAA'}
  \phi^\dagger_{nA}( A|\mathcal{O}|A' )\phi^{\hfill}_{A'n}\nonumber\\
 & =\mbox{Tr}(\phi^\dagger \mathcal{O} \phi)
  =\mbox{Tr}(U_\mathcal{A} U_\mathcal{A}^\dagger\phi^\dagger
  U_\mathcal{A} U_\mathcal{A}^\dagger\mathcal{O} U_\mathcal{A}
  U_\mathcal{A}^\dagger\phi)\nonumber\\ &
  =\mbox{Tr}(U_\mathcal{A}^\dagger\phi^\dagger U_\mathcal{A}
  U_\mathcal{A}^\dagger\mathcal{O} U_\mathcal{A}
  U_\mathcal{A}^\dagger\phi U_\mathcal{A})
  =\mbox{Tr}(\bar{\phi}^\dagger \bar{\mathcal{O}} \bar{\phi})\;.
  \label{eqn:tr-gen-risb-form}
\end{align}
The renormalization matrices from eqn.~(\ref{eq:cop}) are generalized
in the multi-orbital mean-field framework via
\begin{equation}
c_\alpha \rightarrow \sumBeta R_{\beta\alpha} f_\beta
\quad,\qquad c_\alpha^\dagger \rightarrow \sumBeta R_{\alpha\beta}^*
f_\beta^\dagger\quad,
\end{equation}
whereby
\begin{equation}
R_{\alpha\beta}^* = \sum_\gamma T^*_{\alpha\gamma}w_{\gamma\beta}
\label{eq:rmatmulti}
\end{equation}
with (see appendix~\ref{app:renorm})
\begin{equation}
T_{\alpha\gamma}^* = \mbox{Tr}(\phi^\dagger f_\alpha^\dagger 
\phi c_\gamma)\quad,
\end{equation}
and a normalization matrix $w$, which carries the matrix-square root 
of the product of the local particle- and hole-density matrix 
(see Ref.~\onlinecite{lec07} for details). Introducing that matrix ensures
the correct mean-field regime of RISB.

Finally, the multi-orbital constraints at saddle-point compactly read
\begin{align}
  \mbox{Tr}(\phi^\dagger \phi) &= 1\quad, \\ \mbox{Tr}(\phi f_\alpha^\dagger
  f_\beta \phi^\dagger) &= \Braket{f_\alpha^\dagger f_\beta}.
\end{align}

\subsection{First Glance on Symmetry Reduction}
An obvious way to reduce the number of parameters in the problem is by the
use of the symmetries of the local interaction. Hence, one may cut out 
slave-boson amplitudes, which would otherwise violate a given symmetry.
This renders the slave-boson operator block-diagonal in the allowed
combinations of quantum numbers. For instance, let the particle-number 
conservation be a symmetry of the local Hamiltonian, i.e.
the commutator $[\mathcal{H},N]$ vanishes. Then all slave-boson 
amplitudes $\phi_{AB}$ with $n(A)\neq n(B)$ will also be zero.
This is known a priori, so those amplitudes can be ruled out,
and be excluded in solving the saddle-point problem. This does not
only render the $\Phi$ matrix sparse (and block-diagonal) from the
beginning. It also enables to reduce the number of saddle-point
equations, which are nothing but derivatives of the
free-energy functional with respect to the free parameters of the 
formalism. 

In previous implementations of RISB, all operations where 
iterated over such irreducible quantum-number subspaces, which made it 
hard to change from one set of quantum-numbers (or model) to another. 
A different approach shall be presented here, which not only focuses on 
the sparsley populated structure of the matrices under rotation to an 
adapted basis, but also opens the path to use the lattice point-group 
symmetry to further reduce the number of free parameters.

\subsection{Top-Down Reduction of Free Parameters}
The number of free parameters can be further reduced, if the
point-group symmetry of the lattice is to be imprinted on the 
local-interacting Hilbert space. Just as the Pauli matrices are for 
example a matrix-basis set for the four-dimensional real vector-space of all
complex-hermitian two-by-two matrices, one can find a matrix-basis set
obeying a certain point-group symmetry. This enables then the spanning of
space of all complex matrices complaient with that given symmetry. 
If that point-group symmetry is a symmetry of the lattice under consideration, 
it rules out certain many-body transitions, which in turn are here
represented by the slave-boson operator. This makes it possible to 
represent the slave-boson operator in a basis set which only allows for
point-group-supported transitions in the first place.

The idea of expanding an operator representation in a basis
of orthonormal matrices obeying a finite symmetry group~\cite{koster58} is 
already used in the context of an implementation of the Gutzwiller 
formalism~\cite{lanata12}. There, it is done for the case of
paramagnetic problems~\cite{lanata13} for $d$-orbitals and with further
reduction by symmetry of the involved renormalization matrices. The
latter are then promoted to free variables to treat spin-orbit coupling and
crystal-field splitting in $f$-orbital systems~\cite{lanata15,lan17}. 
In the following, the details of an implementation without the need of 
promoting the renormalization matrices to free variables is wrapped up. 

The goal is a decomposition of the represented slave-boson operator 
$\Phi$ into a number of $Y$ basis matrices $\tilde{\Phi}_i$ within the 
adapted basis $\mathcal{A}$ of the form~\footnote{In the following, the
indeces $i,j$ denote different basis matrices and should not be confused
with lattice-site indeces.}
\begin{equation}
  \Phi = p_1 \tilde{\Phi}_1 + p_2 \tilde{\Phi}_2 + \cdots + p_Y
  \tilde{\Phi}_Y  = \sum_{i=1}^Y p_i\,\tilde{\Phi}_i\quad,
\end{equation}
with coefficients $p_i \in \mathbb{C}$, and the orthonormality relation
\begin{equation}
  \delta_{ij} = \braket{\tilde{\Phi}_i\,\tilde{\Phi}_j} :=
  \mbox{Tr}(\tilde{\Phi}_i^\dagger\,\tilde{\Phi}_j)\quad.
\end{equation}

Their indeces are still labeled by the vectors of the adapted basis
$\mathcal{A}$. The basis matrices $\tilde{\Phi}_i$ can be constructed
in a way that they obey a certain point-group symmetry
$G$. They commutate with all elements $g^\mathcal{A}$ of
the symmetry group, represented in $\mathcal{A}$. That also means,
that this expansion of $\Phi$ projects its action on a subspace which 
is commensurable with the symmetry group. All other action is lost. 
The generation of $\tilde{\Phi}_i$ is described in 
appendix~\ref{app:basmat}.

Inserting the given expansion into the general eq.~(\ref{eqn:tr-gen-risb-form})
for the slave-boson representation of operator $\mathcal{O}$, results
in
\begin{align}
\underline{\mathcal{O}} & = \mbox{Tr}\left( \sum_i p_i^*
\tilde{\Phi}^\dagger_i \bar{\mathcal{O}} \sum_j p_j \tilde{\Phi}_j\right)
= \sum_{ij} p_i^*\, \mbox{Tr}( \tilde{\Phi}^\dagger_i
\bar{\mathcal{O}} \tilde{\Phi}_j )\, p_j \nonumber\\ 
& = \myvec{p}^\dagger\,\mbox{Tr}( \Phi^\dagger \bar{\mathcal{O}} 
\Phi )\, \myvec{p}= \myvec{p}^\dagger\, O\, \myvec{p}
\end{align}

In this form, one precomputes and stores the matrices
\begin{eqnarray}
(T^*_{\alpha\gamma})^{ij}&:=& 
\mbox{Tr}(\tilde{\Phi}^\dagger_{i} f_\alpha^\dagger \tilde{\Phi}_j c_\gamma)\quad,\\
(\Delta^{(p)}_{\alpha\beta})^{ij} &:=& 
\mbox{Tr}(\tilde{\Phi}^\dagger_{i} f_\alpha^\dagger f_\beta\tilde{\Phi}_j)\quad,\\
(\mathcal{H}^{\rm (loc)})^{ij}&:=& 
\mbox{Tr}(\tilde{\Phi}^\dagger_{i}\mathcal{H}^{\rm (loc)} \tilde{\Phi}_j)\quad,\\
(N_{\alpha\beta})^{ij}&:=& \mbox{Tr}(\tilde{\Phi}_{i}
f_\alpha^\dagger f_\beta \tilde{\Phi}_j^\dagger)\quad,
\end{eqnarray}
so that for the constraints follows
\begin{eqnarray}
\mbox{Tr}(\phi^\dagger \phi)&=& 
\sum_{ij} p^*_i \mbox{Tr}(\tilde{\Phi}_{i}^\dagger\tilde{\Phi}_{j}) p_j 
= \sum_i p^*_i p_i= 1\quad,\\
\mbox{Tr}(\phi f_\alpha^\dagger f_\beta \phi^\dagger) 
&=& 
\sum_{ij}p_i (N_{_\alpha\beta})^{ij} p_j^*=\Braket{f_\alpha^\dagger f_\beta}\quad.
\end{eqnarray}

The free-energy functional is then rewritten as
\begin{align}
  \Omega[{p^*_i},\Lambda^*,\lambda_0^*] = &
  -\frac{1}{\beta}\sum_{\myvec{k}}
\mbox{Tr}\mbox{ ln}[1+e^{-\beta(\myvec{R}^\dagger
\epsilon_{\myvec{k}}\myvec{R}+(\Lambda+h.c.))}]\nonumber \\
  &-[\lambda_o^*(1-\sum_i p^*_i p_i)]\nonumber\\
  &-[\Lambda^*_{\beta\alpha}\sum_{ij}p_i(N_{\alpha\beta})^{ij}p_j^*+c.c]\nonumber\\
  &+\sum_{ij}p^*_i (\mathcal{H}^{loc})^{ij} p_j\quad.
\end{align}

We expect the functional to map the complex variables 
$z:=\{p_i, \Lambda, \lambda_0\}$ to a real number $\Omega
\in \mathbb{R}$. So for writing the saddle-point equations, the
formal derivative of the functional $\Omega$ with respect to the complex conjugate 
of all variables $\partial \Omega/\partial z^*$ is taken. By using
Wirtinger calculus~\cite{hjor11}, the saddle-point equations for the real- and 
imaginary-part of the functional read
\begin{align}
\frac{\partial \Omega}{\partial \mbox{Re }z^*}&=
2\,\mbox{Re}\frac{\partial \Omega}{\partial z^*} \stackrel{!}{=} 0\quad, \\
\frac{\partial \Omega}{\partial \mbox{Im }z^*}&=
2\,\mbox{Im}\frac{\partial \Omega}{\partial z^*} \stackrel{!}{=} 0\quad.
\end{align}

Technical remarks on the top-down approach are as follows.
One starts with an exhaustive list of quantum numbers for labeling the occuring
states. For a general $p$-shell problem, the problem consists of
labeling 64 states, which can be unambigously described by the five
quantum numbers $\nu$, $s$, $s_z$, $l$ and $l_z$. For a $d$-shell, 
the seniority quantum number $\xi$, which counts the number of
unpaired spins in a given state, has to be added to label the occuring
$2^{10}=1024$ states unambigously.
Since we exclude the effect of superconductivity, we first abandon
mixing between states of different particle number, rendering $\nu$ a
good quantum number. We then apply the point-group symmetry, to assort
states with point-group-compliant combinations of $l$ and
$l_z$. They are not straightforwardly good quantum numbers, since e.g. the
local interaction may mix different states of $l_z$. But
due to the underlying point-group symmetry, only specific mixings are 
allowed. In the next step, the mixing of different values of $s$, the 
total spin momentum, shall be preserved. This still leaves $s_z$ as a 
free parameter and enables us at this stage to locally describe magnetism 
along a spin-quantization axis. However since at this stage, we are only 
interested in paramagnetic solutions, the number of parameters may still be
reduced. In a paramagnetic saddle-point solution, we expect the variational
parameter belonging to a spin multiplet $\varphimat_{s_z}$ with
$s_z=-s,...,s$ to be identical for all $s_z$ values. Using this 
requirement, it is sufficient to average the connection matrices beforehand
via $\varphimat=\frac{1}{\sqrt{2s+1}}\sum_{s_z=-s}^{s}\varphimat_{s_z}$ and
compute the saddle-point solution only for the symmetrized parameter.

\subsection{Double Groups for Spin-Orbit Coupling}
For the inclusion of spin-orbit coupling, it is necessary to expand the
local symmetry group from a normal point group to a double group.
Describing also the spin-direction change after a rotation about
$2\pi$. In addition, a new commutating set of local operators is used, 
$S_z$ and $L_z$ are replaced by the total angular momentum operator $J^2$
and its z-component $J_z$. This is put into practise by introducing a new 
rotation element $\bar{E}$ and extending the group by doubling its elements,
i.e.
\begin{equation}
\mathfrak{D}\,G=\{g,\bar{E}\,g\}\quad \forall\; g \in G
\end{equation}
The new double group consists of all group elements $g$ of the previous
group $G$ plus all previous elements multiplied by $\bar{E}$. This changes also 
the available equivalence classes and amounts to non-integer values in the 
character tables.

\subsection{Local Correlation Functions}
The local-correlation operator on the full orbital space in
magnetic-quantum-number representation (spherical harmonics) is of the form
\begin{eqnarray}
  \bar{O}_{\alpha\beta\gamma\delta}\hspace*{-0.0cm}&:=&\hspace*{-0.0cm} 
\mbox{Tr}(\phi^\dagger (\bar{O})_{\alpha\beta\gamma\delta} \phi)\nonumber\\
\hspace*{-0.0cm}&=&\hspace*{-0.0cm} \mbox{Tr}(\phi^\dagger ( (O^2)_{\alpha\beta\gamma\delta} - 
(O)_{\alpha\beta}(O)_{\gamma\delta}) \phi )\quad.
\end{eqnarray}
It is rotated to the space of cubic harmonics by the transformation ${\bf K}$ with 
extraction of the diagonal elements for the physical interpretation, i.e.
\begin{equation}
  \bar{O}_{\alpha'\beta'} := \bar{O}_{\alpha'\alpha'\beta'\beta'} = 
K^\dagger_{\alpha'\alpha} K^\dagger_{\beta'\gamma} \bar{O}_{\alpha\beta\gamma\delta} 
K_{\beta\alpha'} K_{\delta\beta'}\quad.
\end{equation}

\subsection{Further Details on Implementation and Computation}
\begin{table}[b]
  \centering
  \begin{tabular}{ c c c }
    \toprule
    $M$ & symmetry group $G$ & $n^{\rm (red)}_{\phi}$\\[0.1cm]
    3 & $O$ & 16 \\
    3 & $\mathfrak{D}\,O$ & 50 \\
    5 & $O$ & 873 \\
    5 & $\mathfrak{D}\,O$ & 2064 \\
    5 & $D_4$ & 2516 \\
  \end{tabular}
  \caption{ \label{tab:num-of-variables} Examples for the number of parameters for 
different combinations of the orbital-manifold size $M$ with point-group $G$ 
after reduction by symmetry. Note that for the here chosen applications, the different
cubic groups $O$ and $O_{\rm h}$ yield identical results.}
\end{table}
The basis matrices depend only on the number of orbitals and
point-group symmetry involved, so they can be precomputed and
stored. For a given problem, also matrices like
$(T_{\alpha\beta})^{ij}$ and $(N_{\alpha\beta})^{ij}$ can be
precomputed. They depend on the geometry of the given model, but not
on the values of the interaction parameters. Since all these matrices
are very large but sparse, they are stored in \textit{compressed
  column/row storage}. For every set of interaction parameters, only
$(H^{\rm(loc)})^{ij}$ needs to be recomputed.

Table~\ref{tab:num-of-variables} lists the number $n^{\rm (red)}_{\phi}$ of free parameters 
after symmetry reduction for selected cases of orbital problems of size $M=\{3,5\}$. Let
us also quickly mention the difference in the numerical effort compared to a standard 
density-density slave-boson calculation (e.g. via a straightforward 
Kotliar-Ruckenstein~\cite{kot86} (KR) implementation). In RISB without symmetry 
constraints, the number of slave-bosons amounts for a $M$-orbital problem to $n_\phi=2^{4M}$, 
whereas there are $n_\phi=2^{2M}$ bosons in the standard KR scheme focussing on the 
orbital-density labelling of the local states. Furthermore, RISB operates with matrix 
objects, while KR works with scalar quantities (e.g. an orbital-diagonal QP weight). 
Thus there is a quadratic gain (with some problem-specific prefactor) when going from 
KR- to RISB calculations. Note that in addition the memory demands are much more serious 
in a RISB computation

In order to find the actual numerical solution of the RISB saddle-point equations, we use a 
parallelized and data-distributed implementation of the nonlinear-equation solver from 
Dennis and Schnabel with backtracking~\cite{denns83} as well as a problem-size adapted 
pertubation of the Hessian matrix.

The first-principles DFT data used in section~\ref{sec:realistic} stems from an 
implementation~\cite{mbpp_code} of the mixed-basis pseudopotential formalism.

\section{Combination with Density Functional Theory\label{sec:dftrisb}}
The RISB approach is not limited to sole model problems. It can be combined with
density functional theory (DFT) to directly address correlated materials within a realistic
first-principles setting. There are in principle two ways to facilitate such a DFT+RISB
framework. First, in the so-called 'one-shot' or 'post-processing' scheme, the DFT 
Kohn-Sham Hamiltonian, expressed in a localized basis, replaces the non-interacting part of 
the model Hamiltonian. 
The interacting part is then again provided by a suitable Hubbard-like
form, e.g., by the Slater-Condon Hamiltonian. Thus the hoppings, crystal fields and also 
spin-orbit terms may be taken over from the DFT calculations, and the complete problem is
converged in the RISB formalism. However, there is no feedback of the correlation effects
onto the electronic charge density. To achieve this, the second option, the so-called
charge self-consistent (CSC) scheme has to be employed. 
In the following, we want to briefly mark the essential steps of the DFT+RISB approach.
Essentially, the general structure is very similar as for the known DFT+DMFT formalism and 
we hence refer to Refs.~\onlinecite{ama08,gri12} for further details.

\begin{figure*}[htb]
\centering
\includegraphics*[width=12cm]{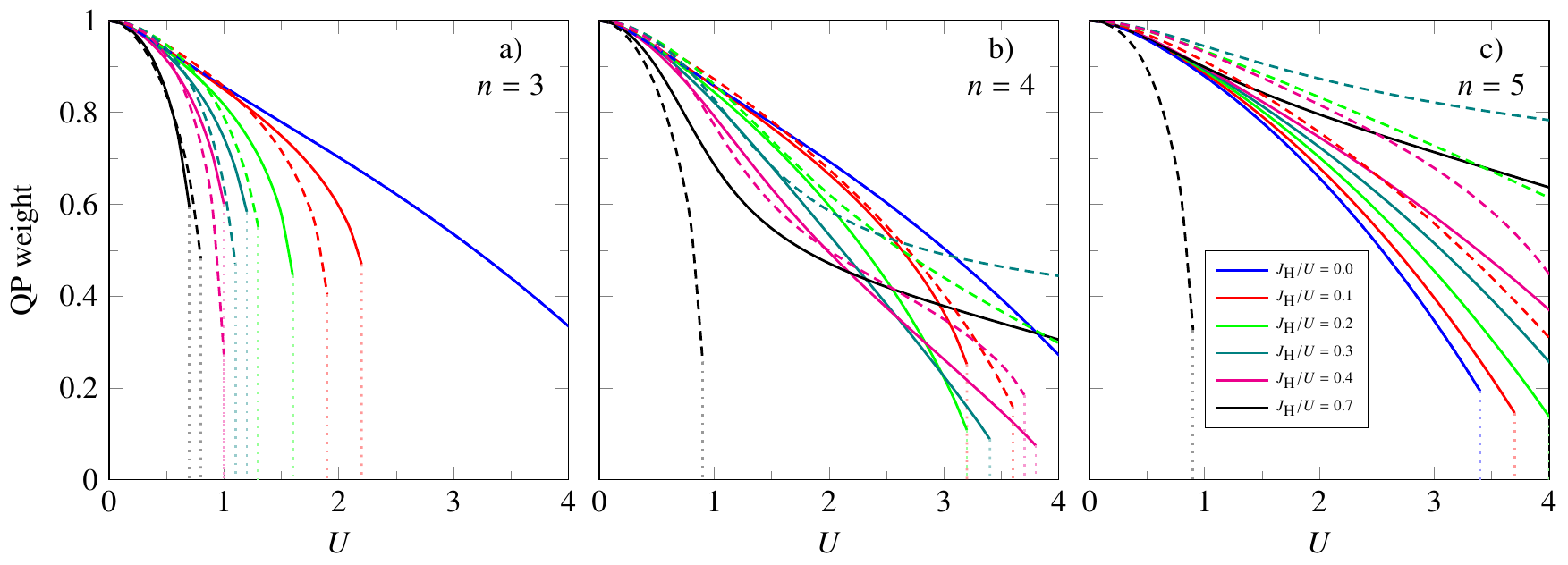}
\caption{\label{fig:pshell-z} 
Quasiparticle (QP) weight with respect to the Hubbard-interaction strength $U$ for the
degenerate 3-orbital $p$-shell model at different fillings $n$ and with
different $J_{\rm H}/U$ ratios. Solid lines: Slater-Condon interaction, dashed lines: 
Slater-Kanamori interaction. The light-dotted vertical lines are guides to the eyes
for the critical $U_c$ marking the Mott transition.}
\end{figure*}%
\begin{figure*}[htb]
\centering
\includegraphics*[width=12cm]{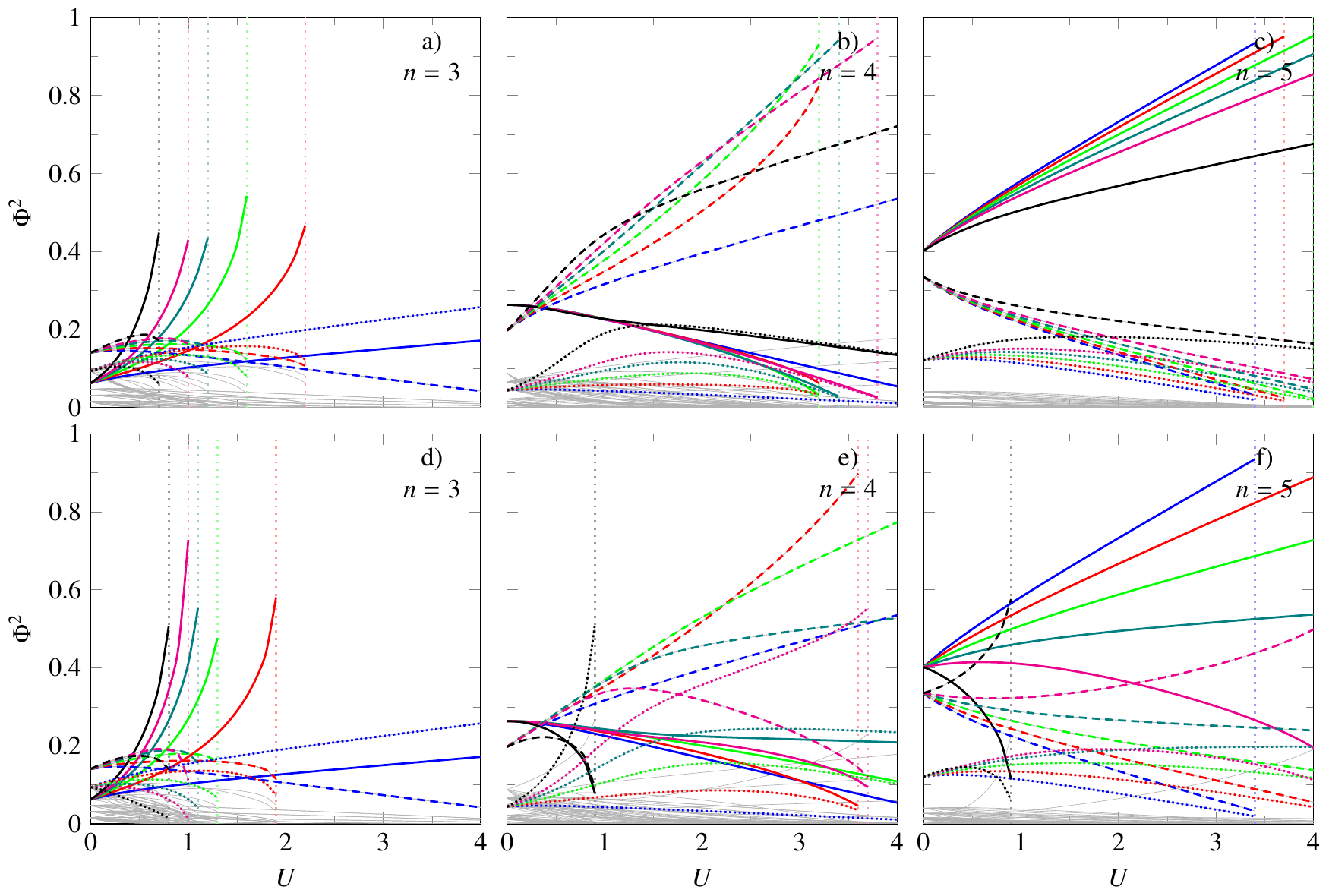}
\caption{\label{fig:pshell-phiamps} 
Slave-boson amplitudes of dominant multiplets in the 3-orbital model at 
different filling $n$. Top row: SC Hamiltonian, bottom row: SK Hamiltonian.
Colors mark different $J_{\rm H}/U$, with the coding as in Fig.~\ref{fig:pshell-z}.
$(a,d)$: solid: $\phi_{10}$, dashed: $\phi_{6}$, dotted: $\phi_{9}$. 
$(b,e)$: solid: $\phi_{15}$, dashed: $\phi_{14}$, dotted: $\phi_{10}$.
$(c,f)$: solid: $\phi_{15}$, dashed: $\phi_{16}$, dotted: $\phi_{14}$. 
Classification of the multiplets $\phi_\Gamma$ can be found in 
Tab.~\ref{tab:pshell-phiamps}.}%
\end{figure*}%
A converged Kohn-Sham self-consistency cycle of a DFT calculation for a periodic crystal 
yields the eigenenergies $\varepsilon_{{\bf k}\nu}$ and eigenfunctions $\psi_{{\bf k}\nu}$ 
for wave vector ${\bf k}$ and band $\nu$ in reciprocal space. A projection operator $P({\bf k})$
enables the mapping of the Bloch (bl) states onto localized orbitals within a chosen energy
window ${\cal W}$. 
The projection operator allows us to define the ${\cal W}$-restricted Kohn-Sham Hamiltonian 
in a localized basis, i.e.
\begin{equation}
\mathcal{H}'({\bf k}) := P({\bf k})\,\mathcal{H}_{\rm bl}({\bf k})\,P^\dagger({\bf k})
\quad.
\end{equation}
We separate the onsite terms from the truly $k$-dependent ones by defining
\begin{equation}
\mathcal{H}_0 := \frac{1}{N_k}\sum_{\bf k} \mathcal{H}'({\bf k})
\end{equation}
and extract the realistic kinetic Hamiltonian in the localized basis via
\begin{equation}
\mathcal{H}^{\rm (kin)}({\bf k}):=\mathcal{H}'({\bf k}) - \mathcal{H}_0\quad.
\end{equation}
Solving the RISB saddle-point equations yields the renormalized Hamiltonian
\begin{equation}
\mathcal{H}''({\bf k})=R^\dagger\,\mathcal{H}^{\rm (kin)}({\bf k})\,R + \Lambda\quad,
\end{equation}
which describes the 'one-shot' solution of the combination with DFT. 

In order to proceed to the CSC solution, the feedback onto the Bloch level is needed. 
This is achieved by replacing the interested DFT-correlated part with the RISB correlated 
one in $\mathcal{H}_{\rm bl}$, reading
\begin{eqnarray}
\mathcal{H}_{\rm bl}^{\rm (new)}(\myvec{k})=&&\mathcal{H}_{\rm bl}(\myvec{k})\\
&&+ P^\dagger(\myvec{k})\,\left( \mathcal{H}''(\myvec{k}) - \mathcal{H}'(\myvec{k})
- \mathcal{H}_{dc}\right)\,P(\myvec{k})\;.\nonumber
\end{eqnarray}
As in the DFT+DMFT framework~\cite{gri12}, the double-counting term $\mathcal{H}_{dc}$ 
takes care of the fact that part of the correlation is already included in DFT from the 
exchange-correlation functional. The fully-localized double-counting~\cite{ani93} is 
used in this work. Note that in the 'one-shot' scheme the double counting
may be absorbed in the chemical potential.

The particle number in the CSC scheme is fixed in the larger space of
$\mathcal{H}_{\rm bl}(\myvec{k})$ at every iteration. At each CSC step, one Kohn-Sham
iteration and a self-consistent RISB calculation is performed. After each latter 
convergence, a charge-shifting matrix $\Delta N_{\rm bl}$ is extracted as
\begin{equation}
\Delta N_{\rm bl} := P^\dagger\, ( N_{\rm RISB} - N_{\rm Kohn-Sham} )\, P\quad,
\end{equation}
which is then fed back to the DFT charge-density calculation to recompute the Kohn-Sham
potential for the next iteration step~\cite{gri12}. This procedure is repeated, until
the CSC cycle converges. Note that $\Delta N_{\rm bl}$ is a traceless matrix which  
reshuffles the charge due to the many-body correlations~\cite{gri12}.

There are other CSC implementations with a similar many-body quality available, namely
a DFT+Gutzwiller approach~\cite{wan08,bor14} as well as the so-called 
Gutzwiller-DFT~\cite{ho08} method. 

\section{Results\label{sec:results}}
The following subsections are devoted to the application of the symmetry-adapted
RISB formalism in the context of an multi-orbital onsite-interacting Hamiltonian. 
Main focus is on the interplay between the onsite Hubbard $U$ and the Hund's 
exchange $J_{\rm H}$ with respect to the overall electron filling $n$. Whereas we first 
deal with minimal lattice models, the attention is shifted to the realistic context of the 
iron-chalcogenide systems FeSe and FeTe in the final part of this section.

As noted in section~\ref{sec:kinham}, all model Hamiltonians in this work are explored
on a simple-cubic lattice with nearest-neighbor hopping. Moreover, a Mott transition 
with increasing Hubbard $U$ is here defined by the disappearance of the metallic state
via a vanishing QP weight $Z$ at a critical $U_c:=U_{c2}$. The issue of metall-Mott phase
coexistence~\cite{fre02} is not investigated in the present work.

\begin{figure*}[htb]
\centering
\includegraphics*[width=12cm]{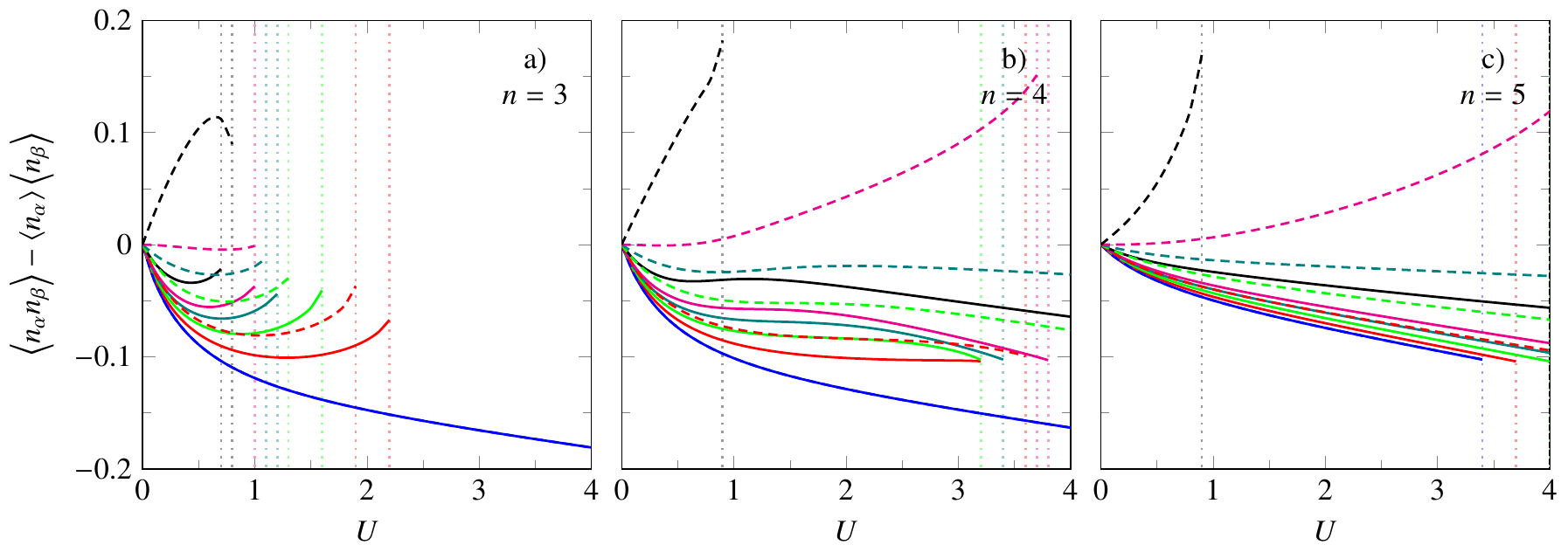}\\
\includegraphics*[width=12cm]{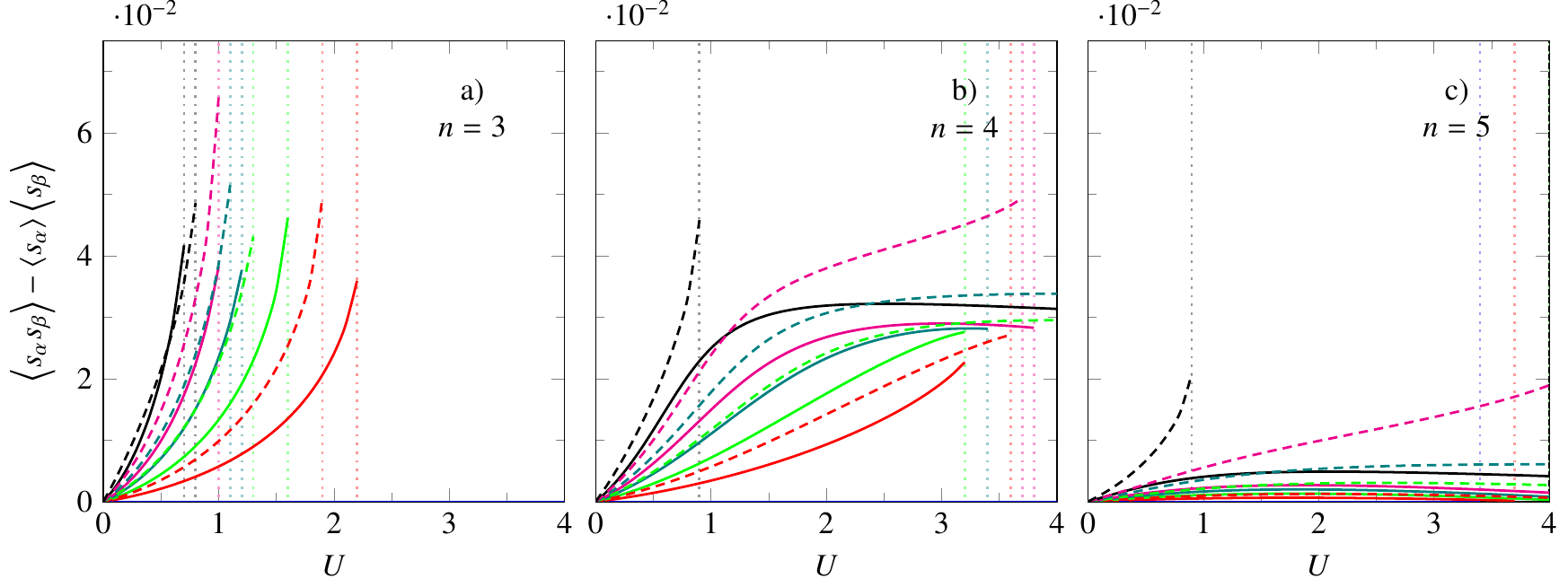}
\caption{ \label{fig:pshell-corr} 
Offdiagonal ($m\neq m'$) onsite correlation function of the 3-orbital model 
for charge (top) and spin (bottom).  Solid lines: Slater-Condon interaction, 
dashed lines: Slater-Kanamori interaction.Colors mark different $J_{\rm H}/U$, 
with the coding as in Fig.~\ref{fig:pshell-z}. }
\end{figure*}%
\begin{figure*}[htb]
\centering
\includegraphics*[width=12cm]{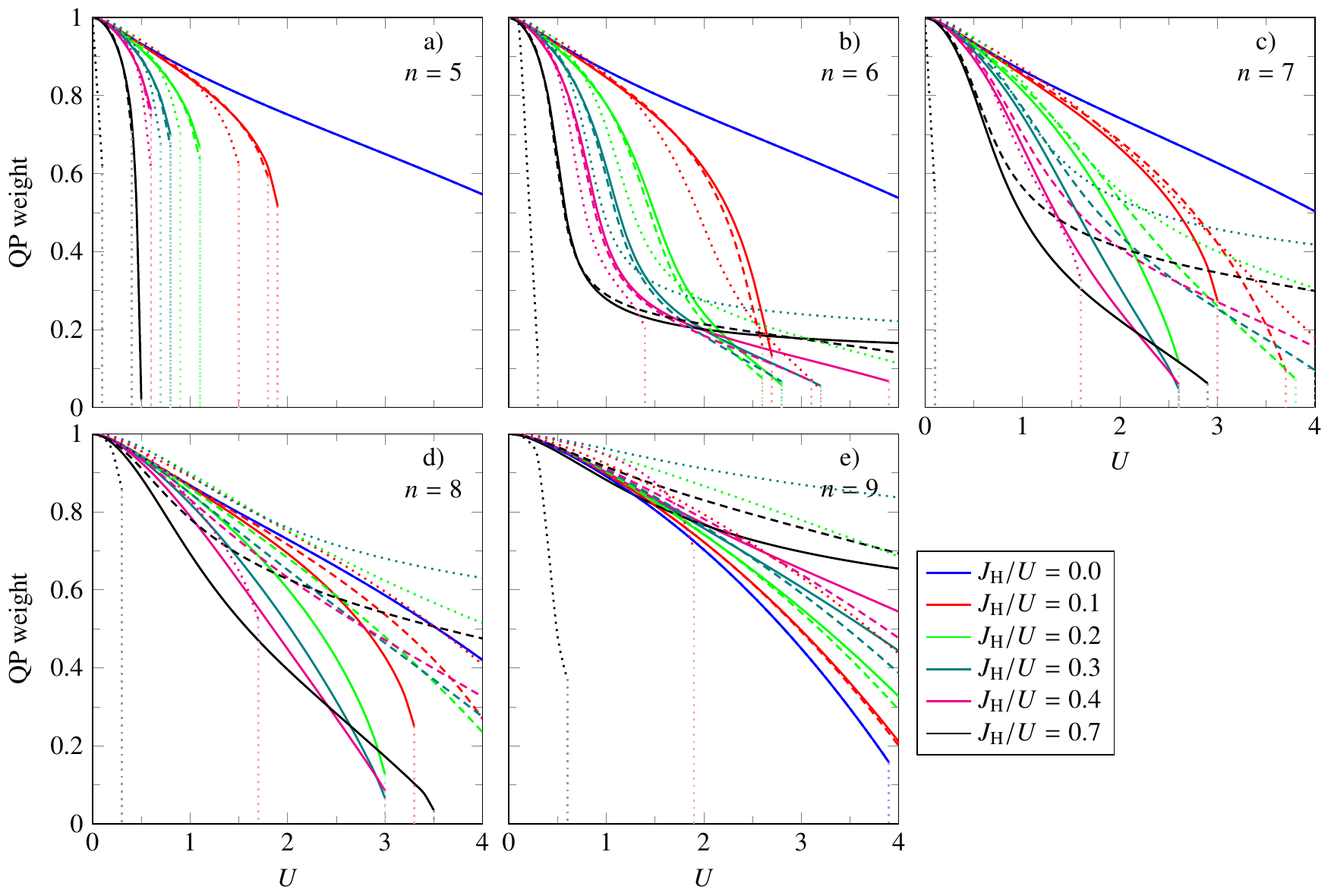}
\caption{\label{fig:dshell-z} 
Quasiparticle (QP) weight with respect to the Hubbard $U$ for the
degenerate 5-orbital $d$-shell model at different fillings $n$ and with
varying $J_{\rm H}/U$ ratio. Solid lines: 
SC Hamiltonian with $r=0.625$, dashed lines: SC Hamiltonian with $r=1.8$ and
dotted lines: SK Hamiltonian. 
Light-dotted vertical lines are guides to the eyes for the critical $U_c$ marking 
the Mott transition.}
\end{figure*}%
\subsection{3-Orbital $p$-Shell Model}
We start with an interacting three-orbital model having the  complete symmetry of a 
$l=1\,(p)$ manifold. Both rotational-invariant local-interaction Hamiltonians, i.e. 
of Slater-Condon and of Slater-Kanamori type, are utilized. The full cubic point group
underlies the present problem.

Figure~\ref{fig:pshell-z} displays the orbital-degenerate QP weight $Z$ for different 
ratios $J_{\rm H}/U$ in the half-filled case $n=3$ as well as the fillings $n=4$ and $n=5$.
Comparing the results for the different electron count on a global level, while for
$J_{\rm H}=0$ the correlation strength increases with rising $n$, for finite Hund's exchange
it weakens in that direction.
At half filling, the critical $U_c$ diminishes with growing $J_{\rm H}/U$ ratio~\cite{fac17},
similar as in the two-orbital case~\cite{lec07,fac17}. On the contrary, for the case of 
$n=4$ electrons in three orbitals, the well-known Janus physics~\cite{med11,geo13} emerges
at sizable $J_{\rm H}/U$: the system becomes strongly correlated with substantially reduced 
$Z$ already at intermediate $U$ and the Mott transition is on the other hand shifted to 
rather large interaction strength. The fast reduction of $Z$ is associated with the 
formation of large local magnetic moments triggered by $J_{\rm H}$. Notably, for the shown 
extreme case of $J_{\rm H}/U=0.7$, the SC Hamiltonian still restores this phenomenology 
whereas the SK form looses it because of the too strong negativity in some interaction 
terms.
For $n=5$, the Janus physics mostly disappears again and the differences between SC and SK
are significant for any size of the Hund's exchange.
\begin{table}[t]
  \centering
  \begin{tabular}{ c l c }
    \toprule
    $\Gamma$ & symmetry & nonzero\\
    6 & $T_1:\sum_{\sigma=-1}^1 (2, 1, \sigma, 2, 1, \{-1,0,1\} )$ & 3 \\[0.1cm]
    9 & $T_2:\sum_{\sigma=-\nicefrac{1}{2}}^{\nicefrac{1}{2}} (3, \nicefrac{1}{2}, \sigma, 3, 2, \{-2,-1,0,1,2\} )$ & 6 \\[0.1cm]
    10 & $A_1:\sum_{\sigma=-\nicefrac{3}{2}}^{\nicefrac{3}{2}} (3, \nicefrac{3}{2}, \sigma, 3, 0, 0 )$ & 1 \\[0.1cm]
    14 & $T_1:\sum_{\sigma=-1}^1 (4, 1, \sigma, 2, 1, \{-1,0,1\} )$ & 3 \\[0.1cm]
    15 & $T_1:\sum_{\sigma=-\nicefrac{1}{2}}^{\nicefrac{1}{2}} (5, \nicefrac{1}{2}, \sigma, 1, 1, \{-1,0,1\} )$ & 3 \\[0.1cm]
    16 & $A_1:(6, 0,0,0,0,0)$ & 1 \\
  \end{tabular}
  \caption{\label{tab:pshell-phiamps} 
Classification of dominant slave-boson amplitudes 
associated with local multiplets as plotted in Fig.~\ref{fig:pshell-phiamps} for the 
3-orbital model. Notation is as follows: $\Gamma$ is an internal label for the slave-boson 
number. $\mathcal{C}:\sum_{\sigma=-s}^s(N,s,\sigma,\{\nu\},\{l\},\{l_z\})$, where curly 
brackets $\{\}$ indicate sets as imprinted by the point-group symmetry class $\mathcal{C}$. 
All sums are normalized by a factor $\sqrt{2s+1}$. The number of nonzero elements refers to 
each spin-sum term in the classification symbol. For the whole number of nonzero elements in 
$\phi_\Gamma$ it has to multiplied by $2s+1$.}
\end{table}
\begin{figure*}[htb]
  \centering
\includegraphics*[width=12cm]{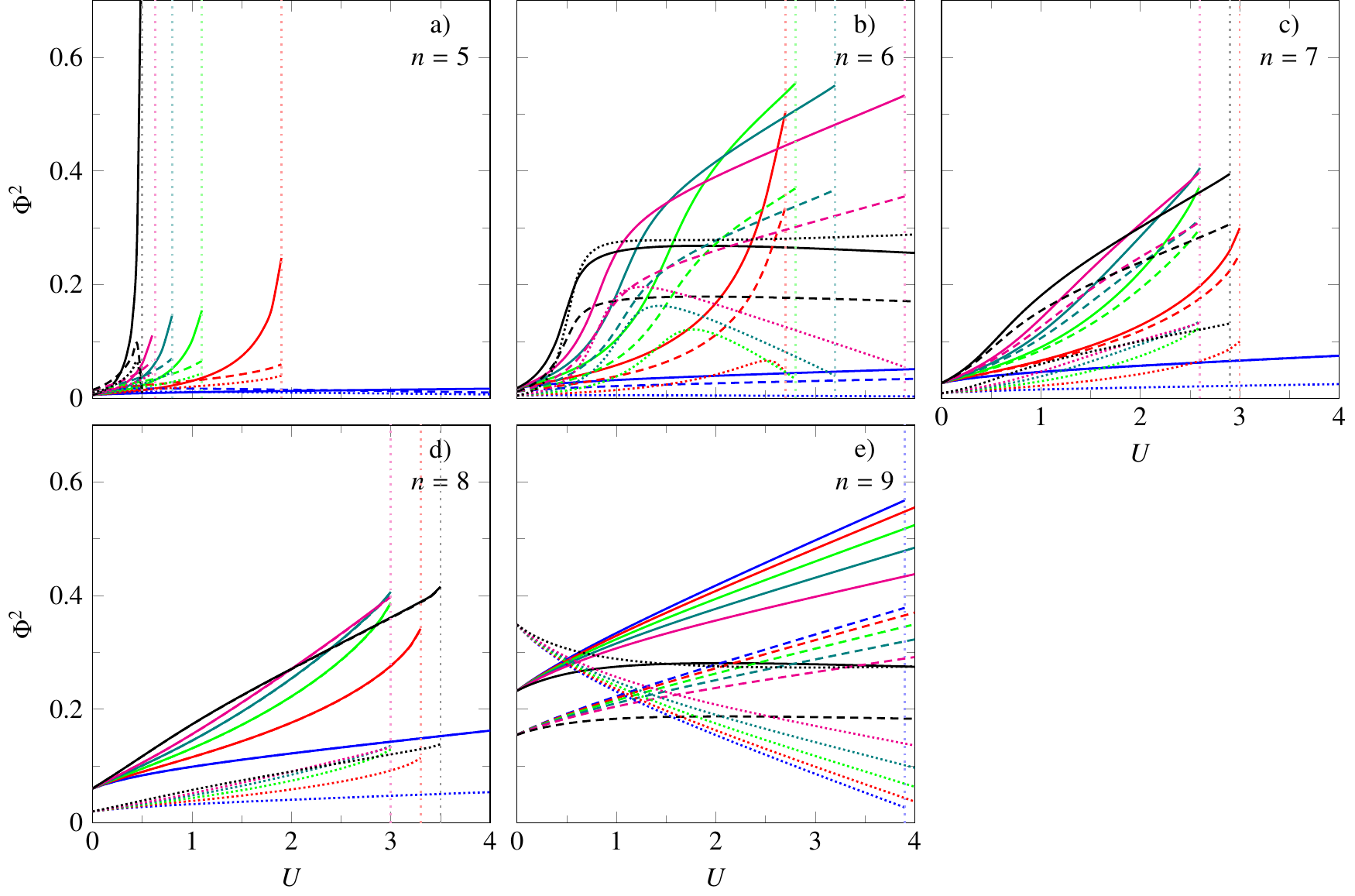}
\caption{\label{fig:phiamps-5deg} 
Slave-boson amplitudes of dominant multiplets in the degenerate 5-orbital model at 
different filling $n$ using the SC Hamiltonian.
Colors mark different $J_{\rm H}/U$, with the coding as in Fig.~\ref{fig:dshell-z}.
(a) solid: $\phi_{568}$, dashed: $\phi_{305}$, dotted: $\phi_{304}$. 
(b) solid: $\phi_{777}$, dashed: $\phi_{776}$, dotted: $\phi_{568}$.
(c) solid: $\phi_{850}$, dashed: $\phi_{846}$, dotted: $\phi_{851}$.
(d) solid: $\phi_{869}$, dashed: $\phi_{868}$, dotted: $\phi_{872}$. 
(e) solid: $\phi_{872}$, dashed: $\phi_{871}$, dotted: $\phi_{873}$. 
The symmetry/quantum number classification of the respective amplitudes is given 
in Tab.~\ref{tab:dshell-phiamps}.}
\end{figure*}%
Importantly, the RISB framework not only provides information on the QP weight which triggers
the band renormalization in reciprocal space. It also enables insight in the local real-space
competition between relevant many-body multiplets via an analysis of the occupation hierachy
of the associated slave-boson amplitudes. As a first observation, for both interaction types, 
i.e. SC and SK, the same multiplets govern the physics described by RISB. This is expected, 
because in the 3-orbital model, both parametrizations obey the same complete symmetry, 
commutating with the set $\{N,S^2,S_z,\Xi,L^2,L_z\}$. 

Figure~\ref{fig:pshell-phiamps} shows the slave-boson amplitudes for the dominant multiplet 
states at different fillings. In the case of $n=3$, the Mott transition is driven by a 
one-dimensional spin quartet of symmetry class $A_1$, resembling Hund's rule. At the 
full-localization transition, this local many-body state with three unpaired spins is the 
only remaining one. For $n=4$ and $n=5$, the local physics of the system is dominated by the 
three-dimensional $T_1$ multiplets. In the four-particle sector, it corresponds to a spin 
triplet with two unpaired spins, while in the five-particle sector to a spin doublet with 
only one unpaired spin. The $n=4$ Janus-face behaviour with increasing $U$, is associated 
with a stronger flattening of the four-particle $T_1$ in combination with a near-degeneracy 
of the three-particle $A_1$ and the five-particle $T_1$. Best illustrated for 
$J_{\rm H}/U=0.7(0.3)$ in the SC(SK) case. Hence symmetric fluctuations from the 
four-particle sector to the three/five particle are uniquely underlying the Janus-face 
physics.
\begin{figure*}[!htb]
\centering
\includegraphics*[width=12cm]{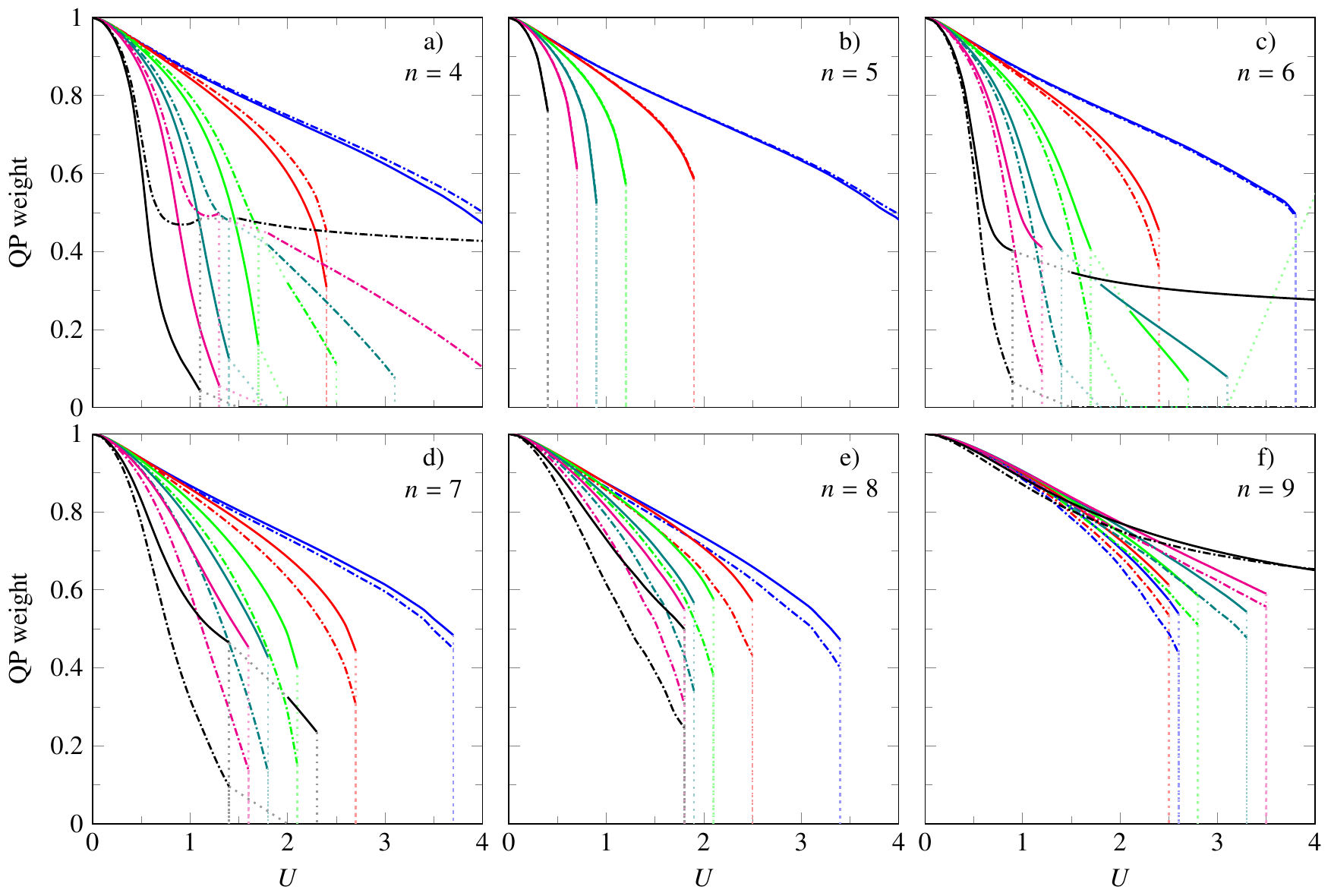}
\caption{ \label{fig:z-cfs}
Quasiparticle weight with respect to the Hubbard $U$ for the
5-orbital $d$-shell model with cubic crystal field, at different fillings $n$ and 
with varying $J_{\rm H}/U$ ratio. The SC Hamiltonian with $r=0.625$ is employed.
Solid lines: $t_{2g}$ states, dashed lines: $e_g$ states. Colors mark different 
$J_{\rm H}/U$, with the coding as in Fig.~\ref{fig:dshell-z}.
Light-dotted vertical lines are guides to the eyes for the critical $U_c$ marking 
the Mott transition. }
\end{figure*}%

To shed further light on the the fluctuations, we plot the onsite charge- and spin-correlation
function between the degenerate orbitals in Fig.~\ref{fig:pshell-corr}. Charge correlations
are negative for the proper realistic repulsion physics of the multi-orbital Hubbard
Hamiltonian, which is always ensured by the SC Hamiltonian. Due to the restricted Hilbert 
space, the fluctuations are smallest in magnitude for filling $n=5$. Because of the Hund's
rule, onsite spin fluctuations between orbitals are positive in sign. Only for $J_{\rm H}=0$
the pathological case of uncorrelated spins sets in.
For $n=4$ the correlation functions exhibit plateau-like appearance for $U$ values in the
Janus-face regime, again most obviously realized for $J_{\rm H}/U=0.7(0.3)$ in the SC(SK) 
case. This enables $J_{\rm H}$ as the dominant relevant energy scale in that regime, and
a rather irrelevant influence of moderate changes of $U$.

\subsection{5-Orbital $d$-Shell Model}
Let us turn to the case of five interacting orbitals on the cubic lattice. Because of
the highly enlarged Hilbert space compared to the 3-orbital scenario, the introduced
symmetry considerations are now truly indispensable for a computational study of such a
manifold. Three cases are explored, namely the orbital-degenerate, the more realistic 
problem of electrons within an octahedral crystal field, and finally the orbital-degenerate
case with spin-orbit coupling.

\subsubsection{Degenerate Case}
\begin{table}[t]
\centering
\begin{tabular}{ c l c }
  \toprule
  $\Gamma$ & symmetry & nonzero \\
  304 & $E:   \sum (4,2,\sigma,4,2,\{-2,0,2\})$ & 5 \\
  305 & $T_2: \sum (4,1,\sigma,4,2,\{-2,-1,1,2\})$ & 6 \\
  568 & $A_1: \sum (5,\nicefrac{5}{2},\sigma,5,0,0)$ & 1 \\
  776 & $E:   \sum (6,2,\sigma,4,2,\{-2,0,2\})$ & 5 \\
  777 & $T_2: \sum (6,2,\sigma,4,2,\{-2,-1,1,2\})$ & 6 \\
  846 & $T_1: \sum (7,\nicefrac{3}{2},\sigma,3,\{1,3\},\{-3,-1,0,1,3\})$ & 22 \\
  850 & $T_2: \sum (7,\nicefrac{3}{2},\sigma,3,3,\{-3,-2,-1,1,2,3\})$ & 12 \\
  868 & $T_1: \sum (8,1,\sigma,2,\{1,3\},\{-3,-1,1,3\})$ & 22 \\
  869 & $T_2: \sum (8,1,\sigma,2,3,\{-3,-2,-1,1,2,3\})$ & 12 \\
  871 & $E:   \sum (9,\nicefrac{1}{2},\sigma,1,2,\{-2,0,2\})$ & 5 \\
  872 & $E:   \sum (9,\nicefrac{1}{2},\sigma,1,2,\{-2,-1,1,2\})$ & 6 \\
  873 & $A_1:      (10,0,0,0,0,0)$ & 1 \\        
\end{tabular}
\caption{\label{tab:dshell-phiamps} 
Classification of dominant slave-boson-amplitudes 
associated with local multiplets as plotted in Fig.~\ref{fig:phiamps-5deg} for the 
degenerate 5-orbital model. See Tab.~\ref{tab:pshell-phiamps} for a detailed explanation
of the symmetry labelling. All sums are normalized by a factor of $\sqrt{2s+1}$.}
\end{table}
Starting with degenerate orbitals, we again apply the full cubic point group for the 
underlying symmetry analysis. The SC as well as the SK Hamiltonian are put into
practise. The Slater-Condon form is employed for the two parametrizations $r=0.625$ 
and $r=1.8$.
The resulting QP weights with respect to the Hubbard $U$ are shown in Fig.~\ref{fig:dshell-z}.
As in the 3-orbital case, globally, the correlation strength increases with growing $n$ for
$J_{\rm H}=0$ and weakens for finite $J_{\rm H}$. There is no clear trend concerning the
size of the differences between the Hamiltoninan forms, but those dependent strongly on 
filling and magnitude of the Hund's exchange. Characteristic features are nonetheless
observable for all three local-interaction forms. The most prominent feature is again
the Janus-face physics, which is most dominant for $n=6$, but occurs also for $n=7$ and
with minor fingerprints also for $n=8$. In comparison to the 3-orbital case, the Janus-face
signature especially for $n=6$ appears more manifest. Again $J_{\rm H}/U=0.7$ in the 
SC case with $r=0.625$ and $J_{\rm H}/U=0.3$ in the SK display the strongest signature.
\begin{figure*}[htb]
\centering
\includegraphics*[width=12cm]{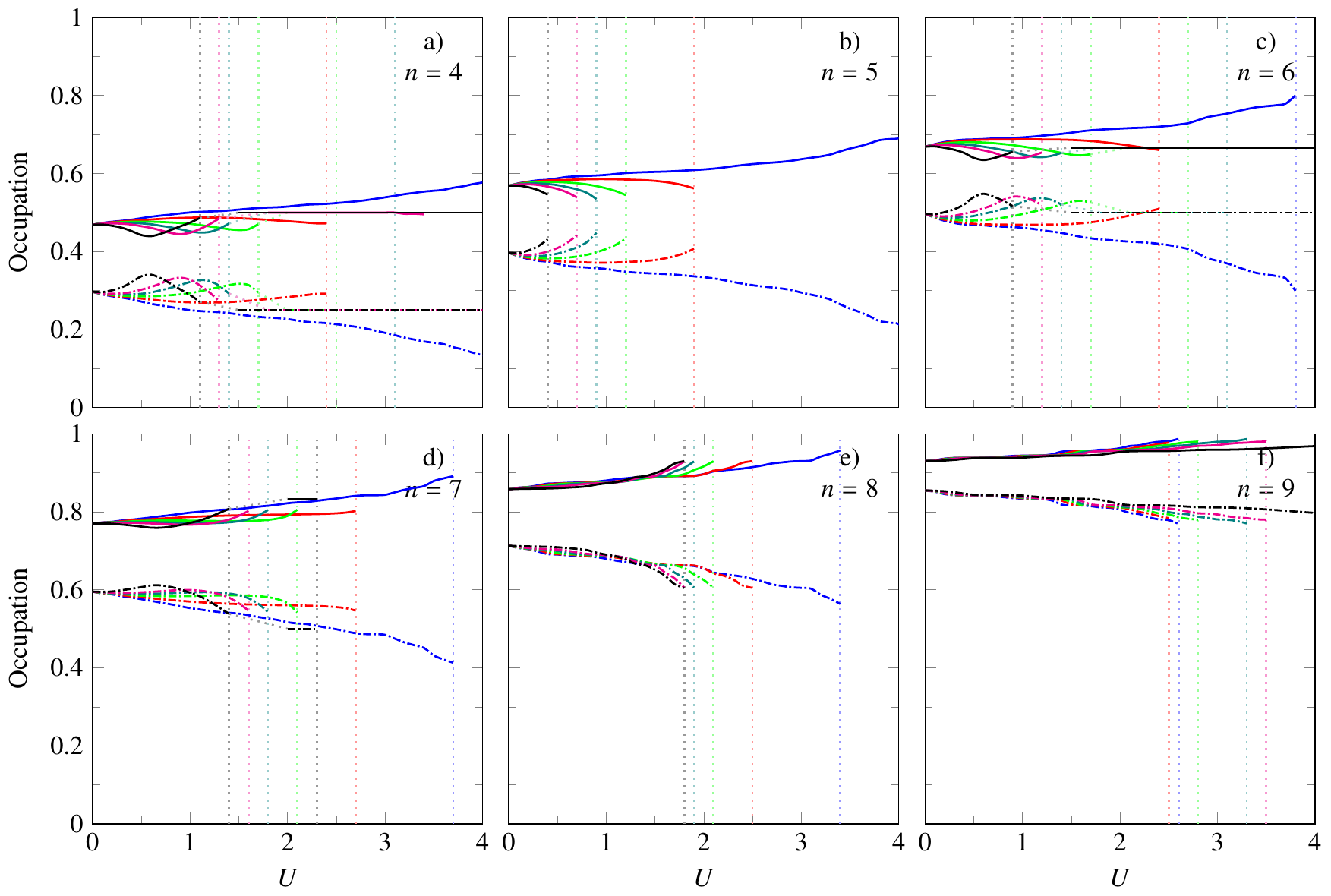}
\caption{\label{fig:occ-cfs}
Occupation of the $e_g$ (solid) and the $t_{2g}$ (dashed) orbitals with respect to the 
Hubbard $U$ for the 5-orbital $d$-shell model with cubic crystal field, at different 
fillings $n$ and with varying $J_{\rm H}/U$ ratio. The data is given per orbital and per
spin, i.e. the total $e_g$($t_{2g}$) occupation is obtained by multiplication with 4(6).
The SC Hamiltonian with $r=0.625$ is employed. Light-dotted vertical lines are guides to 
the eyes for the critical $U_c$ marking the Mott transition.}
\end{figure*}%

For the discussion of the slave-boson amplitudes we restrict the data to the most adequate
case of the SC($r=0.625$) Hamiltonian. The amplitudes for the dominant multiplet states 
are shown in Fig.~\ref{fig:phiamps-5deg}, with the explanation of the $\phi_\Gamma$ labelling
in Tab.~\ref{tab:dshell-phiamps}. Nonsurprisingly, an one-dimensional $A_1$ spin sextet 
with five unpaired spins triggers the Mott transition at half filling. On the other hand, 
the Janus-face phenomenology at $n=6$ seems more intriguing than in the 3-orbital case. Though
again a seemingly concerted behavior of the dominant multiplets takes place, the 
higher seven-particle sector is not majorly involved. Instead, the named $A_1$ five-particle
spin sextet and a six-particle $s=2$ multiplet with $T_2$ symmetry are the key competitors,
accompanied with another six-particle $s=2$ multiplet of $E$ symmetry. For smaller $U$,
in particular the $A_1$- and $E$-symmetry state have similar weight, thus fluctuations
between the five- and six-particle sector occurs via those symmetry channels. The energy
separation between the relevant multiplets increases at larger fillings, giving rise to
an increasingly decoupled behavior with less relevance of fluctuations.

\subsubsection{Cubic Crystal-Field Splitting\label{5orb-cf}}
Degenerate $p$(-like) models may be adequate for some materials problems, e.g. for
the threefold correlated $t_{2g}$ electrons in SrVO$_3$~\cite{pav05}. However, there is 
always a crystal-field splitting between the states of a $d$-shell atom on realistic 
lattices. Thus we like to include an example application of our advanced RISB scheme to
the simplest case, given by a 5-orbital model with cubic crystal-field splitting, i.e. the 
term $\mathcal{H}^{\rm (cf)}$ in the local Hamiltonian (\ref{eq:locham}) becomes now nonzero.
To facilitate the crystal field, a splitting $\Delta=0.2$ 
between the $e_g=\{z^2, x^2-y^2\}$ and the $t_{2g}=\{xz, yz, xy\}$ states is used, here
explicitly reading $\Delta_{e_g}=3/5\Delta$ and $\Delta_{t_{2g}}=-2/5\Delta$ in 
view of eq.~(\ref{eq:cfham}) (cf. Fig~\ref{fig:comic-o-cfs}). This resembles a simplistic
model for an oxide perovskite where the key transition-metal site is located on a simple-cubic
lattice and explicit oxygen degrees of freedom in octahedral position are integrated out.
Note that the hopping for the different orbitals within the $d$-shell are here kept identical.

Albeit the problem of competing high- and low-spin states in 5-orbital
manifolds with crystal field is a prominent one, in our basic application we do not 
investigate such physics. The present crystal-field size is rather small compared
to the band width as well as the local Coulomb interations. Hence the theoretical 
treatment does not result in obvious low-spin states, and furthermore an explicit effort 
to reveal high-to-low-spin transitions is not undertaken.

Figure~\ref{fig:z-cfs} exhibits the ($e_g$,$t_{2g}$) QP weight for different fillings and
varying $J_{\rm H}/U$ ratio, while Fig.~\ref{fig:occ-cfs} displays the associated orbital
fillings. Due to the chosen crystal-field splitting, the $t_{2g}$ states are lower in energy 
than the $e_g$ ones and therefore have larger occupation. At half-filling $n=5$, the 
correlation strength of the two orbital sectors is nearly degenerate and increases with
growing $J_{\rm H}/U$. The filling difference $n_{t_{2g}}-n_{e_g}$ tends to grow with 
rising $U$, but then decreases again close to the Mott transition. For fixed $U$, a larger
Hund's exchange reduces the filling difference, since $J_{\rm H}$ favors orbital balancing
and is a natural opponent of the crystal-field splitting. As expected, away from half filling
the correlation and filling scenario is more sophisticated. For the electron-doped $n=6$ case
the Hund's physics results in a manifest interplay of the Janus-face behavior with 
orbital-selective mechanisms. The less-filled $e_g$ states are close to half filling and
become more strongly correlated than $t_{2g}$. For $J_{\rm H}/U>0.1$, while both orbital
sectors develop Janus-face signature, the $e_g$ orbitals become localized at larger $U$
with the $t_{2g}$ orbitals still metallic~\cite{lanata13}. The localization of the 
$t_{2g}$ states takes then place at even larger $U$. Within the orbital-selective 
$e_g$-localized sector, the occupation is fixed to half-filled orbitals. Further electron
doping leads to a quick vanishing of the orbital-selective behavior. For $n=7$ it only
occurs in a small $U$ window for the extreme case $J_{\rm H}/U=0.7$. For $n=7,8$ the
($e_g$,$t_{2g}$) sectors, though with different QP weight, enter the Mott-insulating regime
at the same critical interaction strength via a first-order transition.
\begin{figure*}[htb]
\centering
\includegraphics*[width=12cm]{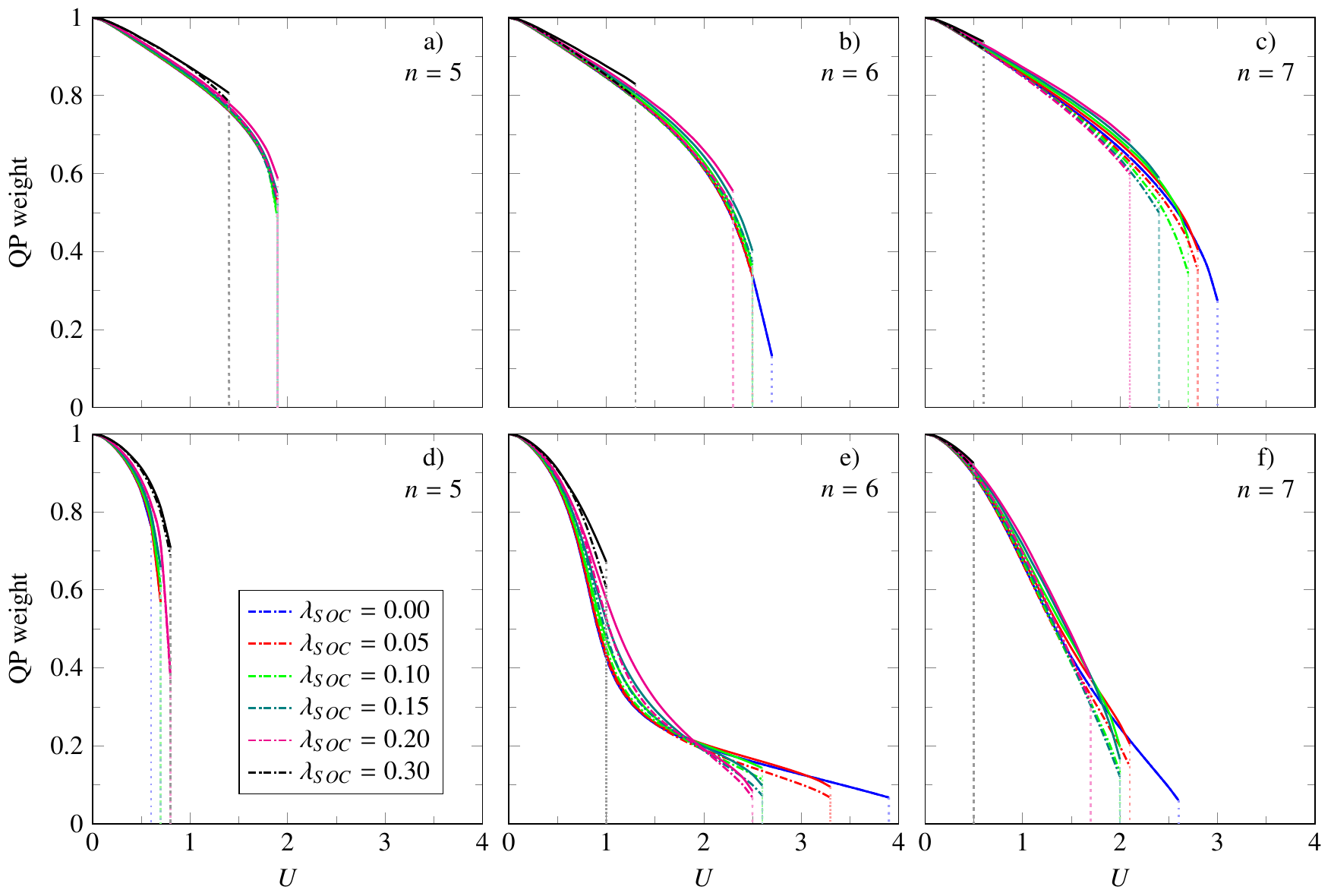}
\caption{ \label{fig:5degsoc-z} 
Quasiparticle weight with respect to the Hubbard $U$ for the degenerate
5-orbital $d$-shell model with spin-orbit coupling, at different fillings $n$ and
two different regimes for the Hund's exchange: (a-c): $J_{\rm H}/U=0.1$ and
(d-f): $J_{\rm H}/U=0.4$. The SC Hamiltonian with $r=0.625$ is employed.
Full lines: $j=3/2$, dashed lines: $j=5/2$.
Light-dotted vertical lines are guides to the eyes for the critical $U_c$ marking 
the metal-insulator transition.}
\end{figure*}%
\begin{figure*}[htb]
\centering
\includegraphics*[width=12cm]{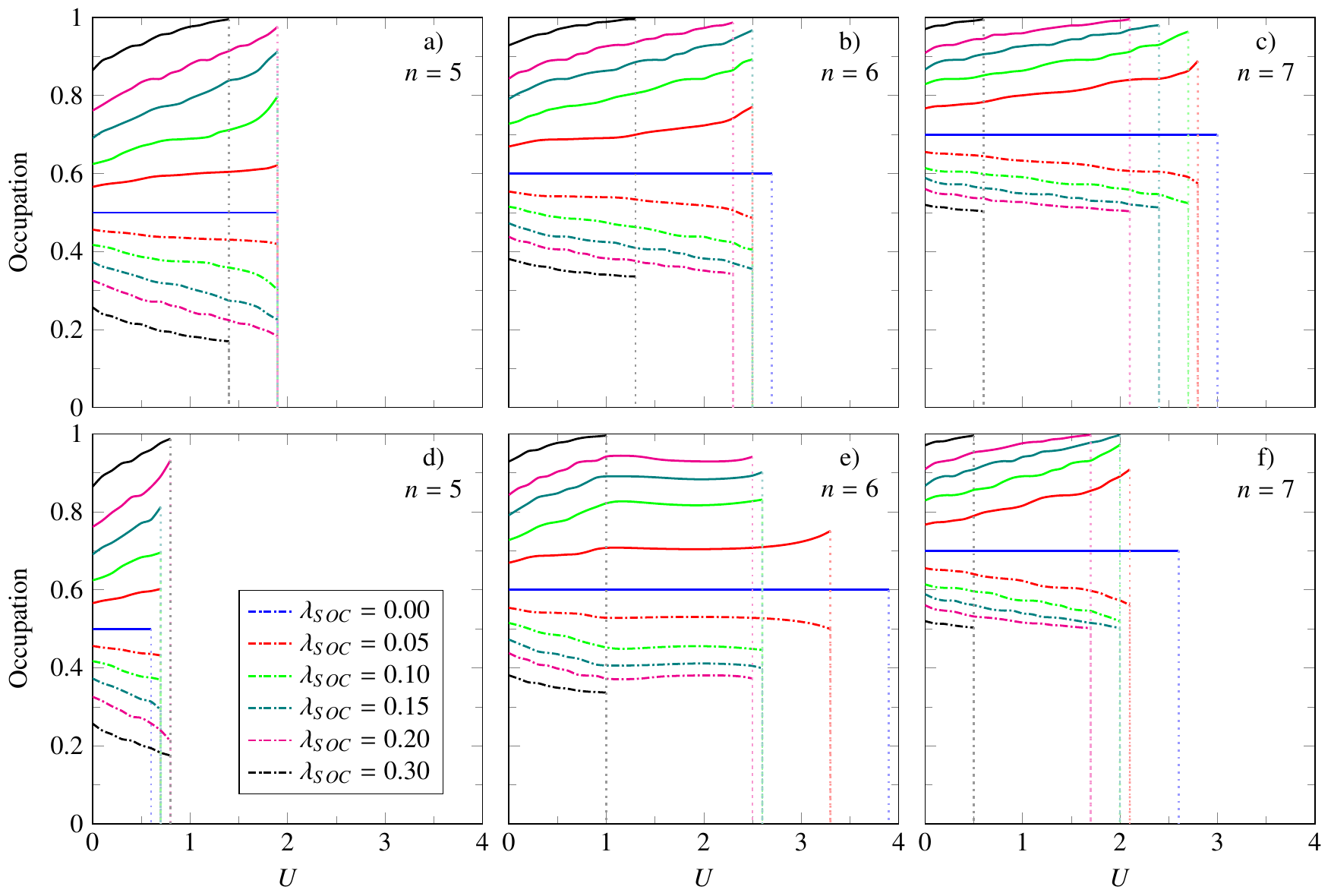}
\caption{ \label{fig:5degsoc-fill} 
Occupation with respect to the Hubbard $U$ in the degenerate
5-orbital $d$-shell model with spin-orbit coupling, at different fillings $n$ and
two different regimes for the Hund's exchange: (a-c): $J_{\rm H}/U=0.1$ and
(d-f): $J_{\rm H}/U=0.4$. The SC Hamiltonian with $r=0.625$ is employed.
Full lines: $j=3/2$, dashed lines: $j=5/2$.
Light-dotted vertical lines are guides to the eyes for the critical $U_c$ marking 
the metal-insulator transition.}
\end{figure*}%

Note that there is an obvious breaking of particle-hole symmetry with crystal-field 
splitting: whereas an additional electron most likely enters the lower-lying crystal-field
level, an additional hole usually favors the higher-lying one. Thus in the hole-doped case 
of $n=4$, the $t_{2g}$ states are now close to half-filling, whereas the $e_g$ ones are
closer to quarter filling. This leads to an inverse orbital-selective behavior in comparison
to $n=6$. Namely, the $t_{2g}$ orbitals become first localized and afterwards the $e_g$ 
orbitals at larger interaction strength. Because of the different size of the orbital
sectors, the $Z(U)$ curves for $n=4,6$ differ not only by an interchange of $e_g$ against
$t_{2g}$.

\subsubsection{Spin-Orbit Coupling}
We conclude the model applications by providing results for the degenerate 5-orbital 
model with finite spin-orbit coupling. The interplay between electron correlations and
SOC has become of seminal interest in the study of strongly correlated 
materials~\cite{jac09,pes10,wat10,mar11,beh12,ham15}. Especially for certain $4d$ and $5d$ 
transition-metal compounds, processes based on interweaved Mott- and SOC physics may lead to 
intriguing phenomena. But also for specific correlated $3d$ compounds, such as e.g. the 
iron pnictides and chalcogenides~\cite{bor16}, the impact of the spin-orbit interaction is 
heavily discussed.

Here, we again provide only a glimpse on this phenomenology, mainly to illustrate the 
power and potential of the advanced RISB scheme to tackle this issue. Albeit the interplay
of crystal-field splitting and spin-orbit coupling is often a relevant physics in materials,
we restrict the basic discussion here to the degenerate 5-orbital model on a cubic lattice.
The Slater-Condon Hamiltonian (with $r=0.625$) for the local interaction is utilized and
two distinct ratios $J_{\rm H}/U$ are employed with rising spin-orbit-interaction paramter
$\lambda_{\rm soc}$. Namely, we study the cases $J_{\rm H}/U=0.1$ and $J_{\rm H}/U=0.4$.
The chosen $\lambda_{\rm soc}$ values put us in the regime of strong spin-orbit coupling,
e.g. as applicable for iridates~\cite{kim08}.
Basically, the additional interaction gives rise to novel states in the 5-orbital manifold, 
given by the quantum number $j=l\pm 1/2$. For the case of a complete $d$-shell with $l=2$, the
$j$ states split into two manifolds: a threefold-degenerate group with $j=5/2$ and a 
twofold-degenerate group with $j=3/2$. Hence, as in the former case of a ($e_g$,$t_{2g}$) 
crystal-field, a splitting of the physical quantities into two sectors is expected. Since
we keep $\lambda_{\rm soc}$ positive, the $j=3/2$ manifold lies lower than the $j=5/2$ one.

Figure~\ref{fig:5degsoc-z} displays the quasiparticle weight $Z$ for different fillings $n$.
Indeed, one may observe the grouping into the both $j=5/2,3/2$ sectors, respectively. From $Z$, 
a finite spin-orbit coupling leads to a further lowering of the critical Hubbard $U$, i.e., 
increases the correlation strength for fixed Coulomb-interaction parameters. With growing
deviation from half filling, this effect also becomes stronger in relative size. Most 
interesting is the case of $n=6$ and sizable $J_{\rm H}/U$, i.e. the setting that results
in the Janus-face regime. There, a rising spin-orbit coupling weakens the Janus-face 
signature, with a final disappearanc at $\lambda_{\rm soc}=0.3$. Note that overall for every
filling and interaction, there is no 'orbital-selective' Mott transition taking place.
Contrary to the crystal-field case, both $j$-sectors become insulating at the identical
interaction strength. Still, the fillings in the two sectors are respectively
rather different (see Fig.~\ref{fig:5degsoc-fill}). The effective spin-orbit splitting
results in a stronger-filled $j=3/2$ manifold, which for rising $U$ and sizable 
$\lambda_{\rm soc}$ rather quickly becomes fully occupied. Only in the original Janus-face
regime, the $j$ polarization may be contained for not-too large $\lambda_{\rm soc}$
within a larger regime of $U$ values.

\subsection{Realistic Application: Iron Chalcogenides\label{sec:realistic}}
In the final application, we tackle a realistic problem and show that the advanced
RISB scheme allied with DFT may serve as a versatile theoretical tool to analyze intricate
correlated materials. Documenting a materials case for the reported
technical advancements is the present main concern. 
The prominent iron chalcogenides FeSe and FeTe as compounds with seemingly important Hund's 
physics are adequate materials examples~\cite{aic10,hau12,gla15,leo15} in line with the previous 
discussions in the model context.
Note that combinations of Gutzwiller techniques and DFT have been applied to the problem of
Fe pnictides and -chalcogenides in previous works~\cite{wan10,schi12,lanata13}.

Figure~\ref{fig:fechalco-struc}a depicts the symmetry-identical $\alpha$-phase 
crystal structure of tetragonal kind (space group $P4/nmm$) of the two compounds. Iron 
square lattices with lattice constant $a$, having Se, Te up and below the square centre in a 
distance $h$, are stacked along the $c$-axis with distance $c$. Based on the available 
experimental data~\cite{mar08,miz09}, we
employed $a^{\rm (FeSe)}=3.77$\,\AA, $a^{\rm (FeTe)}=3.82$\,\AA;
$c^{\rm (FeSe)}=5.52$\,\AA, $c^{\rm (FeTe)}=6.29$\,\AA\; and 
$h^{\rm (FeSe)}=1.47$\,\AA, $h^{\rm (FeTe)}=1.76$\,\AA. Hence as expected, the FeTe compund
has increased lattice parameters compared to FeSe, most significantly a more elongated 
$c$-axis parameter. The nominal filling of the Fe $3d$-shell amounts to $n=6$ electrons,
i.e. as also learned from the previous model results, the systems are good candidates for 
manifest Hund's physics.
There are various theoretical assessments of the local Coulomb integrals
for iron pnictides and -chalcogenides, e.g.~\cite{aic09,miy10}. Therefrom it is agreed that
a value for the Hubbard interaction of $U=4.0$\,eV and a value for the Hund's exchange of 
$J_{\rm H}=0.8$\,eV, i.e. $J_{\rm H}/U=0.2$, are proper choices. In order to study the 
relevance of $J_{\rm H}$, we here take again $U$ as a parameter and allow for two 
different ratios between the Hubbard interaction and the Hund's exchange, 
namely $J_{\rm H}/U=0.15$ and $J_{\rm H}/U=0.224$.

\subsubsection{DFT characterization}
\begin{figure}[b]
\centering
(a)\includegraphics*[width=6cm]{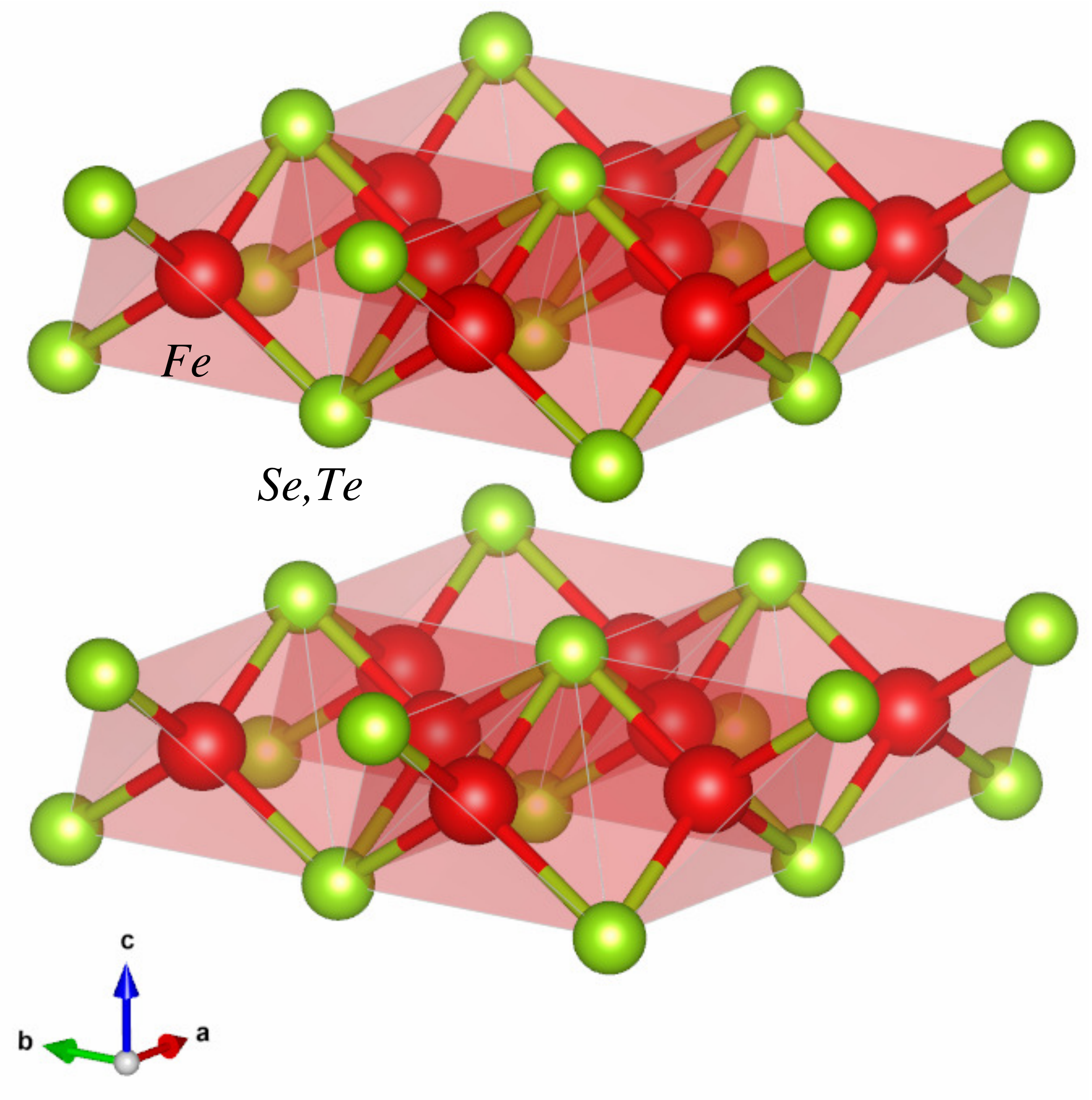}\\[0.2cm]
(b)\includegraphics*[width=8cm]{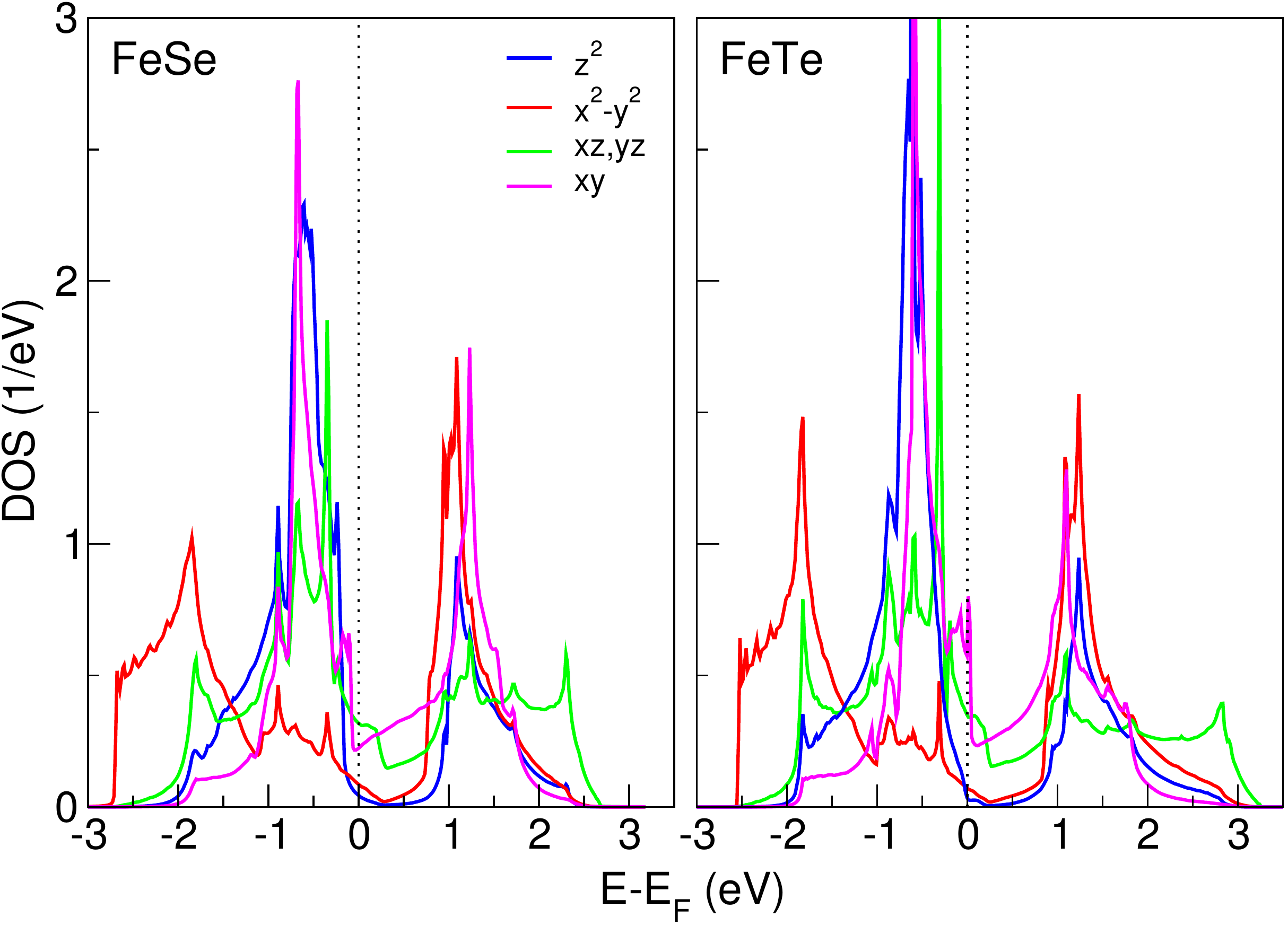}
\caption{\label{fig:fechalco-struc} 
(a) $P4/nmm$ crystal structure and (b) DFT orbital-resolved local density of states 
(lDOS) of FeSe and FeTe.}
\end{figure}%
Let us first report the results within density functional theory (DFT) using the 
generalized-gradient approximation (GGA) based on the Perdew-Burke-Ernzerhof exchange-correlation
functional~\cite{per96}. Original DFT results for FeSe and FeTe have been presented by
Subedi {\sl et al.}~\cite{sub08}. Here, we construct a Wannier-like characterization of
the DFT(GGA) electronic structure based on projected-local orbitals~\cite{ama08}.
The Fe $3d$-shell is split by crystal field, leading to onsite levels 
$\varepsilon_m=\{z^2,x^2-y^2,xz,yz,xy\}$. Those read 
$\varepsilon^{\rm (FeSe)}_m=\{-235,-459,100,100,156\}\,$meV and
$\varepsilon^{\rm (FeTe)}_m=\{-200,-285,177,177,138\}\,$meV. Thus although the internal
($e_g$,$t_{2g}$) degeneracy is mostly lifted within the tetragonal symmetry, the $e_g$ manifold
is energetically favored against $t_{2g}$. This yields also a stronger filling of the $e_g$
states. The orbital-resolved DFT fillings (in the same order as the crystal-field levels) read
$n^{\rm (FeSe)}_m=\{1.50,1.15,1.13,1.13,1.05\}$ and 
$n^{\rm (FeTe)}_m=\{1.48,1.11,1.15,1.15,1.10\}$. Note that the crystal-field level of the 
$x^2-y^2$ orbital is lower than the $z^2$ one. But due to the wider bimodal local density of 
states (lDOS) of the former (cf. Fig.~\ref{fig:fechalco-struc}b), related to the major 
contribution to the inplane bonding, the electron occupation is highest in 
the $z^2$ state. Both compounds have a seemingly very similar DFT electronic structure, e.g.,
the dominant inplane bonding between $x^2-y^2$ orbitals is mediated by a nearest-neighbor 
hopping amplitude of $t_{x^2-y^2}=-438$\,meV for FeSe and of $t_{x^2-y^2}=-433$\,meV for FeTe.
But some differences are observable. FeTe has a slightly larger bandwidth, i.e.,
5.6\,eV compare to the 5.3\,eV of FeSe. Furthermore, FeTe has a sharper $z^2$- as well as 
$xz,yz$-lDOS and higher density of states at the Fermi level compared to FeSe. In comparison
to the previous model studies, an effective Hubbard interation normalized to the half
bandwidth reads $U_{\rm eff}\sim 4.0/2.7\sim 1.5$.

\subsubsection{RISB quasiparticle weights and orbital occupations}
We first employ the DFT Kohn-Sham Hamiltonian obtained within the projected-local-orbital 
formalism in the 'one-shot' or 'post-processing' combinational scheme with RISB, similar 
to the previous approaches by Schickling {\sl et al.}~\cite{schi12} and 
Lanat{\`a}~{\sl et al.}~\cite{lanata13}. To facilitate this, the proper symmetry relations 
invoked by the tetragonal point group are implemented and utilized.

Before discussing the concrete results, it is important to realize the differences of the
present realistic problem compared to the model-5-orbital problem with crystal field from
section~\ref{5orb-cf}. First, the reduced tetragonal symmetry lifts the degeneracy within the 
$e_g$ and $t_{2g}$ subshell, and the crystal-field splitting is now also active 
between these non-degenerate levels. Second, the intra-orbital hopping becomes orbital
dependent and is not bound to the nearest-neighbor term as in the model case. Third, there
are in addition also inter-orbital hopping terms from short to longer range included in the
Kohn-Sham Hamiltonian.

\begin{figure}[t]
\centering
\includegraphics*[width=8.5cm]{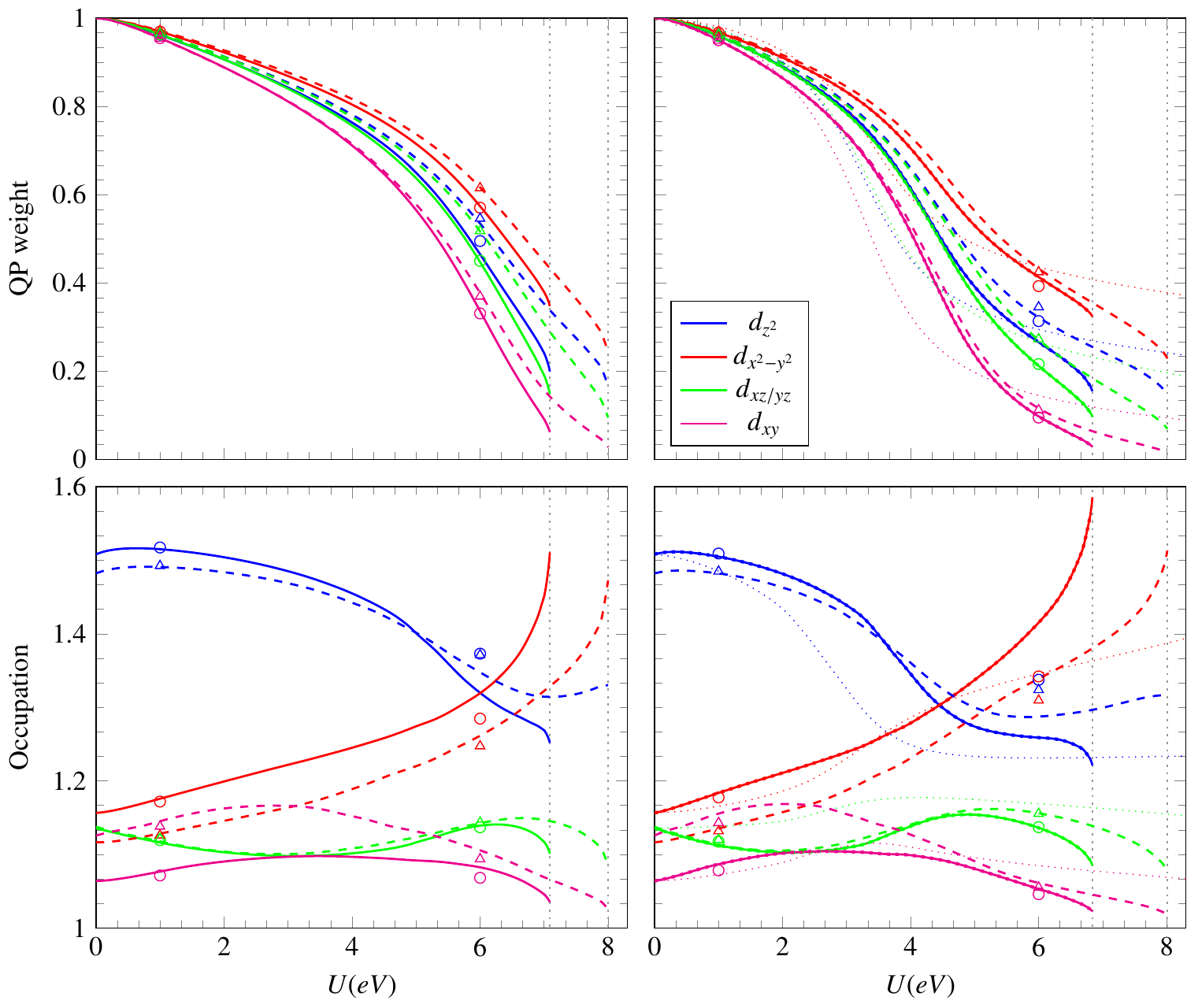}
\caption{\label{fig:tetra-chalo} 
Orbital-resolved quasiparticle weight (top) and occupation (bottom) for FeSe (solid lines) 
and FeTe (dashed lines). For the interacting part, the SC Hamiltonian with $r=0.625$ is used.
Results from charge self-consistent calculations are given by by circles/triangles for 
FeSe/FeTe. Left panel: $J_{\rm H}/U=0.150$, right panel: $J_{\rm H}/U=0.224$. For comparison 
the right panel additionally contains FeSe results for the SC Hamiltonian with $r=1.8$
(thick-dashed lines, mostly on top of the $r=0.625$ curves) and for the SK Hamiltonian 
(light dotted lines).}
\end{figure}

Figure~\ref{fig:tetra-chalo} shows the resulting orbital-dependent quasiparticle weights as 
well as occupation with rising Hubbard $U$. From the QP weights, the FeSe compound seems
slighly more correlated than FeTe, providing also a lower critical $U_{\rm c}$ for the
theoretical Mott transition. The least-filled $xy$ orbital is most strongly correlated as
it resides closest to half filling, similar as the scenario in the model case. For the
case of larger $J_{\rm H}/U=0.224$, which is closer to the assumed realistic interaction,
indeed the Janus-face signature sets in, and is most pronounced for $xy$. The iron 
chalcogenides are hence truly a clear case for dominant Hund's physics. But note that no
explicit orbital-selective behavior is observed, in contrary to the model case with crystal
field. This is probably due to the longer-range- and intra-orbital hoppings that lead to
a stronger entanglement of the orbital correlations.
Interestingly, there is a crossover in the $z^2$ vs. $x^2-y^2$ occupations with rising $U$.
For larger $U$ closer to the Mott transition, the $x^2-y^2$ orbital gains more electrons. This
filling-hierachy change happens for the larger $J_{\rm H}/U$ ratio at smaller $U$, close to
the expected value of $U=4$\,eV. For completeness, we included the FeSe data for the 
Slater-Condon Hamiltonian with $r=1.8$ as well as the Slater-Kanamori Hamiltonian. While 
the former does not yield significant changes, the use of the latter simplified Hamiltonian 
form results in somewhat stronger correlations for intermediate $U$, but shifts the theoretical
Mott transition to larger interaction strengths.
Finally, we incorporated charge self-consistent DFT+RISB results for $Z$ and the orbital
occupations at $U=1$\,eV and $U=6$\,eV. The effect of charge self-consistency for
the QP weight are generally minor, qualitatively heading to a slight correlation-strength
increase. But there are changes for the orbital occupations at larger $U$. Namely, the 
mentioned crossover in the filling hiearchy within the 'one-shot' calculations tends to be 
shifted to larger $U$ or is even absent in the CSC treatment. Thus the orbital occupations
especially in the $e_g$ states of the iron chalcogenides appear to be sensitive to the 
many-body charge-density handling. Note that we do not touch on the prominent issue of 
nematicity~\cite{wat15} in the present context. Since the tetragonal symmetry is hard-cored 
here and spin-orbit effects are neglected in this realistic example, such anisotropic effects 
are excluded. Future studies lifting those restrictions may enable results on this specific
physics.

\section{Summary}
We here documented a rigorous implementation of the mean-field version of the
rotational-invariant slave-boson (RISB) approach in terms of an efficient symmetry-adapted 
handling of multi-orbital degrees of freedom. The point-group symmetry of the underlying
lattice is used to reduce by general means the number of relevant slave-boson amplitudes. 
Complete generality in the form of the local Hamiltonian ensures access to demanding
interacting lattice problems. Spin-orbit coupling is naturally included in the symmetry-faithful
formalism via the introduction of double-group representations. In the present work, the
degree of complexity is limited to cubic- and tetragonal symmetry for up to five local 
orbitals, suitable for generic $d$-shell studies. However, a further advancement of the
present scheme to other (low-)symmetry cases is straightforward and with the present computing
power, seven-orbital problems, i.e. treating a full-interacting $f$-shell, are in principle
within the performance capabilities. The advanced RISB framework is implemented such as to 
be applied to model Hamiltonians, to 'one-shot' combinations with DFT or within a full charge
self-consistent DFT+RISB methodology.

Selected applications within the prominent research field of Hund's physics have been 
presented, however without providing an in-depth survey of the encountered physics. The Hund's
physics with its highlighting influence of the Hund's exchange $J_{\rm H}$ in the presence
of a finite Hubbard $U$ came here across for 3-orbital- and 5-orbital models, as well as in
the realistic context of the FeSe and FeTe compound. The emergence of the Janus-face signature 
and its interplay with orbital-selective physics in the presence of a crystal field 
were reported for the model cases. Moreover, we showed the weakening of this hallmark signature 
with finite spin-orbit interaction. The reliability of the method also for true materials
problems was proven by verifying FeSe and FeTe as the known Hund's materials. Thereby, advanced
RISB is capable of dealing with the subtle differences in the electronic structure. We 
showed that the $e_g$ polarizations of $z^2$- and $x^2-y^2$ kind in these iron 
chalcogenides are prone to an orbital crossover, which might be relevant for fluctuation-driven
processes in the given materials.

\acknowledgments
Computations were performed at the University of Hamburg and at the North-German 
Supercomputing Alliance (HLRN) under Grant No. hhp00035.

\appendix
\section{Computation of the renormalization matrix\label{app:renorm}}
The renormalization matrix ${\bf R}$ may be computed in the multi-orbital framework as follows.
Equation~(\ref{eq:rmatmulti}) provides the standard form for the renormalization matrix,
reading
\begin{equation}
R_{\alpha\beta}^* = \sum_\gamma T^*_{\alpha\gamma}w_{\gamma\beta}\quad.
\end{equation}
The matrix {\bf T} can be written as
\begin{equation}
T_{\alpha\gamma}^* = \mbox{Tr}(\phi^\dagger f_\alpha^\dagger 
\phi c_\gamma)\quad,
\end{equation} 
since
\begin{eqnarray}
\mbox{Tr}(\phi^\dagger f_\alpha^\dagger \phi c_\gamma)
&=&\sum_{AB,CD}\phi^\dagger_{AB}(f_\alpha^\dagger)_{BC}\phi^{\hfill}_{CD}
(c_\gamma)_{DA}\nonumber\\
&=&\sum_{AB,CD}\phi^*_{BA}(f_\alpha^\dagger)_{BC}\phi^{\hfill}_{CD}
(c_\gamma)_{DA}\nonumber\\
&=&\sum_{AB,CD}\langle B|f_\alpha^\dagger|C\rangle
\langle D|c_\gamma|A\rangle\phi^*_{BA}\phi^{\hfill}_{CD}\nonumber\\
&=&\sum_{AB,CD}\langle B|f_\alpha^\dagger|C\rangle
\langle A|c^\dagger_\gamma|D\rangle^*\phi^*_{BA}\phi^{\hfill}_{CD}\quad.
\end{eqnarray}
As the element $\langle A|c^\dagger_\gamma|D\rangle$ is a real number when using the Fock basis,
this form can also be written in the following way,
\begin{equation}
\mbox{Tr}(\phi^\dagger f_\alpha^\dagger \phi c_\gamma)
=\sum_{AB,nm}\langle A|f_\alpha^\dagger|B\rangle
\langle n|c^\dagger_\gamma|m\rangle^*\phi^*_{An}\phi^{\hfill}_{Bm}\quad,
\end{equation}
when changing notation, i.e. $B\rightarrow A$, $C\rightarrow B$, $A\rightarrow n$,
$D\rightarrow m$. Therewith, the final expression coincides with the formula for the renormalization
matrix in the original RISB paper~\cite{lec07}.

\section{Generation of basis matrices\label{app:basmat}}
\noindent
1. We start with a representation $\Theta$ of the point group $G$
on the local many body states (i.e., the adapted basis)
$\mathcal{A}$. It is constructed by taking the Euler (angular
momentum) representation of every group element and repeating it for
all other combinations of quantum numbers others than the total angular
momentum. This creates offdiagonal matrices, which by construction
contain the point-group representation multiple repeated times. The
number of repetitions ${\cal R}$ can be obtained from the vector product of
the character vector of the representation $\Theta$ ($\chi(\Theta)$) and the
character vector of an irreducible representation $\chi(\Theta^\alpha)$, 
normalized to the number of group elements $n_G$:
\begin{equation}
  r = \frac{1}{n_G} \chi(\Theta)\,\chi(\Theta^\alpha)
\end{equation} 
\noindent
2. Then, Casimir-like operators $C_i$ are built, one for every
equivalence class, by summing up all elements of one equivalence
class in the above representation, via
\begin{equation}
  0=[C_i,g]\quad \forall\; i, g
\end{equation}
They commutate with all elements $g$ of the group and with each
other. The commutating set of $C_i$-operators can be simultaniously
diagonalized. The emerging basis block-diagonalizes all elements of the group
in blocks of dimension of the irreducible representation
times its repetitions, i.e., $dim(\Theta^\alpha)\,{\cal R}$.

\noindent
3. To find a basis $\mathcal{K}$ that diagonalizes all elements of the
group, one has to perform a Koster-phase fixing~\cite{koster58} to
separate repeated irreducible representations from each other by 
means of an irreducible representation $\Theta^\alpha$ of the same
point group.

\noindent
4. Now, identity matrices of the dimension of the irreducible
representation are set up, according to Schur's lemma commutating with
all elements of the group (up to a complex factor, which will play the
role of the variational parameter in our context). All basis matrices
for the actual irreducible representation that commutate with all group
elements in basis $\mathcal{K}$ are now identified by taking each and
every summand of the (offdiagonal) direct sum of the identity matrices
separately, according to the number of repetitions, which amounts to
$n_{\tilde{\phi}}(\Theta^\alpha)={\cal R}^2$ different matrices.

\noindent
5. Rotating these identity matrices back to $\mathcal{A}$ results in the
desired basis matrices.\\

\bibliography{bibextra}

\begin{thebibliography}{86}
\expandafter\ifx\csname natexlab\endcsname\relax\def\natexlab#1{#1}\fi
\expandafter\ifx\csname bibnamefont\endcsname\relax
  \def\bibnamefont#1{#1}\fi
\expandafter\ifx\csname bibfnamefont\endcsname\relax
  \def\bibfnamefont#1{#1}\fi
\expandafter\ifx\csname citenamefont\endcsname\relax
  \def\citenamefont#1{#1}\fi
\expandafter\ifx\csname url\endcsname\relax
  \def\url#1{\texttt{#1}}\fi
\expandafter\ifx\csname urlprefix\endcsname\relax\def\urlprefix{URL }\fi
\providecommand{\bibinfo}[2]{#2}
\providecommand{\eprint}[2][]{\url{#2}}

\bibitem[{\citenamefont{Gutzwiller}(1963)}]{gut63}
\bibinfo{author}{\bibfnamefont{M.~C.} \bibnamefont{Gutzwiller}},
  \bibinfo{journal}{Phys. Rev. Lett.} \textbf{\bibinfo{volume}{10}},
  \bibinfo{pages}{159} (\bibinfo{year}{1963}).

\bibitem[{\citenamefont{B\"{u}nemann and Gebhard}(2007)}]{bue07}
\bibinfo{author}{\bibfnamefont{J.}~\bibnamefont{B\"{u}nemann}}
  \bibnamefont{and} \bibinfo{author}{\bibfnamefont{F.}~\bibnamefont{Gebhard}},
  \bibinfo{journal}{Phys. Rev. B} \textbf{\bibinfo{volume}{76}},
  \bibinfo{pages}{193104} (\bibinfo{year}{2007}).

\bibitem[{\citenamefont{B\"{u}nemann et~al.}(1998)\citenamefont{B\"{u}nemann,
  Weber, and Gebhard}}]{bue98}
\bibinfo{author}{\bibfnamefont{J.}~\bibnamefont{B\"{u}nemann}},
  \bibinfo{author}{\bibfnamefont{W.}~\bibnamefont{Weber}}, \bibnamefont{and}
  \bibinfo{author}{\bibfnamefont{F.}~\bibnamefont{Gebhard}},
  \bibinfo{journal}{Phys. Rev. B} \textbf{\bibinfo{volume}{57}},
  \bibinfo{pages}{6896} (\bibinfo{year}{1998}).

\bibitem[{\citenamefont{Huang et~al.}(2012)\citenamefont{Huang, Du, and
  Dai}}]{hua12}
\bibinfo{author}{\bibfnamefont{L.}~\bibnamefont{Huang}},
  \bibinfo{author}{\bibfnamefont{L.}~\bibnamefont{Du}}, \bibnamefont{and}
  \bibinfo{author}{\bibfnamefont{X.}~\bibnamefont{Dai}},
  \bibinfo{journal}{Phys. Rev. B} \textbf{\bibinfo{volume}{86}},
  \bibinfo{pages}{035150} (\bibinfo{year}{2012}).

\bibitem[{\citenamefont{Lanat\`a et~al.}(2012)\citenamefont{Lanat\`a, Strand,
  Dai, and Hellsing}}]{lanata12}
\bibinfo{author}{\bibfnamefont{N.}~\bibnamefont{Lanat\`a}},
  \bibinfo{author}{\bibfnamefont{H.~U.~R.} \bibnamefont{Strand}},
  \bibinfo{author}{\bibfnamefont{X.}~\bibnamefont{Dai}}, \bibnamefont{and}
  \bibinfo{author}{\bibfnamefont{B.}~\bibnamefont{Hellsing}},
  \bibinfo{journal}{Phys. Rev. B} \textbf{\bibinfo{volume}{85}},
  \bibinfo{pages}{035133} (\bibinfo{year}{2012}).

\bibitem[{\citenamefont{Barnes}(1976)}]{bar76}
\bibinfo{author}{\bibfnamefont{S.~E.} \bibnamefont{Barnes}},
  \bibinfo{journal}{J. Phys. F: Metal Phys.} \textbf{\bibinfo{volume}{6}},
  \bibinfo{pages}{1375} (\bibinfo{year}{1976}).

\bibitem[{\citenamefont{Coleman}(1983)}]{col83}
\bibinfo{author}{\bibfnamefont{P.}~\bibnamefont{Coleman}},
  \bibinfo{journal}{Phys. Rev. B} \textbf{\bibinfo{volume}{28}},
  \bibinfo{pages}{5255} (\bibinfo{year}{1983}).

\bibitem[{\citenamefont{Read and Newns}(1983)}]{rea83}
\bibinfo{author}{\bibfnamefont{N.}~\bibnamefont{Read}} \bibnamefont{and}
  \bibinfo{author}{\bibfnamefont{D.}~\bibnamefont{Newns}}, \bibinfo{journal}{J.
  Phys. C} \textbf{\bibinfo{volume}{16}}, \bibinfo{pages}{3273}
  (\bibinfo{year}{1983}).

\bibitem[{\citenamefont{Kotliar and Ruckenstein}(1986)}]{kot86}
\bibinfo{author}{\bibfnamefont{G.}~\bibnamefont{Kotliar}} \bibnamefont{and}
  \bibinfo{author}{\bibfnamefont{A.~E.} \bibnamefont{Ruckenstein}},
  \bibinfo{journal}{Phys. Rev. Lett.} \textbf{\bibinfo{volume}{57}},
  \bibinfo{pages}{1362} (\bibinfo{year}{1986}).

\bibitem[{\citenamefont{Arovas and Auerbach}(1988)}]{aro88}
\bibinfo{author}{\bibfnamefont{D.~P.} \bibnamefont{Arovas}} \bibnamefont{and}
  \bibinfo{author}{\bibfnamefont{A.}~\bibnamefont{Auerbach}},
  \bibinfo{journal}{Phys. Rev. B} \textbf{\bibinfo{volume}{38}},
  \bibinfo{pages}{316} (\bibinfo{year}{1988}).

\bibitem[{\citenamefont{Kotliar and Liu}(1988)}]{kot88}
\bibinfo{author}{\bibfnamefont{G.}~\bibnamefont{Kotliar}} \bibnamefont{and}
  \bibinfo{author}{\bibfnamefont{J.}~\bibnamefont{Liu}},
  \bibinfo{journal}{Phys. Rev. B} \textbf{\bibinfo{volume}{38}},
  \bibinfo{pages}{5142(R)} (\bibinfo{year}{1988}).

\bibitem[{\citenamefont{Li et~al.}(1989)\citenamefont{Li, W\"olfle, and
  Hirschfeld}}]{li89}
\bibinfo{author}{\bibfnamefont{T.}~\bibnamefont{Li}},
  \bibinfo{author}{\bibfnamefont{P.}~\bibnamefont{W\"olfle}}, \bibnamefont{and}
  \bibinfo{author}{\bibfnamefont{P.~J.} \bibnamefont{Hirschfeld}},
  \bibinfo{journal}{Phys. Rev. B} \textbf{\bibinfo{volume}{40}},
  \bibinfo{pages}{6817} (\bibinfo{year}{1989}).

\bibitem[{\citenamefont{Hasegawa}(1997)}]{has97}
\bibinfo{author}{\bibfnamefont{H.}~\bibnamefont{Hasegawa}},
  \bibinfo{journal}{J. Phys. Soc. Jpn.} \textbf{\bibinfo{volume}{66}},
  \bibinfo{pages}{1391} (\bibinfo{year}{1997}).

\bibitem[{\citenamefont{Fr\'{e}sard and Kotliar}(1997)}]{fre97}
\bibinfo{author}{\bibfnamefont{R.}~\bibnamefont{Fr\'{e}sard}} \bibnamefont{and}
  \bibinfo{author}{\bibfnamefont{G.}~\bibnamefont{Kotliar}},
  \bibinfo{journal}{Phys. Rev. B} \textbf{\bibinfo{volume}{56}},
  \bibinfo{pages}{12909} (\bibinfo{year}{1997}).

\bibitem[{\citenamefont{Florens and Georges}(2004)}]{flo04}
\bibinfo{author}{\bibfnamefont{S.}~\bibnamefont{Florens}} \bibnamefont{and}
  \bibinfo{author}{\bibfnamefont{A.}~\bibnamefont{Georges}},
  \bibinfo{journal}{Phys. Rev. B} \textbf{\bibinfo{volume}{70}},
  \bibinfo{pages}{035114} (\bibinfo{year}{2004}).

\bibitem[{\citenamefont{de' Medici et~al.}(2005)\citenamefont{de' Medici,
  Georges, and Biermann}}]{med05}
\bibinfo{author}{\bibfnamefont{L.}~\bibnamefont{de' Medici}},
  \bibinfo{author}{\bibfnamefont{A.}~\bibnamefont{Georges}}, \bibnamefont{and}
  \bibinfo{author}{\bibfnamefont{S.}~\bibnamefont{Biermann}},
  \bibinfo{journal}{Phys. Rev. B} \textbf{\bibinfo{volume}{72}},
  \bibinfo{pages}{205124} (\bibinfo{year}{2005}).

\bibitem[{\citenamefont{Lechermann et~al.}(2007)\citenamefont{Lechermann,
  Georges, Kotliar, and Parcollet}}]{lec07}
\bibinfo{author}{\bibfnamefont{F.}~\bibnamefont{Lechermann}},
  \bibinfo{author}{\bibfnamefont{A.}~\bibnamefont{Georges}},
  \bibinfo{author}{\bibfnamefont{G.}~\bibnamefont{Kotliar}}, \bibnamefont{and}
  \bibinfo{author}{\bibfnamefont{O.}~\bibnamefont{Parcollet}},
  \bibinfo{journal}{Phys. Rev. B} \textbf{\bibinfo{volume}{76}},
  \bibinfo{pages}{155102} (\bibinfo{year}{2007}).

\bibitem[{\citenamefont{Isidori and Capone}(2009)}]{isi09}
\bibinfo{author}{\bibfnamefont{A.}~\bibnamefont{Isidori}} \bibnamefont{and}
  \bibinfo{author}{\bibfnamefont{M.}~\bibnamefont{Capone}},
  \bibinfo{journal}{Phys. Rev. B} \textbf{\bibinfo{volume}{80}},
  \bibinfo{pages}{115120} (\bibinfo{year}{2009}).

\bibitem[{\citenamefont{B\"{u}nemann}(2011)}]{bue11}
\bibinfo{author}{\bibfnamefont{J.}~\bibnamefont{B\"{u}nemann}},
  \bibinfo{journal}{phys. stat. sol. (b)} \textbf{\bibinfo{volume}{248}},
  \bibinfo{pages}{203} (\bibinfo{year}{2011}).

\bibitem[{\citenamefont{Georgescu and Ismail-Beigi}(2015)}]{geo15}
\bibinfo{author}{\bibfnamefont{A.~B.} \bibnamefont{Georgescu}}
  \bibnamefont{and}
  \bibinfo{author}{\bibfnamefont{S.}~\bibnamefont{Ismail-Beigi}},
  \bibinfo{journal}{Phys. Rev. B} \textbf{\bibinfo{volume}{92}},
  \bibinfo{pages}{235117} (\bibinfo{year}{2015}).

\bibitem[{\citenamefont{Facio et~al.}(2017)\citenamefont{Facio, Vildosola,
  Garc{\'i}a, and Cornaglia}}]{fac17}
\bibinfo{author}{\bibfnamefont{J.~I.} \bibnamefont{Facio}},
  \bibinfo{author}{\bibfnamefont{V.}~\bibnamefont{Vildosola}},
  \bibinfo{author}{\bibfnamefont{D.~J.} \bibnamefont{Garc{\'i}a}},
  \bibnamefont{and} \bibinfo{author}{\bibfnamefont{P.~S.}
  \bibnamefont{Cornaglia}}, \bibinfo{journal}{Phys. Rev. B}
  \textbf{\bibinfo{volume}{95}}, \bibinfo{pages}{085119}
  (\bibinfo{year}{2017}).

\bibitem[{\citenamefont{Raczkowski et~al.}(2006)\citenamefont{Raczkowski,
  Fr{\'e}sard, and Ole{\'s}}}]{rac06}
\bibinfo{author}{\bibfnamefont{M.}~\bibnamefont{Raczkowski}},
  \bibinfo{author}{\bibfnamefont{R.}~\bibnamefont{Fr{\'e}sard}},
  \bibnamefont{and} \bibinfo{author}{\bibfnamefont{A.~M.}
  \bibnamefont{Ole{\'s}}}, \bibinfo{journal}{Phys. Rev. B}
  \textbf{\bibinfo{volume}{73}}, \bibinfo{pages}{174525}
  (\bibinfo{year}{2006}).

\bibitem[{\citenamefont{Lechermann}(2009)}]{lec09}
\bibinfo{author}{\bibfnamefont{F.}~\bibnamefont{Lechermann}},
  \bibinfo{journal}{Phys. Rev. Lett.} \textbf{\bibinfo{volume}{102}},
  \bibinfo{pages}{046403} (\bibinfo{year}{2009}).

\bibitem[{\citenamefont{Ferrero et~al.}(2009)\citenamefont{Ferrero, Cornaglia,
  Leo, Parcollet, Kotliar, and Georges}}]{fer092}
\bibinfo{author}{\bibfnamefont{M.}~\bibnamefont{Ferrero}},
  \bibinfo{author}{\bibfnamefont{P.~S.} \bibnamefont{Cornaglia}},
  \bibinfo{author}{\bibfnamefont{L.~D.} \bibnamefont{Leo}},
  \bibinfo{author}{\bibfnamefont{O.}~\bibnamefont{Parcollet}},
  \bibinfo{author}{\bibfnamefont{G.}~\bibnamefont{Kotliar}}, \bibnamefont{and}
  \bibinfo{author}{\bibfnamefont{A.}~\bibnamefont{Georges}},
  \bibinfo{journal}{Phys. Rev. B} \textbf{\bibinfo{volume}{80}},
  \bibinfo{pages}{064501} (\bibinfo{year}{2009}).

\bibitem[{\citenamefont{Mazin et~al.}(2014)\citenamefont{Mazin, Jeschke,
  Lechermann, Lee, Fink, Thomale, and Valent{\'i}}}]{maz14}
\bibinfo{author}{\bibfnamefont{I.~I.} \bibnamefont{Mazin}},
  \bibinfo{author}{\bibfnamefont{H.~O.} \bibnamefont{Jeschke}},
  \bibinfo{author}{\bibfnamefont{F.}~\bibnamefont{Lechermann}},
  \bibinfo{author}{\bibfnamefont{H.}~\bibnamefont{Lee}},
  \bibinfo{author}{\bibfnamefont{M.}~\bibnamefont{Fink}},
  \bibinfo{author}{\bibfnamefont{R.}~\bibnamefont{Thomale}}, \bibnamefont{and}
  \bibinfo{author}{\bibfnamefont{R.}~\bibnamefont{Valent{\'i}}},
  \bibinfo{journal}{Nat. Commun.} \textbf{\bibinfo{volume}{5}},
  \bibinfo{pages}{4261} (\bibinfo{year}{2014}).

\bibitem[{\citenamefont{Behrmann et~al.}(2012)\citenamefont{Behrmann, Piefke,
  and Lechermann}}]{beh12}
\bibinfo{author}{\bibfnamefont{M.}~\bibnamefont{Behrmann}},
  \bibinfo{author}{\bibfnamefont{C.}~\bibnamefont{Piefke}}, \bibnamefont{and}
  \bibinfo{author}{\bibfnamefont{F.}~\bibnamefont{Lechermann}},
  \bibinfo{journal}{Phys. Rev. B} \textbf{\bibinfo{volume}{86}},
  \bibinfo{pages}{045130} (\bibinfo{year}{2012}).

\bibitem[{\citenamefont{Schuwalow et~al.}(2012)\citenamefont{Schuwalow, Piefke,
  and Lechermann}}]{schu12}
\bibinfo{author}{\bibfnamefont{S.}~\bibnamefont{Schuwalow}},
  \bibinfo{author}{\bibfnamefont{C.}~\bibnamefont{Piefke}}, \bibnamefont{and}
  \bibinfo{author}{\bibfnamefont{F.}~\bibnamefont{Lechermann}},
  \bibinfo{journal}{Phys. Rev. B} \textbf{\bibinfo{volume}{85}},
  \bibinfo{pages}{205132} (\bibinfo{year}{2012}).

\bibitem[{\citenamefont{Hampel et~al.}(2015)\citenamefont{Hampel, Piefke, and
  Lechermann}}]{ham15}
\bibinfo{author}{\bibfnamefont{A.}~\bibnamefont{Hampel}},
  \bibinfo{author}{\bibfnamefont{C.}~\bibnamefont{Piefke}}, \bibnamefont{and}
  \bibinfo{author}{\bibfnamefont{F.}~\bibnamefont{Lechermann}},
  \bibinfo{journal}{Phys. Rev. B} \textbf{\bibinfo{volume}{92}},
  \bibinfo{pages}{085141} (\bibinfo{year}{2015}).

\bibitem[{\citenamefont{Behrmann and Lechermann}(2015{\natexlab{a}})}]{beh15}
\bibinfo{author}{\bibfnamefont{M.}~\bibnamefont{Behrmann}} \bibnamefont{and}
  \bibinfo{author}{\bibfnamefont{F.}~\bibnamefont{Lechermann}},
  \bibinfo{journal}{Phys. Rev. B} \textbf{\bibinfo{volume}{91}},
  \bibinfo{pages}{075110} (\bibinfo{year}{2015}{\natexlab{a}}).

\bibitem[{\citenamefont{Behrmann and Lechermann}(2015{\natexlab{b}})}]{beh15_2}
\bibinfo{author}{\bibfnamefont{M.}~\bibnamefont{Behrmann}} \bibnamefont{and}
  \bibinfo{author}{\bibfnamefont{F.}~\bibnamefont{Lechermann}},
  \bibinfo{journal}{Phys. Rev. B} \textbf{\bibinfo{volume}{92}},
  \bibinfo{pages}{125148} (\bibinfo{year}{2015}{\natexlab{b}}).

\bibitem[{\citenamefont{Schir{\'o} and Fabrizio}(2010)}]{schi10}
\bibinfo{author}{\bibfnamefont{M.}~\bibnamefont{Schir{\'o}}} \bibnamefont{and}
  \bibinfo{author}{\bibfnamefont{M.}~\bibnamefont{Fabrizio}},
  \bibinfo{journal}{Phys. Rev. Lett.} \textbf{\bibinfo{volume}{105}},
  \bibinfo{pages}{076401} (\bibinfo{year}{2010}).

\bibitem[{\citenamefont{Behrmann et~al.}(2013)\citenamefont{Behrmann, Fabrizio,
  and Lechermann}}]{beh13}
\bibinfo{author}{\bibfnamefont{M.}~\bibnamefont{Behrmann}},
  \bibinfo{author}{\bibfnamefont{M.}~\bibnamefont{Fabrizio}}, \bibnamefont{and}
  \bibinfo{author}{\bibfnamefont{F.}~\bibnamefont{Lechermann}},
  \bibinfo{journal}{Phys. Rev. B} \textbf{\bibinfo{volume}{88}},
  \bibinfo{pages}{035116} (\bibinfo{year}{2013}).

\bibitem[{\citenamefont{Behrmann et~al.}(2016)\citenamefont{Behrmann,
  Lichtenstein, Katsnelson, and Lechermann}}]{beh16}
\bibinfo{author}{\bibfnamefont{M.}~\bibnamefont{Behrmann}},
  \bibinfo{author}{\bibfnamefont{A.~I.} \bibnamefont{Lichtenstein}},
  \bibinfo{author}{\bibfnamefont{M.~I.} \bibnamefont{Katsnelson}},
  \bibnamefont{and}
  \bibinfo{author}{\bibfnamefont{F.}~\bibnamefont{Lechermann}},
  \bibinfo{journal}{Phys. Rev. B} \textbf{\bibinfo{volume}{94}},
  \bibinfo{pages}{165120} (\bibinfo{year}{2016}).

\bibitem[{\citenamefont{Werner et~al.}(2008)\citenamefont{Werner, Gull, Troyer,
  and Millis}}]{wer08}
\bibinfo{author}{\bibfnamefont{P.}~\bibnamefont{Werner}},
  \bibinfo{author}{\bibfnamefont{E.}~\bibnamefont{Gull}},
  \bibinfo{author}{\bibfnamefont{M.}~\bibnamefont{Troyer}}, \bibnamefont{and}
  \bibinfo{author}{\bibfnamefont{A.~J.} \bibnamefont{Millis}},
  \bibinfo{journal}{Phys. Rev. Lett.} \textbf{\bibinfo{volume}{101}},
  \bibinfo{pages}{166405} (\bibinfo{year}{2008}).

\bibitem[{\citenamefont{Haule and Kotliar}(2009)}]{hau09}
\bibinfo{author}{\bibfnamefont{K.}~\bibnamefont{Haule}} \bibnamefont{and}
  \bibinfo{author}{\bibfnamefont{G.}~\bibnamefont{Kotliar}},
  \bibinfo{journal}{New J. Phys.} \textbf{\bibinfo{volume}{11}},
  \bibinfo{pages}{025021} (\bibinfo{year}{2009}).

\bibitem[{\citenamefont{de' Medici et~al.}(2011)\citenamefont{de' Medici,
  Mravlje, and Georges}}]{med11}
\bibinfo{author}{\bibfnamefont{L.}~\bibnamefont{de' Medici}},
  \bibinfo{author}{\bibfnamefont{J.}~\bibnamefont{Mravlje}}, \bibnamefont{and}
  \bibinfo{author}{\bibfnamefont{A.}~\bibnamefont{Georges}},
  \bibinfo{journal}{Phys. Rev. Lett.} \textbf{\bibinfo{volume}{107}},
  \bibinfo{pages}{256401} (\bibinfo{year}{2011}).

\bibitem[{\citenamefont{Georges et~al.}(2013)\citenamefont{Georges, de' Medici,
  and Mravlje}}]{geo13}
\bibinfo{author}{\bibfnamefont{A.}~\bibnamefont{Georges}},
  \bibinfo{author}{\bibfnamefont{L.}~\bibnamefont{de' Medici}},
  \bibnamefont{and} \bibinfo{author}{\bibfnamefont{J.}~\bibnamefont{Mravlje}},
  \bibinfo{journal}{Annu. Rev. Condens. Matter Phys.}
  \textbf{\bibinfo{volume}{4}}, \bibinfo{pages}{137} (\bibinfo{year}{2013}).

\bibitem[{Note1()}]{Note1}
Note1, \bibinfo{note}{in order not to overburden the notation, we do not mark
  quantum operators explicitly. The difference to c-numbers should be obvious
  from the context, or is mentioned explicitly when needed.}

\bibitem[{Note2()}]{Note2}
Note2, \bibinfo{note}{onedimensional physics is in principle also accessible
  via RISB by invoking a cluster description (see section~\ref {sec:mf}). Yet
  within the scope of the present work, we are not discussing such variations
  of the method.}

\bibitem[{\citenamefont{W\"olfle}(1995)}]{wol95}
\bibinfo{author}{\bibfnamefont{P.}~\bibnamefont{W\"olfle}},
  \bibinfo{journal}{J. Low Temp. Phys} \textbf{\bibinfo{volume}{99}},
  \bibinfo{pages}{625} (\bibinfo{year}{1995}).

\bibitem[{\citenamefont{Nayak}(2000)}]{nay00}
\bibinfo{author}{\bibfnamefont{C.}~\bibnamefont{Nayak}},
  \bibinfo{journal}{Phys. Rev. Lett.} \textbf{\bibinfo{volume}{85}},
  \bibinfo{pages}{178} (\bibinfo{year}{2000}).

\bibitem[{\citenamefont{Lee et~al.}(2006)\citenamefont{Lee, Nagaosa, and
  Wen}}]{leenag06}
\bibinfo{author}{\bibfnamefont{P.~A.} \bibnamefont{Lee}},
  \bibinfo{author}{\bibfnamefont{N.}~\bibnamefont{Nagaosa}}, \bibnamefont{and}
  \bibinfo{author}{\bibfnamefont{X.-G.} \bibnamefont{Wen}},
  \bibinfo{journal}{Rev. Mod. Phys.} \textbf{\bibinfo{volume}{78}},
  \bibinfo{pages}{17} (\bibinfo{year}{2006}).

\bibitem[{\citenamefont{Fr\'{e}sard and Kopp}(2012)}]{fre12}
\bibinfo{author}{\bibfnamefont{R.}~\bibnamefont{Fr\'{e}sard}} \bibnamefont{and}
  \bibinfo{author}{\bibfnamefont{T.}~\bibnamefont{Kopp}},
  \bibinfo{journal}{Ann. Phys. (Berlin)} \textbf{\bibinfo{volume}{524}},
  \bibinfo{pages}{175} (\bibinfo{year}{2012}).

\bibitem[{\citenamefont{Lanat{\`a} et~al.}(2017)\citenamefont{Lanat{\`a}, Yao,
  Deng, Dobrosaljevi{\'c}, and Kotliar}}]{lan17}
\bibinfo{author}{\bibfnamefont{N.}~\bibnamefont{Lanat{\`a}}},
  \bibinfo{author}{\bibfnamefont{Y.}~\bibnamefont{Yao}},
  \bibinfo{author}{\bibfnamefont{X.}~\bibnamefont{Deng}},
  \bibinfo{author}{\bibfnamefont{V.}~\bibnamefont{Dobrosaljevi{\'c}}},
  \bibnamefont{and} \bibinfo{author}{\bibfnamefont{G.}~\bibnamefont{Kotliar}},
  \bibinfo{journal}{Phys. Rev. Lett.} \textbf{\bibinfo{volume}{118}},
  \bibinfo{pages}{126401} (\bibinfo{year}{2017}).

\bibitem[{\citenamefont{Georges et~al.}(1996)\citenamefont{Georges, Kotliar,
  Krauth, and Rozenberg}}]{geo96}
\bibinfo{author}{\bibfnamefont{A.}~\bibnamefont{Georges}},
  \bibinfo{author}{\bibfnamefont{G.}~\bibnamefont{Kotliar}},
  \bibinfo{author}{\bibfnamefont{W.}~\bibnamefont{Krauth}}, \bibnamefont{and}
  \bibinfo{author}{\bibfnamefont{M.~J.} \bibnamefont{Rozenberg}},
  \bibinfo{journal}{Rev. Mod. Phys.} \textbf{\bibinfo{volume}{68}},
  \bibinfo{pages}{13} (\bibinfo{year}{1996}).

\bibitem[{\citenamefont{Kotliar and Vollhardt}(2004)}]{kot04}
\bibinfo{author}{\bibfnamefont{G.}~\bibnamefont{Kotliar}} \bibnamefont{and}
  \bibinfo{author}{\bibfnamefont{D.}~\bibnamefont{Vollhardt}},
  \bibinfo{journal}{Physics Today} \textbf{\bibinfo{volume}{57}},
  \bibinfo{pages}{53} (\bibinfo{year}{2004}).

\bibitem[{\citenamefont{Raimondi and Castellani}(1993)}]{rai93}
\bibinfo{author}{\bibfnamefont{R.}~\bibnamefont{Raimondi}} \bibnamefont{and}
  \bibinfo{author}{\bibfnamefont{C.}~\bibnamefont{Castellani}},
  \bibinfo{journal}{Phys. Rev. B} \textbf{\bibinfo{volume}{48}},
  \bibinfo{pages}{11453(R)} (\bibinfo{year}{1993}).

\bibitem[{\citenamefont{Zimmermann et~al.}(1997)\citenamefont{Zimmermann,
  Fr{\'e}sard, and W{\"o}lfle}}]{zim97}
\bibinfo{author}{\bibfnamefont{W.}~\bibnamefont{Zimmermann}},
  \bibinfo{author}{\bibfnamefont{R.}~\bibnamefont{Fr{\'e}sard}},
  \bibnamefont{and}
  \bibinfo{author}{\bibfnamefont{P.}~\bibnamefont{W{\"o}lfle}},
  \bibinfo{journal}{Phys. Rev. B} \textbf{\bibinfo{volume}{56}},
  \bibinfo{pages}{10097} (\bibinfo{year}{1997}).

\bibitem[{\citenamefont{Maier et~al.}(2005)\citenamefont{Maier, Jarrell,
  Pruschke, and Hettler}}]{mai05}
\bibinfo{author}{\bibfnamefont{T.}~\bibnamefont{Maier}},
  \bibinfo{author}{\bibfnamefont{M.}~\bibnamefont{Jarrell}},
  \bibinfo{author}{\bibfnamefont{T.}~\bibnamefont{Pruschke}}, \bibnamefont{and}
  \bibinfo{author}{\bibfnamefont{M.~H.} \bibnamefont{Hettler}},
  \bibinfo{journal}{Rev. Mod. Phys.} \textbf{\bibinfo{volume}{77}},
  \bibinfo{pages}{1027} (\bibinfo{year}{2005}).

\bibitem[{\citenamefont{Drake}(2006)}]{drake06}
\bibinfo{author}{\bibfnamefont{G.~W.~F.} \bibnamefont{Drake}},
  \emph{\bibinfo{title}{Springer Handbook of Atomic, Molecular, and Optical
  Physics}} (\bibinfo{publisher}{Springer-Verlag New York},
  \bibinfo{year}{2006}).

\bibitem[{\citenamefont{Rudzikas}(2007)}]{rudz07}
\bibinfo{author}{\bibfnamefont{Z.}~\bibnamefont{Rudzikas}},
  \emph{\bibinfo{title}{Theoretical Atomic Spectroscopy}}
  (\bibinfo{publisher}{Cambridge University Press}, \bibinfo{year}{2007}).

\bibitem[{Note3()}]{Note3}
Note3, \bibinfo{note}{the trace operation 'Tr' is throughout this work
  generally understood as the summation of the diagonal elements of the
  involved objects.}

\bibitem[{\citenamefont{Koster}(1958)}]{koster58}
\bibinfo{author}{\bibfnamefont{G.~F.} \bibnamefont{Koster}},
  \bibinfo{journal}{Phys. Rev.} \textbf{\bibinfo{volume}{109}},
  \bibinfo{pages}{227} (\bibinfo{year}{1958}).

\bibitem[{\citenamefont{Lanat\`a et~al.}(2013)\citenamefont{Lanat\`a, Strand,
  Giovannetti, Hellsing, de' Medici, and Capone}}]{lanata13}
\bibinfo{author}{\bibfnamefont{N.}~\bibnamefont{Lanat\`a}},
  \bibinfo{author}{\bibfnamefont{H.~U.~R.} \bibnamefont{Strand}},
  \bibinfo{author}{\bibfnamefont{G.}~\bibnamefont{Giovannetti}},
  \bibinfo{author}{\bibfnamefont{B.}~\bibnamefont{Hellsing}},
  \bibinfo{author}{\bibfnamefont{L.}~\bibnamefont{de' Medici}},
  \bibnamefont{and} \bibinfo{author}{\bibfnamefont{M.}~\bibnamefont{Capone}},
  \bibinfo{journal}{Phys. Rev. B} \textbf{\bibinfo{volume}{87}},
  \bibinfo{pages}{045122} (\bibinfo{year}{2013}).

\bibitem[{\citenamefont{Lanat\`a et~al.}(2015)\citenamefont{Lanat\`a, Yao,
  Wang, Ho, and Kotliar}}]{lanata15}
\bibinfo{author}{\bibfnamefont{N.}~\bibnamefont{Lanat\`a}},
  \bibinfo{author}{\bibfnamefont{Y.}~\bibnamefont{Yao}},
  \bibinfo{author}{\bibfnamefont{C.-Z.} \bibnamefont{Wang}},
  \bibinfo{author}{\bibfnamefont{K.-M.} \bibnamefont{Ho}}, \bibnamefont{and}
  \bibinfo{author}{\bibfnamefont{G.}~\bibnamefont{Kotliar}},
  \bibinfo{journal}{Phys. Rev. X} \textbf{\bibinfo{volume}{5}},
  \bibinfo{pages}{011008} (\bibinfo{year}{2015}).

\bibitem[{Note4()}]{Note4}
Note4, \bibinfo{note}{in the following, the indeces $i,j$ denote different
  basis matrices and should not be confused with lattice-site indeces.}

\bibitem[{\citenamefont{Hj{\o}rungnes}(2011)}]{hjor11}
\bibinfo{author}{\bibfnamefont{A.}~\bibnamefont{Hj{\o}rungnes}},
  \emph{\bibinfo{title}{Complex-Valued Matrix Derivatives}}
  (\bibinfo{publisher}{Cambridge University Press}, \bibinfo{year}{2011}).

\bibitem[{\citenamefont{Dennis~Jr. and Schnabel}(1996)}]{denns83}
\bibinfo{author}{\bibfnamefont{J.~R.} \bibnamefont{Dennis~Jr.}}
  \bibnamefont{and} \bibinfo{author}{\bibfnamefont{R.~B.}
  \bibnamefont{Schnabel}}, \emph{\bibinfo{title}{Numerical Methods for
  Unconstraint Optimization and Nonlinear Equations}}
  (\bibinfo{publisher}{Society for Industrial and Applied Mathematics},
  \bibinfo{year}{1996}).

\bibitem[{\citenamefont{Meyer et~al.}(unpublished)\citenamefont{Meyer,
  Els\"{a}sser, Lechermann, and F\"{a}hnle}}]{mbpp_code}
\bibinfo{author}{\bibfnamefont{B.}~\bibnamefont{Meyer}},
  \bibinfo{author}{\bibfnamefont{C.}~\bibnamefont{Els\"{a}sser}},
  \bibinfo{author}{\bibfnamefont{F.}~\bibnamefont{Lechermann}},
  \bibnamefont{and}
  \bibinfo{author}{\bibfnamefont{M.}~\bibnamefont{F\"{a}hnle}},
  \emph{\bibinfo{title}{FORTRAN 90 Program for Mixed-Basis-Pseudopotential
  Calculations for Crystals}}, \bibinfo{organization}{Max-Planck-Institut
  f\"{u}r Metallforschung, Stuttgart} (\bibinfo{year}{unpublished}).

\bibitem[{\citenamefont{Amadon et~al.}(2008)\citenamefont{Amadon, Lechermann,
  Georges, Jollet, Wehling, and Lichtenstein}}]{ama08}
\bibinfo{author}{\bibfnamefont{B.}~\bibnamefont{Amadon}},
  \bibinfo{author}{\bibfnamefont{F.}~\bibnamefont{Lechermann}},
  \bibinfo{author}{\bibfnamefont{A.}~\bibnamefont{Georges}},
  \bibinfo{author}{\bibfnamefont{F.}~\bibnamefont{Jollet}},
  \bibinfo{author}{\bibfnamefont{T.~O.} \bibnamefont{Wehling}},
  \bibnamefont{and} \bibinfo{author}{\bibfnamefont{A.~I.}
  \bibnamefont{Lichtenstein}}, \bibinfo{journal}{Phys. Rev. B}
  \textbf{\bibinfo{volume}{77}}, \bibinfo{pages}{205112}
  (\bibinfo{year}{2008}).

\bibitem[{\citenamefont{Grieger et~al.}(2012)\citenamefont{Grieger, Piefke,
  Peil, and Lechermann}}]{gri12}
\bibinfo{author}{\bibfnamefont{D.}~\bibnamefont{Grieger}},
  \bibinfo{author}{\bibfnamefont{C.}~\bibnamefont{Piefke}},
  \bibinfo{author}{\bibfnamefont{O.~E.} \bibnamefont{Peil}}, \bibnamefont{and}
  \bibinfo{author}{\bibfnamefont{F.}~\bibnamefont{Lechermann}},
  \bibinfo{journal}{Phys. Rev. B} \textbf{\bibinfo{volume}{86}},
  \bibinfo{pages}{155121} (\bibinfo{year}{2012}).

\bibitem[{\citenamefont{Anisimov et~al.}(1993)\citenamefont{Anisimov, Solovyev,
  Korotin, Czy$\dot{\text{z}}$yk, and Sawatzky}}]{ani93}
\bibinfo{author}{\bibfnamefont{V.~I.} \bibnamefont{Anisimov}},
  \bibinfo{author}{\bibfnamefont{I.~V.} \bibnamefont{Solovyev}},
  \bibinfo{author}{\bibfnamefont{M.~A.} \bibnamefont{Korotin}},
  \bibinfo{author}{\bibfnamefont{M.~T.} \bibnamefont{Czy$\dot{\text{z}}$yk}},
  \bibnamefont{and} \bibinfo{author}{\bibfnamefont{G.~A.}
  \bibnamefont{Sawatzky}}, \bibinfo{journal}{Phys. Rev. B}
  \textbf{\bibinfo{volume}{48}}, \bibinfo{pages}{16929} (\bibinfo{year}{1993}).

\bibitem[{\citenamefont{Wang et~al.}(2008)\citenamefont{Wang, Dai, and
  Fang}}]{wan08}
\bibinfo{author}{\bibfnamefont{G.-T.} \bibnamefont{Wang}},
  \bibinfo{author}{\bibfnamefont{X.}~\bibnamefont{Dai}}, \bibnamefont{and}
  \bibinfo{author}{\bibfnamefont{Z.}~\bibnamefont{Fang}},
  \bibinfo{journal}{Phys. Rev. Lett.} \textbf{\bibinfo{volume}{101}},
  \bibinfo{pages}{066403} (\bibinfo{year}{2008}).

\bibitem[{\citenamefont{Borghi et~al.}(2014)\citenamefont{Borghi, Fabrizio, and
  Tosatti}}]{bor14}
\bibinfo{author}{\bibfnamefont{G.}~\bibnamefont{Borghi}},
  \bibinfo{author}{\bibfnamefont{M.}~\bibnamefont{Fabrizio}}, \bibnamefont{and}
  \bibinfo{author}{\bibfnamefont{E.}~\bibnamefont{Tosatti}},
  \bibinfo{journal}{Phys. Rev. B} \textbf{\bibinfo{volume}{90}},
  \bibinfo{pages}{125102} (\bibinfo{year}{2014}).

\bibitem[{\citenamefont{Ho et~al.}(2008)\citenamefont{Ho, Schmalian, and
  Wang}}]{ho08}
\bibinfo{author}{\bibfnamefont{K.~M.} \bibnamefont{Ho}},
  \bibinfo{author}{\bibfnamefont{J.}~\bibnamefont{Schmalian}},
  \bibnamefont{and} \bibinfo{author}{\bibfnamefont{C.~Z.} \bibnamefont{Wang}},
  \bibinfo{journal}{Phys. Rev. B} \textbf{\bibinfo{volume}{77}},
  \bibinfo{pages}{073101} (\bibinfo{year}{2008}).

\bibitem[{\citenamefont{Fr\'{e}sard and Lamboley}(2002)}]{fre02}
\bibinfo{author}{\bibfnamefont{R.}~\bibnamefont{Fr\'{e}sard}} \bibnamefont{and}
  \bibinfo{author}{\bibfnamefont{M.}~\bibnamefont{Lamboley}},
  \bibinfo{journal}{J. Low Temp. Phys} \textbf{\bibinfo{volume}{126}},
  \bibinfo{pages}{1091} (\bibinfo{year}{2002}).

\bibitem[{\citenamefont{Pavarini et~al.}(2005)\citenamefont{Pavarini, Yamasaki,
  Nuss, and Andersen}}]{pav05}
\bibinfo{author}{\bibfnamefont{E.}~\bibnamefont{Pavarini}},
  \bibinfo{author}{\bibfnamefont{A.}~\bibnamefont{Yamasaki}},
  \bibinfo{author}{\bibfnamefont{J.}~\bibnamefont{Nuss}}, \bibnamefont{and}
  \bibinfo{author}{\bibfnamefont{O.~K.} \bibnamefont{Andersen}},
  \bibinfo{journal}{New J. Phys.} \textbf{\bibinfo{volume}{7}},
  \bibinfo{pages}{188} (\bibinfo{year}{2005}).

\bibitem[{\citenamefont{Jackeli and Khaliullin}(2009)}]{jac09}
\bibinfo{author}{\bibfnamefont{G.}~\bibnamefont{Jackeli}} \bibnamefont{and}
  \bibinfo{author}{\bibfnamefont{G.}~\bibnamefont{Khaliullin}},
  \bibinfo{journal}{Phys. Rev. Lett.} \textbf{\bibinfo{volume}{102}},
  \bibinfo{pages}{017205} (\bibinfo{year}{2009}).

\bibitem[{\citenamefont{Pesin and Balents}(2010)}]{pes10}
\bibinfo{author}{\bibfnamefont{D.}~\bibnamefont{Pesin}} \bibnamefont{and}
  \bibinfo{author}{\bibfnamefont{L.}~\bibnamefont{Balents}},
  \bibinfo{journal}{Nature Phys.} \textbf{\bibinfo{volume}{6}},
  \bibinfo{pages}{376} (\bibinfo{year}{2010}).

\bibitem[{\citenamefont{Watanabe et~al.}(2010)\citenamefont{Watanabe,
  Shirakawa, and Yunoki}}]{wat10}
\bibinfo{author}{\bibfnamefont{H.}~\bibnamefont{Watanabe}},
  \bibinfo{author}{\bibfnamefont{T.}~\bibnamefont{Shirakawa}},
  \bibnamefont{and} \bibinfo{author}{\bibfnamefont{S.}~\bibnamefont{Yunoki}},
  \bibinfo{journal}{Phys. Rev. Lett.} \textbf{\bibinfo{volume}{105}},
  \bibinfo{pages}{216410} (\bibinfo{year}{2010}).

\bibitem[{\citenamefont{Martins et~al.}(2011)\citenamefont{Martins, Aichhorn,
  Vaugier, and Biermann}}]{mar11}
\bibinfo{author}{\bibfnamefont{C.}~\bibnamefont{Martins}},
  \bibinfo{author}{\bibfnamefont{M.}~\bibnamefont{Aichhorn}},
  \bibinfo{author}{\bibfnamefont{L.}~\bibnamefont{Vaugier}}, \bibnamefont{and}
  \bibinfo{author}{\bibfnamefont{S.}~\bibnamefont{Biermann}},
  \bibinfo{journal}{Phys. Rev. Lett.} \textbf{\bibinfo{volume}{107}},
  \bibinfo{pages}{266404} (\bibinfo{year}{2011}).

\bibitem[{\citenamefont{Borisenko et~al.}(2016)\citenamefont{Borisenko,
  Evtushinsky, Liu, Morozov, Kappenberger, Wurmehl, B{\"u}chner, Yaresko, Kim,
  Hoesch et~al.}}]{bor16}
\bibinfo{author}{\bibfnamefont{S.~V.} \bibnamefont{Borisenko}},
  \bibinfo{author}{\bibfnamefont{D.~V.} \bibnamefont{Evtushinsky}},
  \bibinfo{author}{\bibfnamefont{Z.-H.} \bibnamefont{Liu}},
  \bibinfo{author}{\bibfnamefont{I.}~\bibnamefont{Morozov}},
  \bibinfo{author}{\bibfnamefont{R.}~\bibnamefont{Kappenberger}},
  \bibinfo{author}{\bibfnamefont{S.}~\bibnamefont{Wurmehl}},
  \bibinfo{author}{\bibfnamefont{B.}~\bibnamefont{B{\"u}chner}},
  \bibinfo{author}{\bibfnamefont{A.~N.} \bibnamefont{Yaresko}},
  \bibinfo{author}{\bibfnamefont{T.~K.} \bibnamefont{Kim}},
  \bibinfo{author}{\bibfnamefont{M.}~\bibnamefont{Hoesch}},
  \bibnamefont{et~al.}, \bibinfo{journal}{Nat. Phys.}
  \textbf{\bibinfo{volume}{12}}, \bibinfo{pages}{311} (\bibinfo{year}{2016}).

\bibitem[{\citenamefont{Kim et~al.}(2008)\citenamefont{Kim, Jin, Moon, Kim,
  Park, Leem, Yu, Noh, Kim, andd J.-H.~Park et~al.}}]{kim08}
\bibinfo{author}{\bibfnamefont{B.~J.} \bibnamefont{Kim}},
  \bibinfo{author}{\bibfnamefont{H.}~\bibnamefont{Jin}},
  \bibinfo{author}{\bibfnamefont{S.~J.} \bibnamefont{Moon}},
  \bibinfo{author}{\bibfnamefont{J.-Y.} \bibnamefont{Kim}},
  \bibinfo{author}{\bibfnamefont{B.-G.} \bibnamefont{Park}},
  \bibinfo{author}{\bibfnamefont{C.~S.} \bibnamefont{Leem}},
  \bibinfo{author}{\bibfnamefont{J.}~\bibnamefont{Yu}},
  \bibinfo{author}{\bibfnamefont{T.~W.} \bibnamefont{Noh}},
  \bibinfo{author}{\bibfnamefont{C.}~\bibnamefont{Kim}},
  \bibinfo{author}{\bibfnamefont{S.-J.~O.} \bibnamefont{andd J.-H.~Park}},
  \bibnamefont{et~al.}, \bibinfo{journal}{Phys. Rev. Lett.}
  \textbf{\bibinfo{volume}{101}}, \bibinfo{pages}{076402}
  (\bibinfo{year}{2008}).

\bibitem[{\citenamefont{Aichhorn et~al.}(2010)\citenamefont{Aichhorn, Biermann,
  Miyake, Georges, and Imada}}]{aic10}
\bibinfo{author}{\bibfnamefont{M.}~\bibnamefont{Aichhorn}},
  \bibinfo{author}{\bibfnamefont{S.}~\bibnamefont{Biermann}},
  \bibinfo{author}{\bibfnamefont{T.}~\bibnamefont{Miyake}},
  \bibinfo{author}{\bibfnamefont{A.}~\bibnamefont{Georges}}, \bibnamefont{and}
  \bibinfo{author}{\bibfnamefont{M.}~\bibnamefont{Imada}},
  \bibinfo{journal}{Phys. Rev. B} \textbf{\bibinfo{volume}{82}},
  \bibinfo{pages}{064504} (\bibinfo{year}{2010}).

\bibitem[{\citenamefont{Yin et~al.}(2012)\citenamefont{Yin, Haule, and
  Kotliar}}]{hau12}
\bibinfo{author}{\bibfnamefont{Z.~P.} \bibnamefont{Yin}},
  \bibinfo{author}{\bibfnamefont{K.}~\bibnamefont{Haule}}, \bibnamefont{and}
  \bibinfo{author}{\bibfnamefont{G.}~\bibnamefont{Kotliar}},
  \bibinfo{journal}{Phys. Rev. B} \textbf{\bibinfo{volume}{86}},
  \bibinfo{pages}{195141} (\bibinfo{year}{2012}).

\bibitem[{\citenamefont{Glasbrenner et~al.}(2015)\citenamefont{Glasbrenner,
  Mazin, Jeschke, Hirschfeld, Fernandes, and Valent{\'i}}}]{gla15}
\bibinfo{author}{\bibfnamefont{J.~K.} \bibnamefont{Glasbrenner}},
  \bibinfo{author}{\bibfnamefont{I.~I.} \bibnamefont{Mazin}},
  \bibinfo{author}{\bibfnamefont{H.~O.} \bibnamefont{Jeschke}},
  \bibinfo{author}{\bibfnamefont{P.~J.} \bibnamefont{Hirschfeld}},
  \bibinfo{author}{\bibfnamefont{R.~M.} \bibnamefont{Fernandes}},
  \bibnamefont{and}
  \bibinfo{author}{\bibfnamefont{R.}~\bibnamefont{Valent{\'i}}},
  \bibinfo{journal}{Nat. Phys.} \textbf{\bibinfo{volume}{11}},
  \bibinfo{pages}{953} (\bibinfo{year}{2015}).

\bibitem[{\citenamefont{Leonov et~al.}(2015)\citenamefont{Leonov, Skornyakov,
  Anisimov, and Vollhardt}}]{leo15}
\bibinfo{author}{\bibfnamefont{I.}~\bibnamefont{Leonov}},
  \bibinfo{author}{\bibfnamefont{S.~L.} \bibnamefont{Skornyakov}},
  \bibinfo{author}{\bibfnamefont{V.~I.} \bibnamefont{Anisimov}},
  \bibnamefont{and}
  \bibinfo{author}{\bibfnamefont{D.}~\bibnamefont{Vollhardt}},
  \bibinfo{journal}{Phys. Rev. Lett.} \textbf{\bibinfo{volume}{115}},
  \bibinfo{pages}{106402} (\bibinfo{year}{2015}).

\bibitem[{\citenamefont{Wang et~al.}(2010)\citenamefont{Wang, Qian, Dai, and
  Fang}}]{wan10}
\bibinfo{author}{\bibfnamefont{G.-T.} \bibnamefont{Wang}},
  \bibinfo{author}{\bibfnamefont{Y.}~\bibnamefont{Qian}},
  \bibinfo{author}{\bibfnamefont{X.}~\bibnamefont{Dai}}, \bibnamefont{and}
  \bibinfo{author}{\bibfnamefont{Z.}~\bibnamefont{Fang}},
  \bibinfo{journal}{Phys. Rev. Lett.} \textbf{\bibinfo{volume}{104}},
  \bibinfo{pages}{047002} (\bibinfo{year}{2010}).

\bibitem[{\citenamefont{Schickling et~al.}(2012)\citenamefont{Schickling,
  Gebhard, B{\"u}nemann, Boeri, Andersen, and Weber}}]{schi12}
\bibinfo{author}{\bibfnamefont{T.}~\bibnamefont{Schickling}},
  \bibinfo{author}{\bibfnamefont{F.}~\bibnamefont{Gebhard}},
  \bibinfo{author}{\bibfnamefont{J.}~\bibnamefont{B{\"u}nemann}},
  \bibinfo{author}{\bibfnamefont{L.}~\bibnamefont{Boeri}},
  \bibinfo{author}{\bibfnamefont{O.~K.} \bibnamefont{Andersen}},
  \bibnamefont{and} \bibinfo{author}{\bibfnamefont{W.}~\bibnamefont{Weber}},
  \bibinfo{journal}{Phys. Rev. Lett.} \textbf{\bibinfo{volume}{108}},
  \bibinfo{pages}{036406} (\bibinfo{year}{2012}).

\bibitem[{\citenamefont{Margadonna et~al.}(2008)\citenamefont{Margadonna,
  Takabayashi, McDonald, Kasperkiewicz, Mizuguchi, Takano, Fitch, Suard, and
  Prassides}}]{mar08}
\bibinfo{author}{\bibfnamefont{S.}~\bibnamefont{Margadonna}},
  \bibinfo{author}{\bibfnamefont{Y.}~\bibnamefont{Takabayashi}},
  \bibinfo{author}{\bibfnamefont{M.~T.} \bibnamefont{McDonald}},
  \bibinfo{author}{\bibfnamefont{K.}~\bibnamefont{Kasperkiewicz}},
  \bibinfo{author}{\bibfnamefont{Y.}~\bibnamefont{Mizuguchi}},
  \bibinfo{author}{\bibfnamefont{Y.}~\bibnamefont{Takano}},
  \bibinfo{author}{\bibfnamefont{A.~N.} \bibnamefont{Fitch}},
  \bibinfo{author}{\bibfnamefont{E.}~\bibnamefont{Suard}}, \bibnamefont{and}
  \bibinfo{author}{\bibfnamefont{K.}~\bibnamefont{Prassides}},
  \bibinfo{journal}{Chem. Commun.} \textbf{\bibinfo{volume}{43}},
  \bibinfo{pages}{5607} (\bibinfo{year}{2008}).

\bibitem[{\citenamefont{Mizuguchi et~al.}(2009)\citenamefont{Mizuguchi,
  Tomioka, Tsuda, Yamaguchi, and Takano}}]{miz09}
\bibinfo{author}{\bibfnamefont{Y.}~\bibnamefont{Mizuguchi}},
  \bibinfo{author}{\bibfnamefont{F.}~\bibnamefont{Tomioka}},
  \bibinfo{author}{\bibfnamefont{S.}~\bibnamefont{Tsuda}},
  \bibinfo{author}{\bibfnamefont{T.}~\bibnamefont{Yamaguchi}},
  \bibnamefont{and} \bibinfo{author}{\bibfnamefont{Y.}~\bibnamefont{Takano}},
  \bibinfo{journal}{Physica C} \textbf{\bibinfo{volume}{469}},
  \bibinfo{pages}{1027} (\bibinfo{year}{2009}).

\bibitem[{\citenamefont{Aichhorn et~al.}(2009)\citenamefont{Aichhorn,
  Pourovskii, Vildosola, Ferrero, Parcollet, Miyake, Georges, and
  Biermann}}]{aic09}
\bibinfo{author}{\bibfnamefont{M.}~\bibnamefont{Aichhorn}},
  \bibinfo{author}{\bibfnamefont{L.}~\bibnamefont{Pourovskii}},
  \bibinfo{author}{\bibfnamefont{V.}~\bibnamefont{Vildosola}},
  \bibinfo{author}{\bibfnamefont{M.}~\bibnamefont{Ferrero}},
  \bibinfo{author}{\bibfnamefont{O.}~\bibnamefont{Parcollet}},
  \bibinfo{author}{\bibfnamefont{T.}~\bibnamefont{Miyake}},
  \bibinfo{author}{\bibfnamefont{A.}~\bibnamefont{Georges}}, \bibnamefont{and}
  \bibinfo{author}{\bibfnamefont{S.}~\bibnamefont{Biermann}},
  \bibinfo{journal}{Phys. Rev. B} \textbf{\bibinfo{volume}{80}},
  \bibinfo{pages}{085101} (\bibinfo{year}{2009}).

\bibitem[{\citenamefont{Miyake et~al.}(2010)\citenamefont{Miyake, Nakamura,
  Arita, and Imada}}]{miy10}
\bibinfo{author}{\bibfnamefont{T.}~\bibnamefont{Miyake}},
  \bibinfo{author}{\bibfnamefont{K.}~\bibnamefont{Nakamura}},
  \bibinfo{author}{\bibfnamefont{R.}~\bibnamefont{Arita}}, \bibnamefont{and}
  \bibinfo{author}{\bibfnamefont{M.}~\bibnamefont{Imada}}, \bibinfo{journal}{J.
  Phys. Soc. Jpn.} \textbf{\bibinfo{volume}{79}}, \bibinfo{pages}{044705}
  (\bibinfo{year}{2010}).

\bibitem[{\citenamefont{Perdew et~al.}(1996)\citenamefont{Perdew, Burke, and
  Ernzerhof}}]{per96}
\bibinfo{author}{\bibfnamefont{J.~P.} \bibnamefont{Perdew}},
  \bibinfo{author}{\bibfnamefont{K.}~\bibnamefont{Burke}}, \bibnamefont{and}
  \bibinfo{author}{\bibfnamefont{M.}~\bibnamefont{Ernzerhof}},
  \bibinfo{journal}{Phys. Rev. Lett.} \textbf{\bibinfo{volume}{77}},
  \bibinfo{pages}{3865} (\bibinfo{year}{1996}).

\bibitem[{\citenamefont{Subedi et~al.}(2008)\citenamefont{Subedi, Zhang, Singh,
  and Du}}]{sub08}
\bibinfo{author}{\bibfnamefont{A.}~\bibnamefont{Subedi}},
  \bibinfo{author}{\bibfnamefont{L.}~\bibnamefont{Zhang}},
  \bibinfo{author}{\bibfnamefont{D.~J.} \bibnamefont{Singh}}, \bibnamefont{and}
  \bibinfo{author}{\bibfnamefont{M.~H.} \bibnamefont{Du}},
  \bibinfo{journal}{Phys. Rev. B} \textbf{\bibinfo{volume}{78}},
  \bibinfo{pages}{134514} (\bibinfo{year}{2008}).

\bibitem[{\citenamefont{Watson et~al.}(2015)\citenamefont{Watson, Kim,
  Haghighirad, Davies, McCollam, Narayanan, Blake, Chen, Schofield, Hoesch
  et~al.}}]{wat15}
\bibinfo{author}{\bibfnamefont{M.~D.} \bibnamefont{Watson}},
  \bibinfo{author}{\bibfnamefont{T.~K.} \bibnamefont{Kim}},
  \bibinfo{author}{\bibfnamefont{A.~A.} \bibnamefont{Haghighirad}},
  \bibinfo{author}{\bibfnamefont{N.~R.} \bibnamefont{Davies}},
  \bibinfo{author}{\bibfnamefont{A.}~\bibnamefont{McCollam}},
  \bibinfo{author}{\bibfnamefont{A.}~\bibnamefont{Narayanan}},
  \bibinfo{author}{\bibfnamefont{S.~F.} \bibnamefont{Blake}},
  \bibinfo{author}{\bibfnamefont{Y.~L.} \bibnamefont{Chen}},
  \bibinfo{author}{\bibfnamefont{S.~G. A.~J.} \bibnamefont{Schofield}},
  \bibinfo{author}{\bibfnamefont{M.}~\bibnamefont{Hoesch}},
  \bibnamefont{et~al.}, \bibinfo{journal}{Phys. Rev. B}
  \textbf{\bibinfo{volume}{91}}, \bibinfo{pages}{155106}
  (\bibinfo{year}{2015}).

\end{thebibliography}

\end{document}